\documentclass{article}
\usepackage{graphicx,epsfig}
\usepackage{graphicx,subfigure}
\usepackage{arxiv}
\usepackage{amsmath}
\usepackage{amssymb}
\usepackage{titlesec}
\usepackage[utf8]{inputenc} % allow utf-8 input
\usepackage[T1]{fontenc}    % use 8-bit T1 fonts
\usepackage{url}            % simple URL typesetting
\usepackage{appendix}
\usepackage{booktabs}       % professional-quality tables
\usepackage{amsfonts,bm}       % blackboard math symbols
\usepackage{microtype}      % microtypography
\usepackage[mathscr]{euscript}
\usepackage[dvipsnames]{xcolor}
\pdfoutput=1
\usepackage{hyperref}
\hypersetup{
    colorlinks=true,
    linkcolor=black,
    citecolor=black,
    filecolor=black,
    urlcolor=black,
}
\makeatother
\usepackage[numbers, compress]{natbib}
\title{Hydrodynamic instability of shear imposed falling film over a uniformly heated inclined undulated substrate}
\author{
  Md. Mouzakkir Hossain\\
  Department of Mathematics\\
  SRM Institute of Science and Technology\\
  Kattankulathur-603203, Tamil Nadu, India\\
  \texttt{mouzakkir123@gmail.com} \\
\And
  Sukhendu Ghosh\\
  Department of Mathematics\\
  Indian Institute of Technology Jodhpur\\
  Rajasthan-342037, India\\
  \texttt{sukhendu.math@gmail.com} \\
     \And
  Harekrushna Behera*\\
  Department of Mathematics\\
  SRM Institute of Science and Technology\\
  Kattankulathur-603203, Tamil Nadu, India\\
  \texttt{hkb.math@gmail.com} \\
  \And
  G.P. Raja Sekhar\\
  Department of Mathematics\\
  Indian Institute of Technology Kharagpur\\
  Kharagpur-721302, West Bengal, India\\
  \texttt{rajas@iitkgp.ac.in} \\
}

\begin{document}
\maketitle

\begin{abstract}
Linear and weakly nonlinear stability analyses of an externally shear-imposed, gravity-driven falling film over a uniformly heated wavy substrate are studied. The longwave asymptotic expansion technique is utilized to formulate a single nonlinear free surface deflection equation. The linear stability criteria for the onset of instability are derived using the normal mode form in the linearized portion of the surface deformation equation. Linear stability theory reveals that the flow-directed sturdy external shear grows the surface wave instability by increasing the net driving force. On the contrary, the upstream-directed imposed shear may reduce the surface mode instability by restricting the gravity-driving force, which has the consequence of weakening the bulk velocity of the liquid film. However, the surface mode can be stabilized/destabilized by increasing the temperature-dependent density/surface-tension variation. Further, the bottom steepness shows dual behaviour on the surface instability depending upon the wavy wall's portion (uphill/downhill). At the downhill portion, the surface wave becomes more unstable than at the bottom substrate's uphill portion.
Moreover, the multi-scale method is incorporated to obtain the complex Ginzburg-Landau equation in order to study the weakly nonlinear stability, confirming the existence of various flow regions of the liquid film. At any bottom portion (uphill/downhill), the flow-directed external shear expands the super-critical stable zones, which causes an amplification in the nonlinear wave amplitude, and the backflow-directed shear plays a counterproductive role. On the other hand, the super-critical stable region decreases or increases as long as the linear variation of density or surface tension increases with respect to the temperature, whereas the sub-critical unstable region exhibits an inverse trend.

\end{abstract}

% keywords can be removed

\section{Introduction}\label{Introduction}

In the last few decades, the hydrodynamic stability of a falling film has been an important topic for researchers due to its vast application in chemical, mechanical, nuclear engineering, and coating industries \cite{weinstein2004coating}. To improve the transport of momentum, heat, and mass across the solid-liquid and liquid-gas interfaces, the finite-amplitude waves of a liquid film are mainly utilized in devices like vertical tube evaporators, chemical reactors, absorption columns, nuclear emergency cooling systems, and heat exchangers \cite{webb2005enhanced}. Hence, it is necessary to study the proper behaviour of liquid films, which will help to tackle the homogeneous film growth that can be constructed for diverse industrial applications. \citet{benjamin1957wave} and \citet{yih1963stability} initiated the theoretical study of falling film over an inclined plane. They derived the critical Reynolds number $Re_c$ for the onset of primary instability, which was further experimentally verified by \citet{liu1993measurements} and \citet{liu1994solitary}. As on date, vast literature is available that has mainly focused on the liquid film flow instability under various flow set-ups (see, for example, \cite{chang2002complex}, \cite{kalliadasis2003marangoni}, \cite{kalliadasis2003thermocapillary}, \cite{sadiq2008thin}, \cite{mukhopadhyay2011stability}, and  \cite{samanta2011falling}).  

Concurrently, it is well known that the instability of the isothermal and non-isothermal fluid flow over an incline is of particular significance in many usages, such as heat exchangers, condensers, and nuclear reactors. \citet{lin1975stability} discussed the Marangoni instability in the non-uniformly heated falling film.  Since then, several researchers \cite{sreenivasan1978surface, joo1991long, miladinova2002long, scheid2002nonlinear, mukhopadhyay2007nonlinear} have shown their interest in this specific field. \citet{miladinova2002long} considered a liquid falling film over a non-uniformly heated inclined bed. They formulated the nonlinear free surface evolution equation by using the longwave theory and observed the significant thermocapillary impact on the surface wave amplitude and phase speed based on linear and weakly nonlinear theories.
\citet{scheid2002nonlinear} further extended the study of \citet{miladinova2002long} and discussed the nonlinear wave dynamics of falling film down a non-uniformly heated vertical plate. They revealed the dual instability mechanisms of thermocapillarity.
The study of the onset of falling flow instability down a uniformly heated inclined bottom substrate was initiated by \citet{goussis1991surface}. They investigated the instability mechanism of thermocapillary instability by utilizing the Orr-Sommerfeld and linearized energy balance equations \cite{kelly1989mechanism}. Later, \citet{kalliadasis2003thermocapillary} performed longwave instability analysis on the thin falling film over a  uniformly heated plate from small to moderate Reynolds numbers. They assured the emergence of three different modes and observed the strong hydrodynamic mode instability due to the thermocapillary effect. Particular attention has been directed towards a clear understanding of thermocapillary instability of liquid film down a heated bed, as is evident by a large collection of works \cite{kalliadasis2003marangoni, d2010film, sarma2018marangoni, patne2021thermocapillary}.

However, it is a known fact that the variation in temperature significantly affects the fluid characteristics viscosity, density, thermal conductivity, and surface tension. Studies of non-isothermal flows often concentrate on the surface tension gradient with temperature and overlook the variations of other fluid characteristics at the onset of instability. This is owing to the presumption that these small variations would not have a substantial influence (for more details, follow the works of \citet{hwang1988non, reisfeld1990nonlinear, kabova2002downward, usha2005dynamics}, and the citations therein). 
\citet{goussis1991surface} were the first to discuss the viscosity gradient effect on the primary instability of surface mode in a non-isothermal falling film by longwave theory. They found the instability nature of the heating and the stabilizing impact of cooling. \citet{reisfeld1990nonlinear} considered a heated fluid flow system with temperature-dependent viscosity. They used viscous scales and longwave theory to develop the nonlinear deformation equation of film thickness and claimed the attenuation in rupture time of the film owing to viscosity fluctuation. 
\citet{kabova2002downward} investigated the variable viscosity effect on the non-isothermal, steady-state film flow dynamics. 

\citet{pascal2013long} considered gravity-driven film flow over a heated inclined plate, where liquid film properties (such as density, dynamical viscosity, thermal conductivity, and specific heat) change linearly with the temperature. Using linear stability theory, they derived the asymptotic expansion of the critical Reynolds number for the onset of surface mode instability in the limit of small parameter values. Later, \citet{mukhopadhyay2018long} performed the longwave asymptotic expansion technique \cite{benney1966long} to obtain the nonlinear deformation equation for the falling film down a uniformly heated plane. They showed that different flow zones near the onset of linear stability are significantly affected by the fluid property variables. The influence of various liquid properties varying with temperature on the overall stability/instability of the flow system is elaborated in detail in \citet{d2014effects}, \citet{d2016thermosolutal}, \citet{pascal2016thermosolutal}, and the citations therein. 

All the above investigations have mainly insisted on studying the fluid flow system over a flat vertical/inclined bed, even though the fluid film does not flow over a perfectly flat substrate in reality. The structural design of the bottom significantly affects the film flow dynamics. Over the last few decades, fluid flows down a wavy wall has drawn great attention in engineering applications such as two-phase heat exchangers \cite{focke1986flow, shah1988plate} and in absorption columns and distillation trays \cite{de1991mechanics}. In most circumstances, the apparatus is constructed deliberately with a wavy surface, while in other applications, corrugation cannot be avoided easily. So, it is important to understand how the gravity-driven falling liquid film is affected physically by deviations from the ideal condition of a flat inclined surface. From the pioneering work of \citet{pozrikidis1988flow}, who investigated a liquid film flows over a periodic solid substrate, several researchers \cite{tougou1978long, wierschem2005effect, trifonov2007stability, trifonov2007stability, d2010film, mogilevskiy2019stability, mukhopadhyay2020hydrodynamics, mukhopadhyay2022long} had shown their interest in the hydrodynamic instability analysis of a falling film over the wavy bottom profile. \citet{wierschem2005effect} investigated theoretically as well as experimentally a gravity-driven falling fluid flow over an inclined undulated substrate of long steepness compared to the film thickness. They identified a larger critical Reynolds number (to onset the instability) than that for a flat bottom surface within the framework of linear stability analysis. Using integral boundary layer theory, \citet{mogilevskiy2019stability} analyzed a falling liquid film over a weakly wavy bottom and revealed the stability/instability impact of bottom topography depending upon the corrugation period and inclination angle. 
% \citet{d2010film}
Later, \citet{mukhopadhyay2020hydrodynamics} studied the linear and weakly nonlinear stability analysis of thin film flow over a non-uniformly heated inclined wavy bed. Based on the weakly nonlinear stability analysis, they discovered the existence of different flow zones predicted by the complex Ginzburg-Landau equation (CGLE). Their investigation revealed that the Marangoni force develops surface wave instability, and the wavy bottom steepness shows dual behaviour in the gravity mode depending upon the uphill and downhill portions. Recently, \citet{mukhopadhyay2022long} have extended the work of \citet{mukhopadhyay2020hydrodynamics} by investigating the interfacial phase change influence on the falling liquid film over an inclined uniformly heated wavy wall. Utilizing the former technique, they observed that evaporation/condensation reduces/enhances the average film thickness due to mass loss/gain.

Moreover, studies of a shear-imposed falling film model are widely used in many industrial and natural set-ups \cite{craik1966wind, smith1990mechanism, wei2005effect}. \citet{craik1966wind} used both experimental and theoretical approaches on the tangential stress acting on a fluid’s interfacial surface when an external air stream was implemented. He observed that the interfacial instability occurred irrespective of the air stream’s intensity
when the fluid film was thin enough. The identified mechanisms indicated that the fluid layer is not only driven by gravity force but may also experience additional interfacial stress generated by airflow, typically acting in either the downstream or upstream direction. The instability mechanism of a shear-imposed film was rendered by \citet{smith1990mechanism}. They predicted that the competition between kinematic and dynamic waves is mainly responsible for the surface mode instability instigated by the external shear. Later, \citet{samanta2014shear} modeled a shear-imposed falling flow system over an inclined plane to observe the external shear effect on both the linear and nonlinear dynamics of surface waves. \citet{bhat2019linear} imposed the external shear at the contaminated surface of a falling fluid film over an inclined wall and studied the primary instability within the framework of the Orr-Sommerfeld eigenvalue problem. They observed the key role of external shear in controlling the instability mechanism by altering the net driving force when the gravitational influence is very weak. \citet{sani2020effect} used the constant additional shear force as an active mechanism at the surface of the contaminated double-layered film falling over a slippery plate to control the surface and interfacial instability in the flow system. Using normal-mode analysis, they derived the Orr-Sommerfeld boundary value problem and solved it using the numerical technique of Chebyshev collocation. Their results revealed that the downstream-directed external shear stabilizes the unstable surface mode but destabilizes the unstable interface mode, whereas the upstream-directed external shear shows the opposite tendency. Consequently, external shear plays a crucial role in controlling the dynamics and instability mechanisms in open-type flows.

The above facts encourage researchers to analyze the external shear effect on different wave dynamics in various fluid flow models, for which homogeneous film growth can be developed for different industrial usages \cite{hossain2022linear, chattopadhyay2023effect}. Thus, we have been motivated to explore the appearance of surface wave instability for an externally shear-imposed free surface film flow down a uniformly heated undulated bottom. This model has the potential to provide a large testing field for the surface mode. For example, solar energy, wind energy, and unwanted heat from industrial waste. In some scenarios, the apparatus is intentionally constructed with a wavy surface, while in other applications, corrugation is simply unavoidable. For solar energy, utilized in solar refrigeration, solar heat storage, and the transportation of heat or cooling over a large distance, the process of absorption or desorption is widely used. Our primary objective in this work is to identify the influence of wavy bottom topography on the external shear-induced unstable Yih mode \cite{yih1967instability}, and to discuss the corresponding impact of various variables due to fluid property.
To do this, we have implemented the longwave asymptotic theory to obtain the nonlinear deflection equation \cite{benney1966long} of film thickness. The normal mode approach is utilized in the linearized part of the Benney equation to discuss the linear instability mechanism of the liquid film. Additionally, the multiple scale method is implemented to compute the CGLE in order to examine the existence of different flow zones in the considered fluid flow model. This manuscript is laid out as follows: the governing equations and the boundary conditions are formulated in Sec.~\ref{MF}. The nonlinear free surface deformation equation is carried out in Sec.~\ref{NEE}. In Secs.~\ref{LSA} and \ref{WNL}, the linear and weakly nonlinear stability analyses are performed, respectively. The detailed discussion about the results obtained by linear and weakly nonlinear analysis is incorporated in Sec.~\ref{results}. Finally, the key outcomes have been concluded in Sec.~\ref{CON}.

\section{Mathematical Formulation}  \label{MF}
\begin{figure}[ht!]
\begin{center}
\includegraphics[width=12cm]{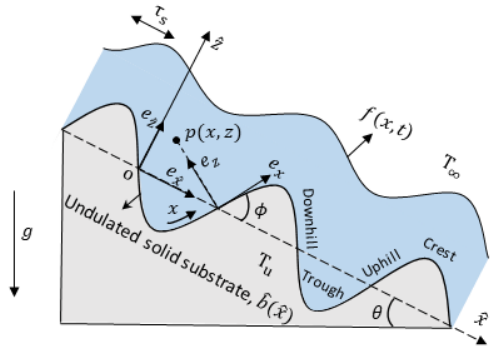}
\end{center}
	\caption{Sketch of Newtonian thin film flows down a uniformly heated wavy substrate. The external shear is imposed at the liquid surface. }\label{f1}
 \end{figure}

A viscous thin Newtonian falling film of thickness $f$ is considered over a wavy bed with an air shear $\tau_s$ imposed at the liquid surface (see, Fig.~\ref{f1}). The wave-like bottom wall is considered to be a good heat conductor. A heat source, such as an electric heating device situated within the undulated bottom, produces heat. The solid substrate is heated in such a way that it is maintained at a constant temperature $T_u$, which is taken to be greater than that of the ambient gas above the
fluid layer, denoted by $T_{\infty}$. Correspondingly, we have assumed that the fluid density $\rho$ and the surface tension $\sigma$ change with temperature linearly as $\rho=\rho_{\infty}-\hat{\alpha}(T-T_{\infty})$ and $\sigma=\sigma_{\infty}-\gamma(T-T_{\infty})$ due to the small temperature variation \cite{pascal2013long} in the liquid film. Here, the parameters $\hat{\alpha}$ and $\gamma$ measure the rate of change with temperature. Further, the terms  $\rho_{\infty}$ and $\sigma_{\infty}$ stand for the value of the fluid's density ($\rho$) and surface tension ($\sigma$) at the reference temperature ($T_{\infty}$), respectively. The other fluid properties are, as discussed in the work of \citet{chattopadhyay2021odd}, assumed to remain fixed. Some of these, such as viscosity, can vary with temperature, but it is only density and surface tension that will realistically depend on the temperature difference. We anticipate that this combined variation will be more pertinent in the context of the current flow problem. 

The Cartesian coordinate system $(e_{\hat{x}},~e_{\hat{z}})$ forms an angle $\theta$ with the horizontal plane, and the periodic bottom profile $\hat{b}(\hat{x})$ is of wavelength $\hat{\lambda}$ and amplitude $\hat{a}$ with fluid flow direction $\hat{x}$. The liquid layer is heated owing to the temperature difference $\Delta T=T_u-T_{\infty}$ between the uniformly heated incline of constant temperature $T_u$ and the air temperature $T_{\infty}$ adjacent to the free surface. The liquid film is supposed to be non-volatile and thin. Consequently, it is possible to avoid the evaporation and buoyancy effect of the liquid. Since the bottom substrate is wavy-type, it is necessary to implement the local curvilinear coordinate system for films
thinner than the radius of curvature of the bottom. Here, $u$ and $w$ are considered streamwise and cross-streamwise velocity components of the fluid flow, respectively. We have taken a local coordinate system $e_x,~e_z$ at arbitrary point $\hat{x}e_{\hat{x}}+\hat{b}({\hat{x}})e_{\hat{z}}$ of the bottom substrate, where $e_x$ is along the tangent and $e_z$ is along the normal to the bottom. 
Therefore, for each point P($x,~z$) in the fluid medium, the curvilinear coordinate $(x,z)$ is defined in the coordinate system $(e_x,e_z)$, where $x$ is the arc length of the bottom and the distance $z$ is along $e_z$ to the bottom. 
% Thus for an arbitrary point P(x,z) in the fluid medium,  the arc length $x$ of the bottom and the distance $z$ along $e_{z}$ to the bottom are taken as the curvilinear coordinate $(x,z)$ in the coordinate system $(e_x,~e_z)$. 
Thus, the arbitrary point becomes $P\equiv(\hat{x}-z~\sin{\phi},\hat{b}(\hat{x})+z~\cos{\phi})$ in the co-ordinate system ($e_{\hat{x}},~e_{\hat{z}}$) (see, Appendix  \ref{coordinate}), where $\phi=\phi(\hat{x})=\arctan (\hat{b}^{'}(\hat{x}))$ (the prime symbol $^{'}$ marks the derivative with respect to the corresponding Cartesian coordinate $\hat{x}$). 
This relation is identical due to the moderately higher steepness of the wavy bottom than the liquid film thickness. Also, the curvature $\kappa$ of the wavy substrate is represented by  $\displaystyle \kappa(\hat{x})=-\frac{\partial^2\hat{b}(\hat{x})}{\partial\hat{x}^2}\bigg[1+\bigg(\frac{\partial\hat{b}(\hat{x})}{\partial\hat{x}}\bigg)^2\bigg]^{-\frac{3}{2}}$. The transformation of the cartesian to the curvilinear coordinate system is properly explained in the work of \citet{wierschem2005effect}.

The dimensional equations of motion \cite{wierschem2005effect,   mukhopadhyay2020hydrodynamics, mukhopadhyay2022long} to describe the shear imposed fluid flow over a uniformly heated wavy bottom substrate are given as follows: 
\allowdisplaybreaks
\begin{align}
& \frac{1}{1+\kappa z}\bigg(u_x+\kappa w\biggr)+w_z=0, \label{e1}\\
&\rho_{\infty}\bigg[u_t+\frac{1}{1+\kappa z}\bigg(u_x+\kappa w\biggr)u+w~u_z\bigg]=-\frac{1}{1+\kappa z} p_x+\rho g\sin{(\theta-\phi)}+\mu\bigg[\frac{1}{(1+\kappa z)^3}\kappa_x\bigg(w-z~u_x\bigg)\nonumber\\
&~~~~~~~~~~~~~+\frac{1}{(1+\kappa z)^2}\bigg(u_{xx}-\kappa^2u+2\kappa w_x\bigg)+\frac{\kappa}{1+\kappa\,z}u_z+u_{zz}\bigg], \label{e2}\\
&\rho_{\infty}\bigg[w_t+\frac{1}{1+\kappa z}\bigg(w_x-\kappa u\biggr)u+w~w_z\bigg]=- p_y-\rho g\cos{(\theta-\phi)}+\mu\bigg[-\frac{1}{(1+\kappa z)^3}\kappa_x\bigg(u+zw_x\bigg)\nonumber\\
&~~~~~~~~~~~~~+\frac{1}{(1+\kappa z)^2}\bigg(w_{xx}-\kappa^2w-2\kappa u_x\bigg)+\frac{\kappa}{1+\kappa\,z}w_z+w_{zz}\bigg], \label{e3}\\
& T_t+\frac{1}{1+\kappa z}uT_x+wT_z=K_c\bigg[\frac{1}{(1+\kappa z)^2}T_{xx}-\frac{1}{(1+\kappa z)^3}z\kappa_xT_x+\frac{\kappa}{1+\kappa z} T_z+T_{zz}\bigg],\label{e4}
\end{align}
where the subscripts in the flow variables ($u,~w,~T,~\text{and}~f$) denote the partial derivatives with respect to the corresponding independent variables. Here, $p$ represents the fluid pressure, $g$ is the gravitational acceleration, $\mu$ is the dynamic viscosity, $T$ denotes the temperature, and $K_c$ signs the thermal diffusivity, which is assumed to be constant. At the free surface $z=f(x,t)$, the kinematic condition is addressed as
\begin{align}
    &w=f_{t}+\frac{1}{1+\kappa f}u~f_{x}.\label{e5}
\end{align}
The dynamic condition suggests continuous stress across the liquid-air interface $z=f(x,t)$. The tangential stress is balanced by the external shear stress at $z=f(x,t)$, which yields \cite{samanta2014effect, hossain2022linear}
\begin{align}
&\frac{\mu}{\sqrt{1+\bigg(\frac{f_x}{1+\kappa f}\bigg)^2}}\bigg[\biggl\{(1+\kappa f)^2-f_x^2\biggr\}\bigg\{\frac{1}{1+\kappa f}(w_x-\kappa u)+u_y\biggr\}+4(1+\kappa f)f_xw_z\bigg]\nonumber\\
&~~~~~~~~~~~~~~~~~~~~~~~~~~~~~~~~~~=\bigg[\tau_s+(1+\kappa~f)\biggl\{\sigma_x+f_x\sigma_z\biggr\}\bigg].\label{e6}
\end{align}
Moreover, because of the capillary force, there exists a normal stress jump across the interface $z=f(x,t)$ with the curvature of the free surface \cite{joo1991long, mukhopadhyay2022long}, resulting   
\begin{align}
&p_{\infty}-p+\frac{2\mu}{1+\bigg(\frac{f_x}{1+\kappa f}\bigg)^2}\bigg[\frac{1}{(1+\kappa f)^3}(u_x+\kappa w)f_x^2+w_z-\frac{1}{1+\kappa f}\biggl\{\frac{1}{1+\kappa f}(w_x-\kappa u)+u_z\biggr\}f_x\bigg]\nonumber\\
&~~~~~~~~~~~~~=\sigma\frac{(1+\kappa f)f_{xx}-ff_x\kappa_x-\kappa\{(1+\kappa f)^2+2f_x^2\}}{\biggl\{(1+\kappa f)^2+f_x^2\biggr\}^{3/2}}, \label{e7}\
\end{align}
where $p_{\infty}$ is the atmospheric pressure.
The energy balance equation at the free surface $z=f(x,t)$ is of the form
\begin{align}
&\frac{K_T}{\sqrt{1+\bigg(\frac{f_x}{1+\kappa f}\bigg)^2}}\bigg[T_z-\frac{f_x}{(1+\kappa f)^2}T_x\bigg]+K_g(T-T_{\infty})=0,\label{e8}
\end{align}
where $K_T$ is the thermal conductivity of the liquid and the heat transfer coefficient $K_g$ measures
the transport rate of heat between the fluid film and the surrounding air.

Along the liquid-bottom interface $z=0$, there is no velocity slip and no penetration of fluid into the wavy substrate, and the uniformly heated bottom will have
\begin{align}
u=0,\quad w=0, \quad \text{and} \quad T=T_u.\label{e9}
\end{align}

To analyze the fluid flow system, it is necessary to introduce the dimensionless governing equations along with the boundary conditions. The thin film that flows over a flat bottom is used as a reference. Hence, the Nusselt velocity of the mean flow $\displaystyle U_c=\frac{\rho_{\infty}g\sin{\theta} \hat{f}^2}{3\mu}$ with the mean film thickness $\hat{f}$ is taken as the characteristic velocity. The normalized form of the dimensional flow variables is as follows: 
\begin{align}
&\bar{x}=\frac{2\pi}{\hat{\lambda}}x,\quad \bar{z}=\frac{z}{\hat{f}},\quad \bar{f}=\frac{f}{\hat{f}},\quad \bar{u}=\frac{u}{U_c},\quad \bar{w}=\frac{\hat{\lambda}}{2\pi\hat{f}U_c}w,\quad \bar{t}=\frac{2\pi U_c }{\hat{\lambda}}t,\quad \bar{\kappa}=\frac{\hat{\lambda}^2}{4\pi^2\hat{a}}\kappa,\nonumber\\ 
& \bar{T}=\frac{T-T_{\infty}}{\Delta T},\quad \bar{p}=\frac{p}{\rho_\infty U_c^2},\quad \bar{\hat{x}}(\bar{x})=\frac{2\pi}{\hat{\lambda}}\hat{x}(x),\quad \bar{\hat{b}}(\bar{\hat{x}})=\frac{1}{\hat{a}}\hat{b}(\frac{\hat{\lambda}}{2\pi}\bar{\hat{x}}),\quad \bar{\phi}=\arctan(\xi\bar{\hat{b}}_{\bar{\hat{x}}}). \label{e10}
\end{align}
Using the normalized form (Eq.~\eqref{e10}) in the governing Eqs.~\eqref{e1}-\eqref{e9}, we obtain the non-dimensional form of the equations of motion associated with the boundary conditions (after suppressing the over-bar symbols) as follows:  
\begin{align}
&u_x+w_z+\epsilon\xi\kappa\bigg(w+zw_z\bigg)=0,\label{e11} \\
&\epsilon Re\bigg[u_t+uu_x+wu_z\bigg]=-\epsilon Re p_x+3\frac{\sin{(\theta-\phi)}}{\sin{\theta}}(1-K_{\rho}T)+u_{zz}+\epsilon\xi \kappa u_z+ O(\epsilon^2)\label{e12},\\
&-\epsilon\,\xi\,Re\,\kappa\, u^2=-Re p_z-3\frac{\cos{(\theta-\phi)}}{\sin{\theta}}(1-K_{\rho}T)+\epsilon w_{zz}+O(\epsilon^2),\label{e13}\\
&\epsilon Re Pr(T_t+uT_x+wT_z)=T_{zz}+\epsilon\,\xi\,\kappa\, T_z+O(\epsilon^2),\label{e14}\\
& w=f_{t}+u~f_{x}-\epsilon\,\xi\,\kappa\,u\,f\,f_x+O(\epsilon^2)\quad \text{at}\quad z=f(x,t),\label{e15}\\
& u_z+\epsilon\,\xi\,\kappa\,(2fu_z-u)=\tau-\epsilon\, Ma\,(T_x+f_xT_z) +O(\epsilon^2)\quad \text{at}\quad z=f(x,t),\label{e16}\\
& p_{\infty}-p+\frac{2\epsilon}{Re}(w_z+u_zf_x)=\epsilon^2 We~(1-K_{\sigma}T)(f_{xx}-\chi\kappa+\xi^2\kappa^2f)\quad \text{at}\quad z=f(x,t),\label{e17}\\
&T_z+Bi~T+O(\epsilon^2)=0\quad \text{at}\quad z=f(x,t),\label{e18} \\
&u=0,\quad w=0, \quad \mbox{and} \quad T=1\quad \text{at}\quad z=0.\label{e19}
\end{align}
Here, the ratio of inertia to viscous force is denoted by the Reynolds number $\displaystyle Re=\frac{\rho_{\infty}U_c \hat{f}}{\mu}$, the scaled gradients of density and surface tension of the fluid with respect to temperature are represented by $\displaystyle K_{\rho}=\frac{\hat{\alpha}\Delta T}{\rho_{\infty}}$ and $\displaystyle K_{\sigma}=\frac{\gamma \Delta T}{\sigma_{\infty}}$, respectively, dimensionless Prandtl number $\displaystyle Pr=\frac{\mu}{\rho_{\infty} K_c}$ balances the viscosity of the fluid in correlation with the thermal conductivity, the dimensionless parameter $\displaystyle  \tau=\frac{\tau_s \hat{f}}{\mu U_c}$ is the externally imposed shear, and the ratio of surface to viscous force is marked by $\displaystyle  We=\frac{\sigma_{\infty}}{\rho_{\infty}U^2_c \hat{f}}$. Further, $\displaystyle  Bi=\frac{K_g \hat{f}}{K_T}$ is the Biot number, which describes the ratio of the heat transfer resistances in the liquid and at the liquid surface, $\displaystyle  \epsilon=\frac{2 \pi \hat{f}}{\hat{\lambda}}<<1$ is the aspect ratio of the fluid flow and $\displaystyle  \xi=\frac{2\pi \hat{a}}{\hat{\lambda}}$ is the bottom steepness. For most of the liquids, $K_{\rho}$ and $k_{\sigma}$ are positive. 
We have considered, $Re$ and $\tau$ are of $O(1)$, the Weber number $We$ is of $O(1/\epsilon^2)$, the Marangoni number $Ma$ is of $O(1/\epsilon)$, the dimensionless parameter $\displaystyle  \chi=\frac{\xi}{\epsilon}$ is of $O(1)$, and both $K_{\rho}$ and $K_{\sigma}$ are assumed to be of $O(1)$. The externally imposed shear can take positive and negative values. The positive imposed shear $\tau>0$ implies the external shear in the flow direction, while the imposed shear in the back-flow direction can be represented by $\tau<0$. The order of wavy bottom steepness $\xi$ is considered to be $1$ or smaller with the thin-film parameter $\epsilon$. Further, the free surface of the liquid film is anticipated to be thermally insulated, so that the Biot number $Bi$ is expected to be $O(\epsilon^2)$ (i.e., very small). A very small Biot number $Bi$ mainly describes the negligible transition of heat energy from the free surface to the surrounding air medium. As a result, the liquid surface is thermally insulated \cite{pascal2016thermosolutal, mukhopadhyay2018long}.   
\section{\bf{Derivation of nonlinear evolution equation}}\label{NEE}
In this subsection, our main objective is to derive the nonlinear free surface deformation equation of the liquid thickness $f(x,t)$ of the shear-imposed falling film down a uniformly heated undulated substrate. To do this, the longwave approximation method \cite{sadiq2008thin, mukhopadhyay2020waves} is used, where the velocity components, pressure, and temperature can be expanded in powers of small aspect ratio $\epsilon<<1$ as 
\begin{align}
(u,w,p,T)=(u_0,w_0,p_0,T_0)+\epsilon~(u_1,w_1,p_1,T_1)+O(\epsilon^2). \label{e20}
\end{align}
Substituting the above approximation Eq.~\eqref{e20} in the non-dimensional governing Eqs.~\eqref{e11}-\eqref{e19} of the flow system and then solving the zeroth-order equations (see, Appendix \ref{Equations11}), we get  \begin{align}
&u_0=-3\frac{\sin{(\theta-\phi)}}{\sin{\theta}}(1-K_{\rho})\bigg[\frac{z^2}{2}-fz\bigg]+\tau~z,\label{eq21}\\
&w_0=-3\frac{\sin{(\theta-\phi)}}{\sin{\theta}}(1-K_{\rho})f_x\frac{z^2}{2},\label{eq22}\\
&p_0=p_{\infty}-\frac{1}{Re}\bigg[3\frac{\cos{(\theta-\phi)}}{\sin{\theta}}(1-K_{\rho})(z-f)\bigg]-\epsilon^2We(1-K_{\sigma})\bigg[f_{xx}-\chi\kappa+\xi^2\kappa^2f\bigg],\label{eq23}\\
&T_0=1,\\\label{eq233}
&q_0=\int_0^fu_0\,dz= \frac{\sin{(\theta-\phi)}}{\sin{\theta}}(1-K_{\rho})f^3+\frac{1}{2}\tau~f^2.
\end{align}

\subsection{\bf{Base state solution}}
To study the linear instability of the unstable surface mode of laminar-type flow, it is essential to derive the unidirectional parallel flow solution, called the base flow solution. The plane-parallel flow solution is obtained by substituting $f=1$ in the above zeroth-order solutions (Eqs.~\eqref{eq21}-\eqref{eq23}) as 
\begin{align}
&U_b(z)= -3\frac{\sin{(\theta-\phi)}}{\sin{\theta}}(1-K_{\rho})\bigg[\frac{z^2}{2}-z\bigg]+\tau\,z,\quad W_b(z)=0\label{e25},\\
&P_b(z)=p_{\infty}-\frac{1}{Re}\bigg[3\frac{\cos{(\theta-\phi)}}{\sin{\theta}}(1-K_{\rho})(z-1)\bigg]-\epsilon^2We(1-K_{\sigma})\bigg[-\chi\kappa+\xi^2\kappa^2\bigg].\label{e26}
\end{align}
\begin{figure}[ht!]
\begin{center}
\subfigure[]{\includegraphics*[width=7.2cm]{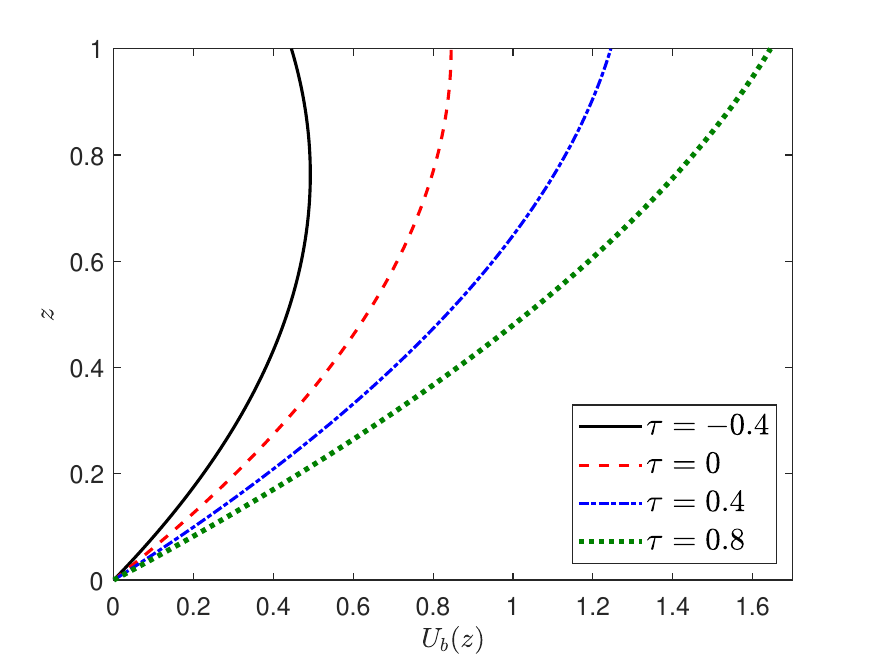}} 
\subfigure[]{\includegraphics*[width=7.2cm]{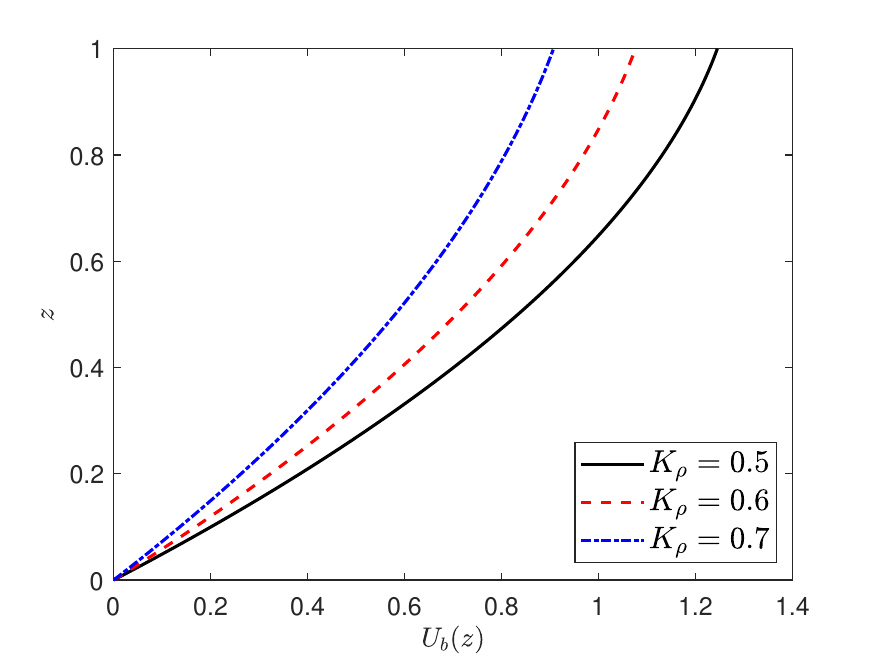}} 
\subfigure[]{\includegraphics*[width=7.2cm]{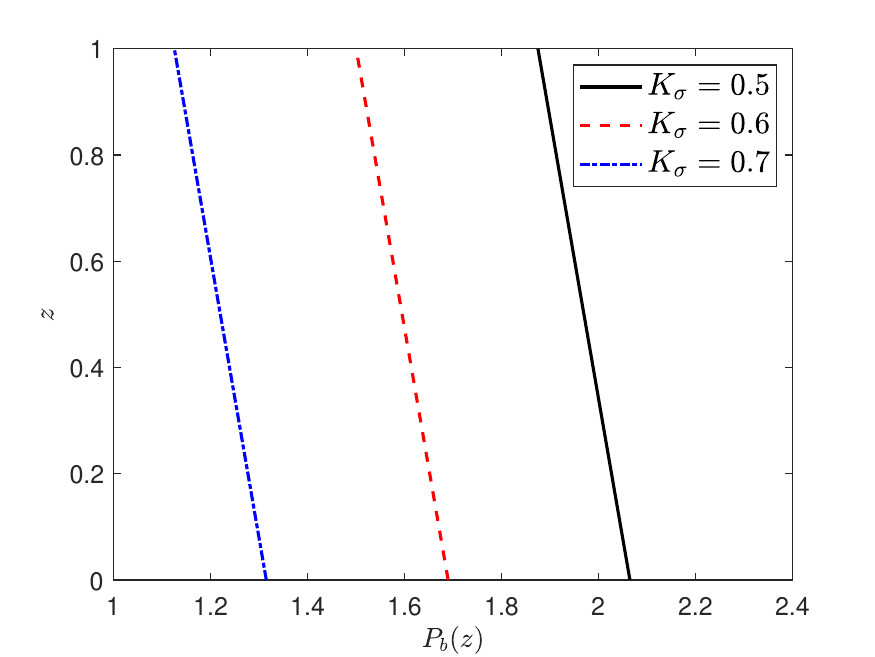}} 
\end{center}
\caption{The effect of (a) external force $\tau$  with $K_{\rho}=0.5$ and (b) the parameter $K_{\rho}$ with $\tau=0.4$ on the base velocity solution. (c) The effect of $K_{\sigma}$ on the base pressure with $K_{\rho}=0.5$, $Re=2$, and $We=450$. The other fixed parameters are $\xi=0.
1\pi$, $x=1.30$ (downhill), $\theta=60^{\circ}$, and $\epsilon=0.1$.}\label{Fig_2}
\end{figure}

It is observed that the semi-parabolic type stream-wise base velocity solution $U_b$ is significantly influenced by the externally imposed shear $\tau$ (Fig.~\ref{Fig_2}(a)) and the parameter $K_{\rho}$ (Figs.~\ref{Fig_2}(b)). It is observed that the flow-directed external shear $\tau$ increases the base velocity, as in Fig.~\ref{Fig_2}(a), and intensifies the base flow rate, whereas the flow suffers retardation due to the imposed shear in the back-flow direction. On the other hand, the parameter $K_{\rho}$ attenuates the base velocity (see, Fig.~\ref{Fig_2}(b)) and weakens the base flow rate of the fluid flow. Further, the base pressure $P_b$ explicitly depends on the parameters $We$  and $K_{\sigma}$ (see, Eq.~\eqref{e26}). It is to be noted that the parameter $K_{\sigma}$ significantly affects the base pressure of liquid film over the wavy bottom, which is not the case for flat bottom \cite{chattopadhyay2021odd}. The base pressure, as shown in Figs.~\ref{Fig_2}(c), is maximum at the wavy bottom boundary ($z=0$) and minimum at the liquid surface ($z=1$). The higher $K_{\sigma}$ value randomly reduces the fluid basic pressure and causes weak hydrostatic pressure in the liquid flow. 

Next, the first-order solutions by solving the first-order equations (see, Appendix \ref{Equations22}) are given by  \allowdisplaybreaks
\begin{align}
&u_1 = Re\frac{\sin{(\theta-\phi)}}{\sin{\theta}}(1-K_{\rho})\bigg[\frac{\sin{(\theta-\phi)}}{\sin{\theta}}(1-K_{\rho})\bigg(\frac{3}{8}fz^4-\frac{3}{2}f^2z^3+3f^4z\bigg)
-\tau\bigg(-\frac{1}{8}z^4+\frac{1}{2}fz^3\nonumber\\
&-f^3z\bigg)\bigg]f_x+Re(\frac{z^2}{2}-fz)\bigg[\frac{3}{Re}\frac{\cos{(\theta-\phi)}}{\sin{\theta}}(1-K_{\rho})f_x-\epsilon^2We(1-K_{\sigma})\bigg(f_{xxx}-\xi\kappa_x+\xi^2\kappa^2f_x\nonumber\\
&+2\xi^2\kappa f \kappa_x\bigg)\bigg]+\xi\kappa\bigg[\frac{\sin{(\theta-\phi)}}{\sin{\theta}}(1-K_{\rho})\bigg(\frac{1}{2}z^3-\frac{3}{2}fz^2+3f^2z\bigg)-\frac{1}{2}\tau\,z^2\bigg],\\
&q_1 = \int_0^f u_1dz= \nonumber\\
&Re\frac{\sin{(\theta-\phi)}}{\sin{\theta}}(1-K_{\rho})\bigg[\frac{6}{5}\frac{\sin{(\theta-\phi)}}{\sin{\theta}}(1-K_{\rho})f+\frac{2}{5}\tau\bigg]f^5f_x\nonumber-\frac{1}{3}Ref^3\bigg[\frac{3}{Re}\frac{\cos{(\theta-\phi)}}{\sin{\theta}}(1-K_{\rho})f_x\nonumber\\  
&-\epsilon^2We(1-K_{\sigma})\bigg(f_{xxx}-\chi\kappa_x+\xi^2\kappa^2f_x+2\xi^2\kappa \kappa_x\,f\bigg)\bigg]+\xi\kappa\,f^3\bigg[\frac{9}{8}\frac{\sin{(\theta-\phi)}}{\sin{\theta}}(1-K_{\rho})f-\frac{1}{6}\tau\bigg].
\end{align}

Now, integrating the continuity Eq.~\eqref{e11} with respect to $z$ from $0$ to $f$ and then using the Leibnitz's rule and the boundary conditions Eqs.~\eqref{e15} and \eqref{e19}, we derive
\begin{align}
    &f_t+(1-\epsilon\xi\kappa f)\bigg[\frac{\partial q_0}{\partial x}+\epsilon\frac{\partial q_1}{\partial x}\bigg]+O(\epsilon^2)=0.\label{e29}
\end{align}
Finally, substituting the expressions for $q_0$ and $q_1$ in the above Eq.~\eqref{e29} and then applying the transformation $(x,t)=\epsilon(\Tilde{x},\Tilde{t})$,  we obtain (after suppressing the $\sim$ symbol) 
\begin{align}
&f_t+a(f)f_x+\bigg[b(f)f_x+c(f)f_{xxx}\bigg]_x+ O(\epsilon^2)=0,\label{e30}
\end{align}  
\begin{align}
&a(f)= \bigg[3\frac{\sin{(\theta-\phi)}}{\sin{\theta}}(1-K_{\rho})f+\tau\bigg]f+\epsilon\bigg[-\xi\kappa f^2\bigg\{3\frac{\sin{(\theta-\phi)}}{\sin{\theta}}(1-K_{\rho})f+\tau\bigg\}\nonumber\\
&~~~~~~+\epsilon^2\, We\,Re\, (1-K_{\sigma}) \bigg\{-\chi+\frac{8}{3}\,\xi^2\,\kappa \,f \bigg\}\kappa_x\,f^2+\frac{1}{6}\,\xi\,\kappa\bigg\{27\frac{\sin{(\theta-\phi)}}{\sin{\theta}}(1-K_{\rho})f-3\tau\bigg\}f^2\bigg],\nonumber\\
& b(f)=\frac{2}{5}Re\frac{\sin{(\theta-\phi)}}{\sin{\theta}}(1-K_{\rho})\bigg[3\frac{\sin{(\theta-\phi)}}{\sin{\theta}}(1-K_{\rho})f+\tau\bigg]f^5-\bigg[\frac{\cos{(\theta-\phi)}}{\sin{\theta}}(1-K_{\rho})\nonumber\\
&~~~~~ -\frac{1}{3}\epsilon^2 We Re (1-K_{\sigma})\xi^2 \kappa^2\bigg]f^3,\nonumber\\
&c(f)=\frac{1}{3}We~Re~(1-K_{\sigma}) f^3,\nonumber
\end{align}
where suffix signifies the derivatives with respect to the corresponding variables and
this nonlinear free surface deflection Eq.~\eqref{e30} is generally named as Benney equation \cite{benney1966long}.

Now, to investigate surface wave instability, the film thickness $f(x,t)$ can be represented as 
\begin{align}
f(x,t)=1+F(x,t), \quad \text{where the deflection part } F(x,t)<<1.\label{e31}
\end{align}

Substituting the above Eq.~\eqref{e31} in the nonlinear deformation Eq.~\eqref{e30} and considering up to $3^{rd}$ order terms, the evolution equation can be rewritten as follows:
 \begin{align}
\nonumber F_t &+ a_1 F_x + b_1 F_{xx} + c_1 F_{xxxx} + a_1^{'} F F_x + b_1^{'}\bigg( F^2_x + F F_{xx} \bigg) + c_1^{'} \bigg( F_x F_{xxx} + F F_{xxxx} \bigg) \\
    & + \frac{1}{2}a_1^{''} F^2 F_x + b_1^{''} \bigg( F F^2_x + \frac{1}{2} F^2 F_{xx} \bigg) + c_1^{''} \bigg( F F_x F_{xxx} + \frac{1}{2} F F_{xxxx} \bigg) + O(F^4) = 0,\label{e32}
\end{align}
where $a_1=a(f=1)$, $b_1=b(f=1)$, $c_1=c(f=1)$. The corresponding derivatives $a_1^{'}=a^{'}(f=1)$, $a_1^{''}=a^{''}(f=1)$, $b_1^{'}=b^{'}(f=1)$, $b_1^{''}=b^{''}(f=1)$, $c_1^{'}=c^{'}(f=1)$, and $c_1^{''}=c^{''}(f=1)$. Here, the prime symbol $'$ denotes the derivative with respect to $f$. 

The unsteady Eq.~\eqref{e32} analyzes the dynamics of infinitesimal disturbances within the film. This can be employed to forecast the time-dependent behaviour of an initially sinusoidal disturbance on a shear-imposed viscous film flowing down a uniformly heated undulated substrate. It is important to highlight that each and every coefficient of the unsteady Eq.~\eqref{e32} conceives the parameters $K_{\sigma}$, $We$, and $Re$, whereas the parameters $\tau$, $K_{\rho}$, and $\xi$ have been involved with the coefficients $a_1$, $a_1^{'}$, $a_1^{''}$, $b_1$, $b_1^{'}$, and $b_1^{''}$. 
This fact ensures that the externally imposed shear, wavy bottom steepness and the coefficients of density and surface tension variations significantly affect the linear as well as the nonlinear stability of the thin film flow problems. Therefore, the objective of the present study is to estimate the combined effects of $\tau$, $K_{\rho}$, $K_{\sigma}$, and $\xi$ in association with $Re$ and $k$ on the linear as well as the nonlinear stability analysis of the shear-imposed liquid film flowing down a uniformly heated undulated substrate.
\section{Linear stability analysis}\label{LSA}
This section mainly focuses on examining the linear response to an infinitesimal disturbance of the liquid surface. 
% The primary goal of this section is to investigate the linear response to an infinitesimal disturbance of the liquid surface. 
To do this, the deformation part $F(x,t)$ of the liquid film is assumed to be of the sinusoidal type, i.e.,
\begin{align}
F(x,t)=\eta~\exp{[i(kx-\omega t)]}+c.c., \label{e33}
\end{align}
where the abbreviation $c.c.$ is for complex conjugate, the real wavenumber $k$ is in the streamwise direction, $\omega=\omega_r+i\omega_i$ represents the complex frequency with oscillating frequency $\omega_r$ and linear growth rate $\omega_i$, and $\eta$ signifies the infinitesimal disturbance amplitude. 

Utilizing the normal mode form Eq.~\eqref{e33} in the linearized part of the nonlinear deformation Eq.~\eqref{e32}, we get the following dispersion relation as
\begin{align} 
    -\mathrm{i} \omega + \mathrm{i} k a_1 + \bigg[ k^4 c_1 - k^2 b_1 \bigg] = 0.\label{e34}
\end{align}
Equating the real and imaginary parts of Eq.~\eqref{e34}, we get
\begin{align} 
&\omega_i =  k^2 \bigg[ b_1-c_1~k^2\bigg]\quad \text{and} \quad \omega_r=a_1\,k.
\end{align}
\begin{figure}[ht!]
\begin{center}
\subfigure[]{\includegraphics*[width=7.2cm]{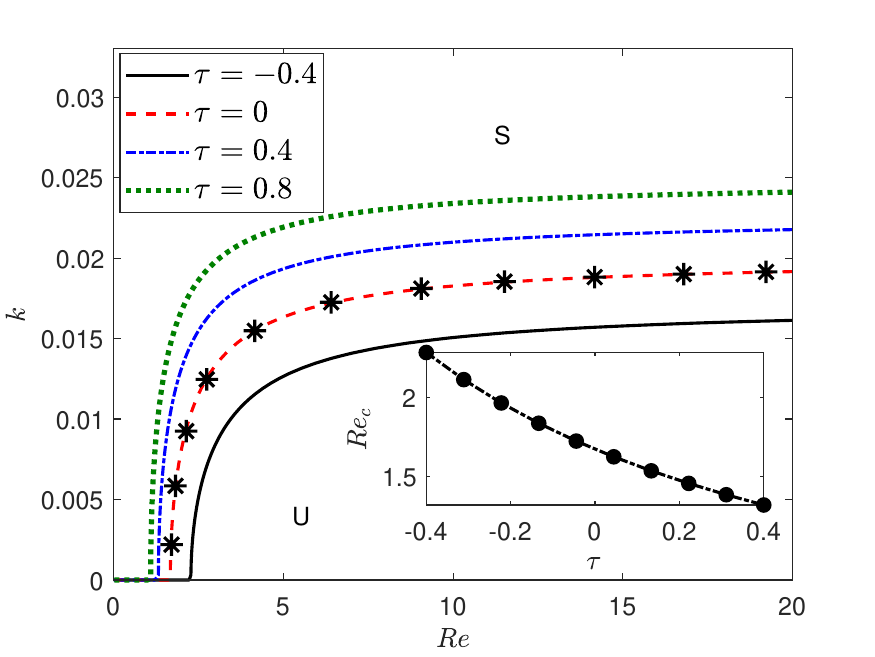}} 
\subfigure[]{\includegraphics*[width=7.2cm]{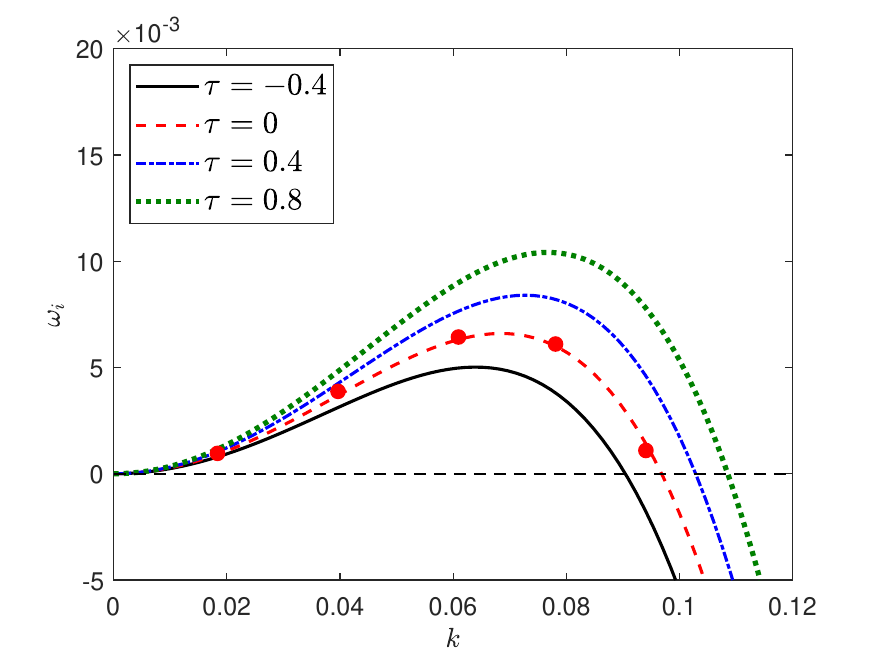}} 
\end{center}
\caption{(a) The impact of varying external shear $\tau$ on the instability boundaries in ($Re-k$) plane when $K_{\rho}=0.5$, $K_{\sigma}=0.5$, $We=4496$, $\xi\rightarrow 0$, $\phi\rightarrow 0$, and $\theta=45^{\circ}$. The result of \citet{chattopadhyay2021odd} (see, Fig.~3(a) of \cite{chattopadhyay2021odd}) is denoted by the black asterisk symbols. (b) The variation in temporal growth rate curves with different external shear $\tau$ when $K_{\rho}=0$, $K_{\sigma}=0$, $We=450$, $Re=2$, $\xi=0.1\pi$, $x=1.30$ (downhill), $\theta=60^{\circ}$, and $\epsilon=0.1$. The red-filled circles mark the outcome of \citet{mukhopadhyay2020hydrodynamics} (see, Fig.~11 of \cite{mukhopadhyay2020hydrodynamics}). }\label{Fig_3}
\end{figure}
The expression for the temporal growth rate of the considered film flow model is given by 
\begin{align}
& \hspace{-1cm}\omega_i=\bigg[ \frac{2}{5}Re\frac{\sin{(\theta-\phi)}}{\sin{\theta}}(1-K_{\rho})\bigg\{3\frac{\sin{(\theta-\phi)}}{\sin{\theta}}(1-K_{\rho})+\tau\bigg\}-\bigg\{\frac{\cos{(\theta-\phi)}}{\sin{\theta}}-\frac{1}{3}\epsilon^2 We~Re~(1-K_{\sigma}) \xi^2\kappa^2\bigg\}\bigg]k^2\nonumber\\
&\hspace{4cm}-\bigg[\frac{1}{3}We\,Re(1-K_{\sigma})\bigg] k^4.\label{e36}    
\end{align}
The film flow is linearly stable if $\omega_i<0$, unstable if $\omega_i>0$, and marginally stable (i.e., neither stable nor unstable) if $\omega_i=0$.
The above Eq.~\eqref{e36} depicts that the parameters $\tau$, $K_{\rho}$, and $K_{\sigma}$ and the bottom steepness $\xi$  significantly affect the temporal growth rate curve of the liquid film over the uniformly heated undulated substrate.

In Fig.~\ref{Fig_3}(a), the marginal stability curves are plotted for different values of $\tau$, when the fixed values are $K_{\rho}=0.5$, $K_{\sigma}=0.5$, $We=4496$, and $\theta=45^{\circ}$, provided $\xi\rightarrow 0$ (i.e., flat inclined plane). The symbols U and S, respectively, mark the unstable and stable regions of the gravity modes. The marginal stability curve for our considered fluid flow model fully recovers the result of the odd-viscosity induced falling film over the uniformly heated inclined plane, studied by \citet{chattopadhyay2021odd} in the absence of $\tau$ and the odd viscosity coefficient. It is found that higher applied shear $\tau$ in the flow direction expands the boundary line of instability by reducing the critical Reynolds number ($Re_c$), which assures the destabilizing behaviour of the positive imposed shear ($+\tau$). The continuous reduction in the critical Reynolds number for the higher $\tau$ value is clearly observed (please refer to the inset of Fig.~\ref{Fig_3}(a)). The sturdy external force in the stream-wise direction enhances the base flow rate, which mainly advances the transition of base flow to the perturbed surface energy.  
On the contrary, the imposed shear $-\tau$ in the back-flow direction shrinks the boundary line of instability and promotes the stabilizing nature of the surface mode. 
On the other hand, the temporal growth rate curves for different imposed shears are demonstrated in Fig.~\ref{Fig_3}(b) when the fixed parameters are $K_{\rho}=0$, $K_{\sigma}=0$, $We=450$, $Re=2$, $\xi=0.1\pi$, $x=1.30$ (downhill), and $\theta=60^{\circ}$. In the limit $\tau=0$, the growth rate curve well matches with the work of \citet{mukhopadhyay2020hydrodynamics} (when $Mn\rightarrow 0$ in \cite{mukhopadhyay2020hydrodynamics}), where a thin Newtonian falling film is considered over a non-uniformly heated undulated substrate. Here the positive external shear enhances the maximum temporal growth rate, while the opposite situation is observed for the negative imposed shear.

Moreover, the phase velocity $c_r=\frac{\omega_r}{k}=a_1$ assures the non-dispersive nature (i.e., independent of $k$) of the surface wave. The condition for the onset of instability can be acquired by the linear growth rate $\omega_i>0$, which yields
\begin{align}
    Re >\frac{\frac{\cos{(\theta-\phi)}}{\sin{\theta}}(1-K_{\rho})-Bo(1-K_{\sigma}) \xi^2\kappa^2}{\frac{2}{5}\frac{\sin{(\theta-\phi)}}{\sin{\theta}}(1-K_{\rho})\bigg(3\frac{\sin{(\theta-\phi)}}{\sin{\theta}}(1-K_{\rho})+\tau\bigg)},\label{e37} 
\end{align}
where the inverse Bond number $Bo=\frac{1}{3}\,\beta\,Re$ with $\beta=\epsilon^2 We$. Note that the critical Reynolds number $Re_c$ of the flow system reduces to $Re_c=\frac{5}{6}\cot{\theta}$, which was derived by \citet{yih1963stability} and \citet{ benjamin1957wave} when the fluid flows over the non-heated inclined plane (i.e., $\xi,~K_{\rho},~\text{and}~K_{\sigma}\rightarrow0$) with negligible external shear $\tau$ at the liquid surface. Also, the $Re_c$ with the limit $K_{\rho},\, K_{\sigma},\,\text{and}\, \tau\rightarrow0$ matches well with the result obtained by \citet{mukhopadhyay2020hydrodynamics} subject to the limit $Mn\rightarrow0$.  
Furthermore, when $\tau$ is absent, the cut-off value of Reynolds number ($Re_c$)  for the flat inclined plane ($\xi\rightarrow0$) becomes identical with the odd viscosity-induced falling film
with variable density studied by \citet{chattopadhyay2021odd} in the absence of the odd viscosity coefficient in their work.

For moderately small wavy bottom steepness, we can write
\begin{align}
    & \kappa=\kappa_0+\xi^2\kappa_2+O(\xi^4)\quad\text{and}\quad\phi=\xi\phi_1+O(\xi^3),\quad~\text{where} \quad \phi_1=\frac{\partial\hat{b}(x)}{\partial\hat{x}}, \quad \kappa_0=-\frac{\partial^2\hat{b}(x)}{\partial\hat{x}^2},\nonumber
\end{align}
and
\begin{align}
    &\frac{\sin{(\theta-\phi)}}{\sin{\theta}}=1-\xi\phi_1\cot{\theta}-\frac{1}{2}\xi^2\phi_1^2+O(\xi^3)
    ,\nonumber\\
    &\frac{\cos{(\theta-\phi)}}{\sin{\theta}}= \cot{\theta}+\xi\phi_1-\frac{1}{2}\xi^2\phi_1^2\cot{\theta}+O(\xi^3).\nonumber
\end{align}
It turns out that the bottom curvature $\kappa$ does not include the first-order terms, whereas the local inclination angle contains leading order $\xi$. The detailed expansion form is elaborated in the work of \citet{hacker2009integral}.
% By substituting the above expansion form in the Eq.~\eqref{e37} results
% \begin{align}
%     &Re_c=\frac{5}{6}\bigg[\cot{\theta}+\xi\bigg(1+2\cot^2{\theta}\bigg)\phi_1+\xi^2\bigg(\frac{5}{2}+3\cot^2{\theta}\bigg)\cot{\theta}\phi^2_1-Bo \kappa_0^2\bigg]\label{crit}
% \end{align}
Also, the cut-off wavenumber below which the fluid flow becomes unstable is derived as   
\begin{align}
&k_c=\sqrt{\frac{b_1}{c_1}},\quad\text{where the growth rate curve ($\omega_i$) vanishes.} \label{ee25} 
\end{align}

The maximum growth rate $\omega_{i,m}$ at $\displaystyle k_m=\frac{k_c}{\sqrt{2}}$ can be computed by $\displaystyle \frac{d\omega_i}{dk}=0$ and the fluid flow becomes linearly unstable for $Re>Re_c$ and $0<k<k_c$. 

\section{\bf{Weakly nonlinear stability analysis}}\label{WNL}
The linear stability analysis can only explore the linear response of the infinitesimal perturbation of the surface waves and is unable to examine the finite-amplitude perturbation. We have observed from linear stability analysis that the marginal line ($k=k_c$, where $\omega_i=0$) creates the boundary line between the stable and unstable zones of the fluid flow. Hence, it would expect that a thin band of width $\delta<<1$ of unstable modes appears in the vicinity of the neutral curve $k=k_c$ over a time $\delta^{-2}$ and over a distance $\delta^{-1}$ so that $\omega_i\sim O(\delta^2)$ \cite{lin1974finite}. Consequently, the unsteady part $F=\eta\exp{[i(kx-\omega\,t)]}+c.c.=\eta\exp{[ik\{x-(\omega_r+i\omega_i)\,t\}]}+c.c.$ reduces to $\eta~\exp{[i(kx-\omega_r t)]}+c.c.$, since $\exp{(\omega_i t)}$ is very small and is absorbed in the deflection amplitude $\eta$. This ensures that the linear stability analysis is not enough to study the actual behaviour of the film flow. Thus, we have used the weakly nonlinear stability analysis \cite{debnath2005nonlinear, sadiq2008thin, mukhopadhyay2020hydrodynamics, hossain2022shear, chattopadhyay2022dynamics}, which can predict the correct behaviour of finite amplitude disturbances of the surface waves. The weakly nonlinear stability analysis can anticipate whether a finite amplitude disturbance can produce instability in the linear stable zone (i.e., sub-critical instability), as well as whether the nonlinear growth of the surface wave perturbation of finite amplitude can cause a new equilibrium state in the linear unstable zone (i.e., super-critical stability) or perturbation enhancing toward instability.

To study weakly nonlinear waves, the  CGLE (i.e., the equation of nonlinear disturbance amplitude $\eta(x_1,t_1,t_2)$) is derived using the method of multiple scales \cite{debnath2005asymptotic} as
\begin{align}
&\eta_{t_2}+ 2ik(b_1-2c_1k^2)\eta_{x_1}+\biggl(b_1-6k^2c_1\biggr)\eta_{x_1x_1}-\delta^{-2}\omega_i\,\eta+\biggl(\Upsilon_2+i \,\Upsilon_4\biggr) \eta^2\bar{\eta}=0,  \label{e46}
\end{align}
where the slow independent variables $x_1,~t_1,~\rm{and}~t_2$ are dependent on the fast independent variables $x~\rm{and}~t$ by the relation \begin{align}
    t_1=\delta t,~~t_2=\delta^2 t,~~~\text{and}~~~x_1=\delta x. \label{e39}
\end{align}
Here, $\delta<<1$ stands for the smallness of the corresponding independent variables.
Further, 
\begin{align}
&\Upsilon_2=\biggl(7k^4G_r\,c_1^{'}-k^2G_rb_1^{'}-\frac{1}{2}k^2b_1^{''}+\frac{1}{2}k^4c_1^{''}\biggr)-k\,G_i\,a_1^{'},\label{e47}
\\
&\Upsilon_4=\biggl(7k^4G_i\,c_1^{'}-k^2G_ib_1^{'}\biggr)+k\,G_r\,a_1^{'}+\frac{1}{2}ka_1^{''}, \label{e48}\\
&G_r=\frac{\biggl(b_1^{'}-k^2c_1^{'}\biggr)}{2\biggl(-b_1+4k^2c_1\biggr)},\quad \text{and}\quad
G_i=\frac{a_1^{'}}{4\,k\,\biggl(b_1-4k^2c_1\biggr)}. \label{e49}
\end{align}
The Eq.~\eqref{e46} mainly characterizes the weak nonlinearity of the externally shear-imposed thin film flows over a uniformly heated wavy bottom.
Further, we have assumed that the wave is filtered, for which there is no spatial modulation. As a result, the diffusion term in Eq.~\eqref{e46} vanishes and the solution of the reduced equation, Eq.~\eqref{e46} can be repressed as
\begin{align}
 \eta=\Lambda_0(t_2)\exp{\biggl[-i\,\phi(t_2)t_2}\biggr], \label{e51}  
\end{align}
which yields
\begin{align}
&\frac{\partial\Lambda_0}{\partial t_2}=\biggr(\delta^{-2}\omega_i-\Upsilon_2\Lambda^2_0\biggr)\Lambda_0, \label{e52}
\\
&\frac{\partial\biggl(\phi(t_2)\,t_2\biggr)}{\partial t_2}=\Upsilon_4\,\Lambda^2_0. \label{e53}
\end{align}
The Eq.~\eqref{e52} is called the Landau equation \cite{pleiner29}, where the Landau coefficient $\Upsilon_2$ provides the nonlinear nature of the flow system. If $\Upsilon_2=0$, the Eq.~\eqref{e52} reduces to the linear equation of the filtered wave amplitude, which grows or decays accordingly as $\omega_i>0$ or $\omega_i<0$. In fact, $\Upsilon_2>0$ assures the saturation of
the perturbed amplitude, and the opposite trend (that means the saturation does not occur for $\Upsilon_2<0$) holds if the value $\Upsilon_2$ is negative. 
This mathematical fact indicates that the flow system is stable for $\Upsilon_2>0$ and unstable for $\Upsilon_2<0$. 
The Eq.~\eqref{e53} modifies the disturbed wave speed caused by the infinitesimal perturbation emerging in the nonlinear system. The nonlinear wave amplitude and phase velocity are determined by
\begin{align}
&\delta\,\Lambda_0=\sqrt{\frac{\omega_i}{\Upsilon_2}} \quad \mbox{and} \quad \phi(t_2)=\frac{\omega_i\Upsilon_4}{\delta^2\Upsilon_2}. \label{e54}
\end{align}
Substituting $\eta=\Lambda_0(t_2)\exp{\biggl[-i\,\phi(t_2)t_2}\biggr]$ in the Eq.~\eqref{e33}, and using $t_2=\delta^2 t$, the nonlinear wave speed $\mathcal{N}_c$ can be formulated by
\begin{align}
&\mathcal{N}_c=\omega_r+\delta^2\,\phi(t_2)=\omega_r+\omega_i\frac{\Upsilon_4}{\Upsilon_2}. \label{e55}
\end{align}
The signs of $\Upsilon_2$ and $\omega_i$ play a vital role in examining the fluid flow system and indicate four different types of regions: (i) supercritical stable zone, which appears in the linear unstable zone $\omega_i>0$ if $\Upsilon_2>0$, where the linear growth due to perturbation will form a new equilibrium state with a finite amplitude, (ii) supercritical unstable zone in the linear unstable region $\omega_i>0$ when $\Upsilon_2<0$, where both linear and nonlinear disturbances grow continuously and make the system more unstable, (iii) unconditional stable zone ($\omega_i<0$, $\Upsilon_2>0$), where both linear and nonlinear finite amplitude disturbances are unconditionally stable, and (iv) subcritical unstable zone in the linear stable region $\omega_i<0$ when $\Upsilon_2<0$, where a finite amplitude disturbance may generate instability in the linear stable zone.

\section{\bf{Flow over a sinusoidal substrate}}\label{results}
The analysis presented so far considered the fluid flow model with an arbitrary wavy bottom. To understand the instability mechanism of the shear-imposed falling film over the uniformly heated wavy bottom, it is beneficial to select a standard sinusoidal bottom profile 
\begin{align}
    &\hat{b}(\hat{x})=\hat{a}\cos{\frac{2\pi\hat{x}}{\hat{\lambda}}}\label{bottom}
\end{align}
with wave amplitude $\hat{a}$. Throughout the study, the wavelength $\hat{\lambda}=300~mm$ and  $\hat{a}=15~mm$ are considered. The downhill region is taken as $0<\hat{x}<150~mm$, whereas the uphill region is taken as $150mm~<\hat{x}<300~mm$ \cite{mukhopadhyay2020hydrodynamics,mukhopadhyay2022long}, and the crest is $\hat{x}=0$ and trough $\hat{x}=150~mm$. Notably, the bottom steepness is assumed to be moderately small, though it is considered a fixed value rather than a perturbation parameter.   
% Note that the bottom steepness is assumed to be moderately small, yet it enters as a fixed quantity and not as a perturbation parameter.
As both the bottom curvature $\kappa(\hat{x})$ and the local inclination $\phi(\hat{x})$ are functions of $\hat{x}$, the critical Reynolds number $Re_c$ will also be a function of $\hat{x}$ (i.e., $Re_c(\hat{x})$). Our entire investigation is based on the periodic type bottom structure of moderate steepness. The dependence of the surface on the local inclination angle $(\theta-\phi)$ strongly restricts the maximum steepness for a given inclination angle $\theta$ \cite{wierschem2005effect}. After critical numerical observations, $0\leq\xi\leq0.4$ is chosen (the bottom steepness range) for the inclination angle $\theta=60^{\circ}$ to make the
considered fluid flow configurations physically and geometrically consistent.

\subsection{\bf{Result based on linear stability analysis}}
\begin{figure}[ht!]
\begin{center}
\subfigure[]{\includegraphics*[width=7.2cm]{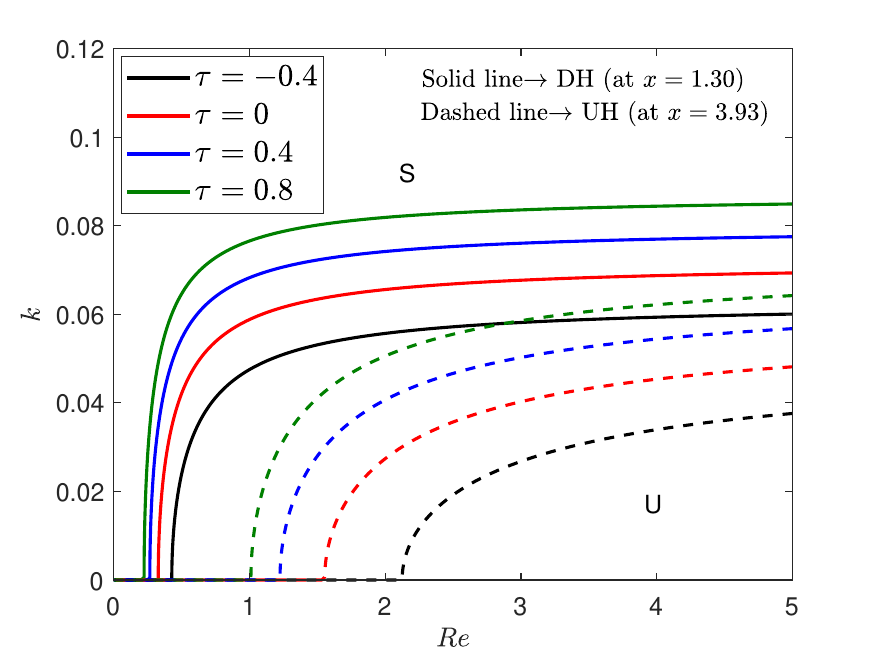}} 
\subfigure[]{\includegraphics*[width=7.2cm]{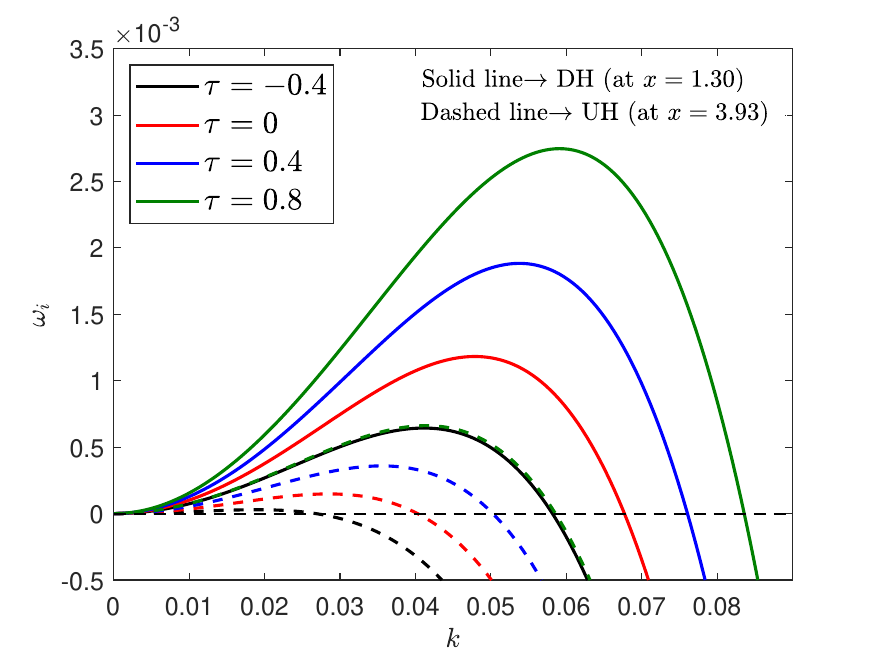}} 
\end{center}
\caption{(a) The impact of varying external shear $\tau$ on the instability boundaries in ($Re-k$) plane and (b) the corresponding growth rate with $Re=3$. The remaining parameters are $K_{\rho}=0.5$, $K_{\sigma}=0.5$, $We=450$, $\xi=0.1\pi$, $\theta=60^{\circ}$, and $\epsilon=0.1$.}\label{Fig_4}
\end{figure}

The core concern of this subsection is to examine the stability/instability behaviour of the infinitesimal perturbation in gravity waves under different fluid characteristic parameters. Fig.~\ref{Fig_4}(a) demonstrates that for all $\tau$ values, the bandwidth of the unstable zone for the bottom's downhill is greater than the uphill of the wavy bottom due to the attenuation in the critical Reynolds number. This fact assures that the liquid film over an uphill portion becomes more stable than the downhill portion of the undulated bottom. Also, for both downhill and uphill portions, the higher external shear in the flow direction promotes surface wave instability by enhancing the unstable bandwidth. This outcome is further strengthened by the corresponding temporal growth rate curves, depicted in Fig.~\ref{Fig_4}(b), where the downstream/upstream-directed potent applied shear force increases/decreases the growth rate of surface mode, no matter whether fluid is over the uphill or downhill portion of the undulated bottom structure.  
\begin{figure}[ht!]
\begin{center}
\subfigure[$\tau=-0.4$]{\includegraphics*[width=5.4cm]{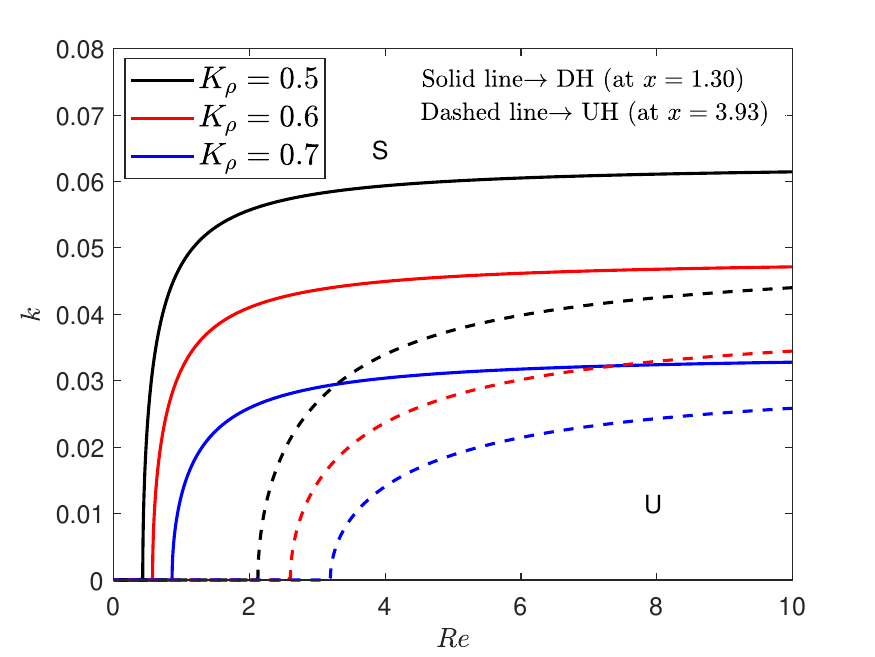}} 
\subfigure[$\tau=0$]{\includegraphics*[width=5.4cm]{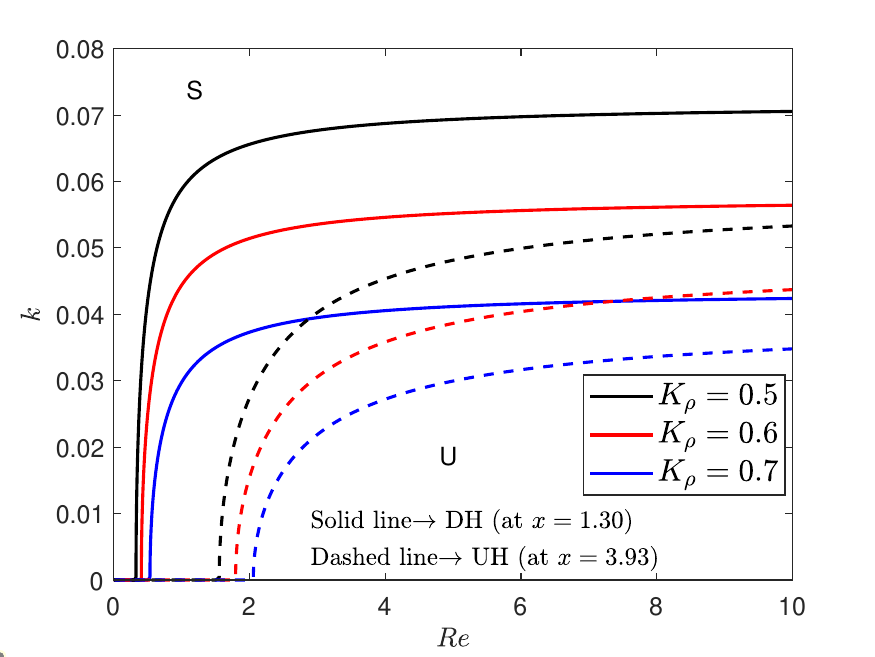}}
\subfigure[$\tau=0.4$]{\includegraphics*[width=5.4cm]{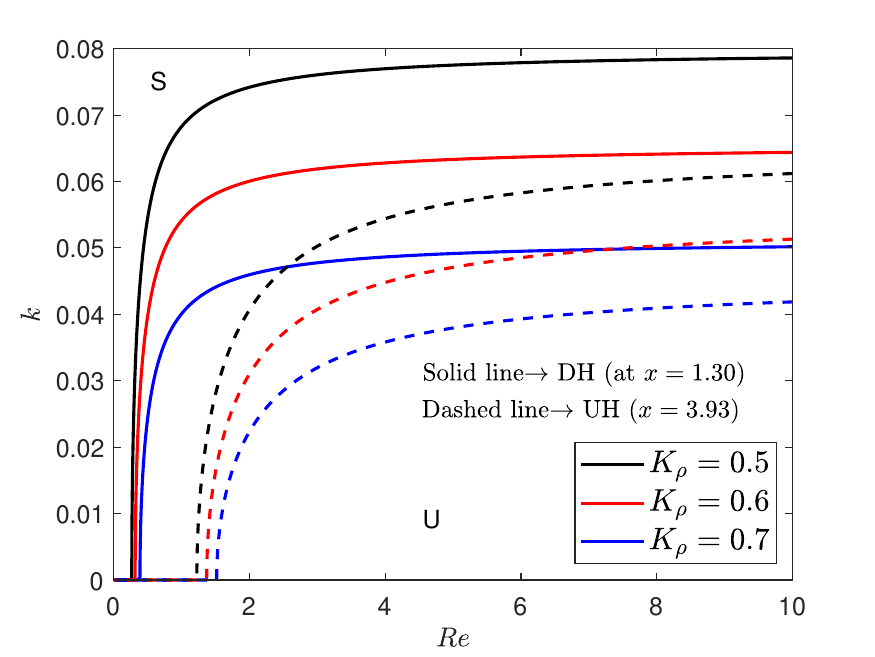}}
\end{center}
\caption{Stability boundaries in ($Re-k$) plane for different $K_{\rho}$ with uphill (UH: $x=3.93$) and downhill (DH: $x=1.30$) when (a) $\tau=-0.4$, (b) $\tau=0$, and (c) $\tau=0.4$. The remaining parameters are $K_{\sigma}=0.5$, $We=450$, $\xi=0.1\pi$, $\theta=60^{\circ}$, and $\epsilon=0.1$. }\label{Fig_5}
\end{figure}

To examine the precise effect of the density gradient $K_{\rho}$ with respect to temperature on the surface mode, the neutral stability curves are displayed in Fig.~\ref{Fig_5} for different $K_{\rho}$ values for both the downhill and uphill portions of the bottom when (a) $\tau=-0.4$, (b) $\tau=0$, and (c) $\tau=0.4$. For each $\tau$ value, the parameter $K_{\rho}$ shrinks the linear unstable region of the fluid flow over both uphill and downhill portions by increasing the critical Reynolds number. The gradual attenuation of the unstable zone at a higher $K_{\rho}$ value yields stabilizing behaviour of the surface mode. As the scaled density gradient $K_{\rho}>0$ increases, the fluid density decreases, which slows the mass flow rate and consequences the dissipation of free surface wave energy. This fact is mainly responsible for the stabilizing nature of the parameter $K_{\rho}$ on the surface wave. Further, it is found that for each $K_
{\rho}$ value, the fluid becomes more stable on the uphill portion as compared to the downhill portion of the inclined wavy bottom. 

\begin{figure}[ht!]
\begin{center}
\subfigure[$\tau=-0.4$]{\includegraphics*[width=5.4cm]{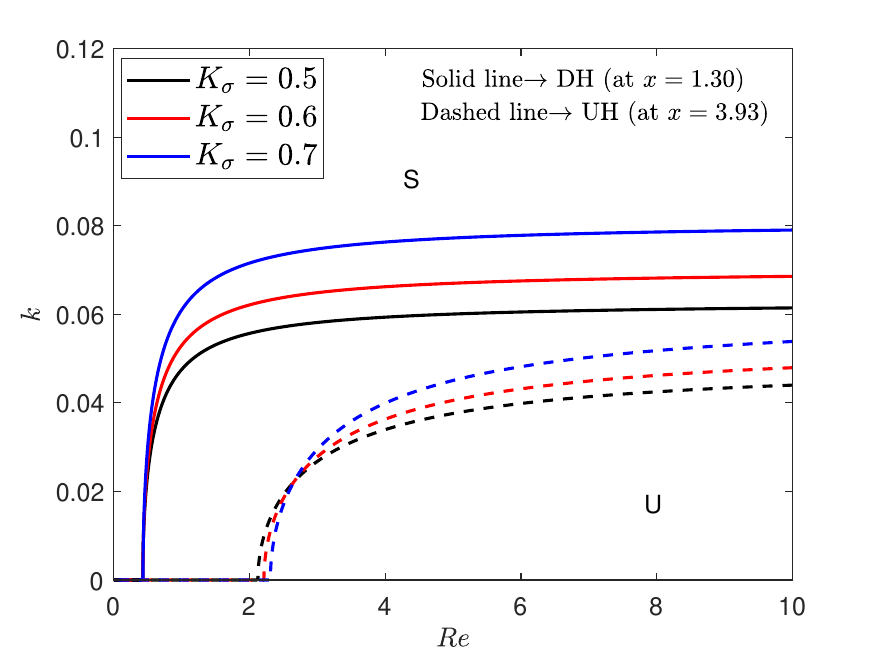}} 
\subfigure[$\tau=0$]{\includegraphics*[width=5.4cm]{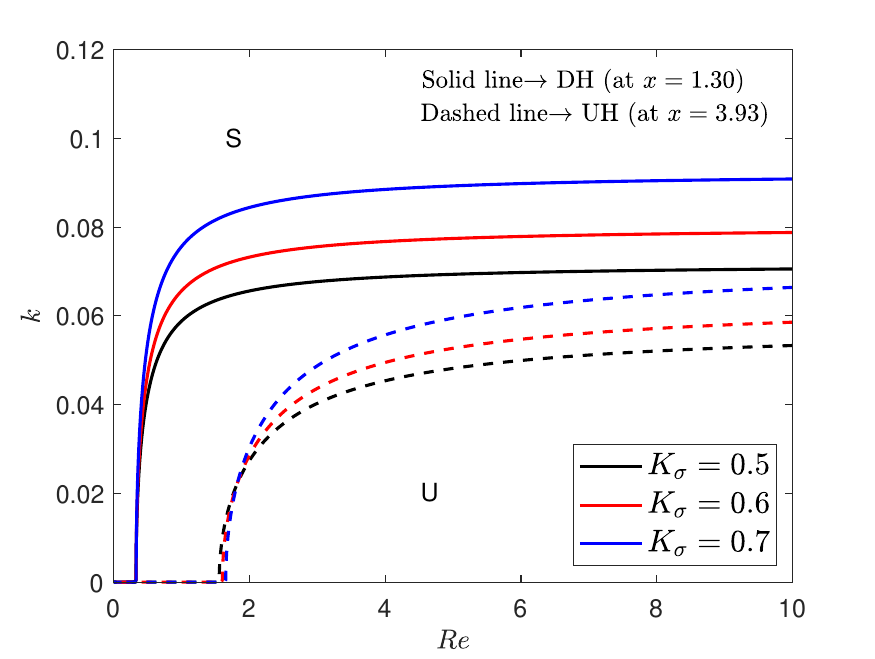}}
\subfigure[$\tau=0.4$]{\includegraphics*[width=5.4cm]{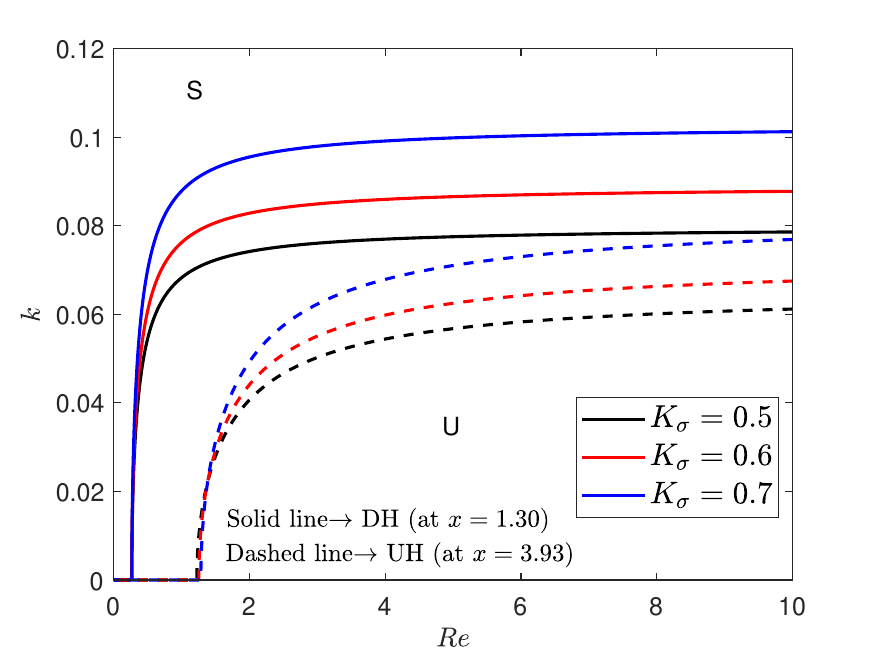}}
\end{center}
\caption{Stability boundaries in ($Re-k$) plane for different $K_{\sigma}$ with uphill (UH: $x=3.93$) and downhill (DH: $x=1.30$) when (a) $\tau=-0.4$, (b) $\tau=0$, and (c) $\tau=0.4$. The remaining parameters are $K_{\rho}=0.5$, $We=450$, $\xi=0.1\pi$, $\theta=60^{\circ}$, and $\epsilon=0.1$. }\label{Fig_6}
\end{figure}

On the other hand, Fig.~\ref{Fig_6} shows the variation of the boundary line of film flow instability over both the downhill and uphill portions with respect to the parameter $K_{\sigma}>0$ (fluid tension gradient per temperature) when the external shear (a) acts in the flow direction ($\tau>0$), (b) is absent ($\tau=0$), and (c) acts along the reverse-flow direction ($\tau<0$). For each $\tau$ value, as $K_{\sigma}$ increases, the unstable zone in the finite wavenumber domain ($k\nrightarrow0$) increases and causes the destabilizing behaviour of the liquid surface. The higher $K_{\sigma}$ weakens the hydrostatic pressure (see, Fig.~\ref{Fig_2}(c)), which increases the flow rate and promotes the destabilization of the liquid surface. In the longwave zone ($k\rightarrow0$), the parameter $K_{\sigma}$ has a negligible effect on the surface mode instability. The temporal growth rate (see, Eq.~\eqref{e36}) of the linear perturbation contains the term $\displaystyle \frac{1}{3}We\,Re\,(1-K_{\sigma})k^4$, which is mainly responsible for the negligible effect in the longwave region.
Besides, an increase in external shear in the downstream direction delays the onset of instability as the $Re_c$ increases, and the size of the unstable region (see, Figs.~\ref{Fig_6}(a) and (b)) shrinks along with the reduction in the range of unstable wavenumbers. Therefore, the downstream-directed imposed shear stabilizes the flow system due to the restriction in the net driving force. In contrast, the opposite trend is observed for the external shear in the upstream direction (see, Figs.~\ref{Fig_6}(b) and (c)).      

\begin{figure}[ht!]
\begin{center}
\subfigure[]{\includegraphics*[width=7.2cm]{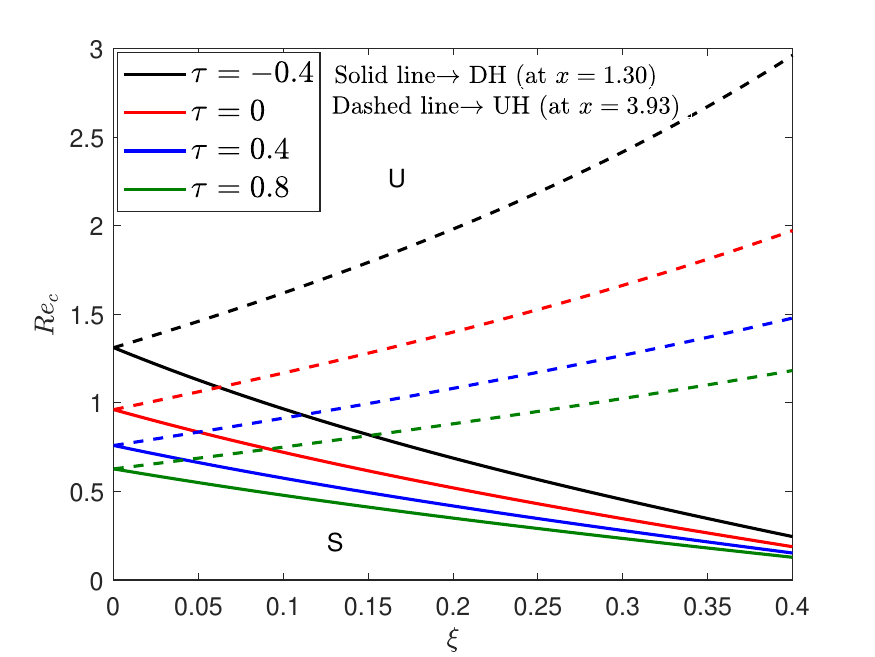}}
\subfigure[]{\includegraphics*[width=7.2cm]{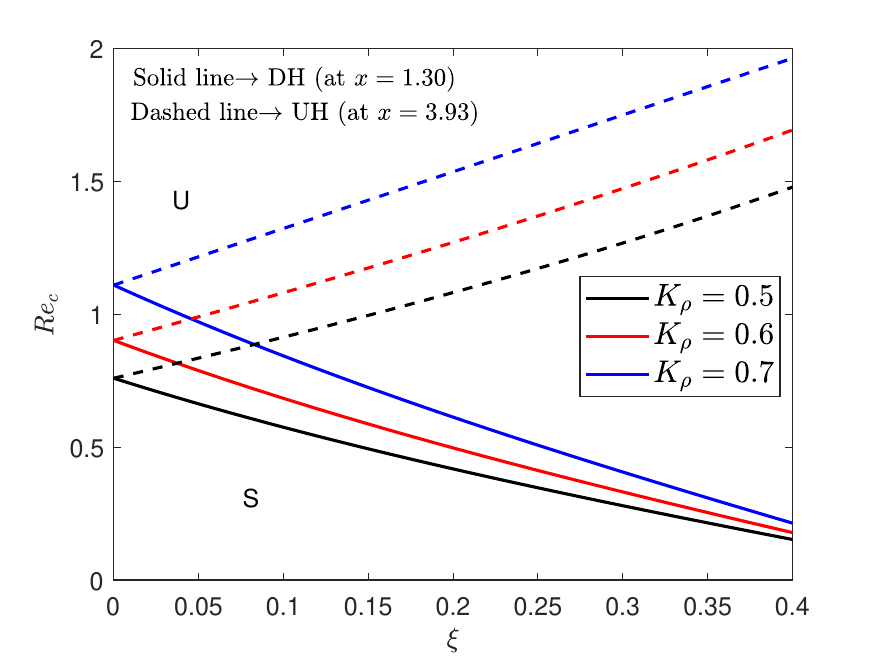}} 
% \subfigure[$\tau=0.4$]{\includegraphics*[width=5.4cm]{FIG/tau_k_sigma_3}}
\end{center}
\caption{(a) The variation of critical Reynolds number $Re_c$ versus the wavy bottom steepness $\xi$ with different $\tau$ for uphill (UH: $x=3.93$) and downhill (DH: $x=1.30$) when $K_{\rho}=0.5$. (b) The variation of critical Reynolds number $Re_c$ versus the wavy bottom steepness $\xi$ with different $K_{\rho}$ for uphill (UH: $x=3.93$) and downhill (DH: $x=1.30$) when $\tau=0.4$. The remaining parameters are $K_{\sigma}=0.5$, $Bo=1$, and $\theta=60^{\circ}$.}\label{Fig_7}
\end{figure}

In Fig.~\ref{Fig_7}, we have demonstrated the critical Reynolds number $Re_c$ for the onset of instability as a function of the wavy bottom steepness $\xi$ with different $\tau$ (Fig.~\ref{Fig_7}(a)) and $K_{\rho}$  (Fig.~\ref{Fig_7}(b)) values in the case of the uphill and downhill portions of the bottom substrate. As in Figs.~\ref{Fig_7}(a) and (b), $ Re_c$ continuously decreases as soon as the bottom steepness $\xi$ increases when the liquid flow approaches the downhill portion of the wavy wall, but in the case of the uphill portion, the function $Re_c$ behaves totally opposite to the downhill portion. Thus, it is evident that when the fluid is on the downhill portion, the wavy bottom steepness enhances the surface wave instability, whereas the bottom steepness dampens the surface wave energy by reducing the instability of the liquid film if it flows over the uphill portion of the wavy bottom. So, the way wall steepness $\xi$ displays a dual impact on linear stability. Further, it is found from Fig.~\ref{Fig_7}(a) that whether the liquid flows over the uphill or downhill of the wavy bed, the external shear ($+\tau$) in the flow direction reduces the critical Reynolds number ($Re_c$), which causes the potent fluid flow instability. On the contrary, the external shear in the opposite direction of the liquid flow reduces the film flow instability by increasing the $Re_c$ value. On the other hand, Fig.~\ref{Fig_7}(b) depicts that for both the uphill and downhill portions of the undulated bottom, $Re_c$ increases as the density gradient $K_{\rho}$ of the fluid with respect to the temperature increases for any bottom steepness and yields the stabilizing nature of the surface wave.

\subsection{\bf{Results based on weakly nonlinear stability analysis}} 
In this section, we have discussed the effect of the physical parameters $\tau$, $K_{\rho}$, and $K_{\sigma}$ on different flow regions in the case of the downhill (DH) and uphill (UH) portions of the wavy bottom with different waviness ($\xi$) based on the weakly nonlinear stability analysis. 
It is evident from Eq.~\eqref{e52} that the positive or negative value of $\omega_i$ and $\Upsilon_2$ mainly provide assurance of the proper instability/stability characteristic of film flow. Note that, the Landau coefficient $\Upsilon_2$ (see, Eq.~\eqref{e47}) has the singularity at $k=k_s=\frac{k_c}{2}(Re,\,We,\,\tau,\,K_{\sigma}, K_{\rho}, \theta)$ obtained by equating the denominator of $G_r$ to be zero. The curve $k=k_s$ bifurcates the linear unstable region ($\omega_i>0$) into two sub-regions: (i) supercritical stable region ($\omega_i>0,\, \Upsilon_2>0$), where the nonlinear waves attain a finite equilibrium amplitude and (ii) supercritical explosive region ($\omega_i>0,\, \Upsilon_2<0$), where the flow attains supercritical
equilibrium. On the other side, the continuous part $\Upsilon_2$ separates the linear stable region ($\omega_i<0$) into two portions: (iii) subcritical unstable region ($\omega_i<0,\, \Upsilon_2<0$), where a finite amplitude disturbances attain instability and (iv) subcritical stable region ($\omega_i<0,\, \Upsilon_2>0$), where finite-amplitude disturbances are unconditionally stable. 
\begin{table}[ht!]
\caption{$k-$ range of flow states for different $Re$ with different external shear $\tau$. $k_{\text{DH}}$ and $k_{\text{UH}}$, respectively, mark the wavenumbers for downhill and uphill. The right boundary of flow zone II is the corresponding singularity $k_s$ of  $\Upsilon_2$ and the left boundary is the critical wavenumber $k_c$, at which $\omega_i=0$. $\Upsilon_2=0$ at left boundary of the flow zone IV.}\label{Table_1}\vspace{0.2cm}
\centering
% \small
\resizebox{16cm}{!}{
\addtolength{\tabcolsep}{5pt}
\begin{tabular}{ c| c| c c c c }
\hline
$\tau$ & \ $Re$  & I & II & III & IV  \\ 
\hline
 & 1 & $0<k_{\text{DH}}<0.0238$ & $0.0238<k_{\text{DH}}<0.0476$ &  $0.0476<k_{\text{DH}}<0.0.1656$ & $0.1656<k_{\text{DH}}\leq0.2$  \\
 %\cline{2-6}
 & & $-$ &$-$  &  $-$ & $0.1406<k_{\text{UH}}\leq0.2$\\
 \cline{2-6} 
 &3 & $0<k_{\text{DH}}<0.0291$& $0.0291<k_{\text{DH}}<0.0582$ & $0.0582<k_{\text{DH}}<0.1167$& $0.1167<k_{\text{DH}}\leq 0.2$ \\
-0.4& &$0<k_{\text{UH}}<0.0134$ & $0.0134<k_{\text{UH}}<0.0268$ &  $0.0268<k_{\text{UH}}<0.0988$ & $0.0988<k_{\text{UH}}\leq0.2$\\ \cline{2-6} 
 & 5  & $0<k_{\text{DH}}<0.0302$ & $0.0302<k_{\text{DH}}<0.0604$ &  $0.0302<k_{\text{DH}}<0.0996$
  & $0.0996<k_{\text{DH}}\leq 0.2$ \\
  & & $0<k_{\text{UH}}<0.0187$ & $0.0187<k_{\text{UH}}<0.0374$ &  $0.0374<k_{\text{UH}}<0.0836$ & $0.0836<k_{\text{UH}}\leq0.2$ \\ \cline{2-6} 
 & 7  & $0<k_{\text{DH}}<0.0304$ & $0.0304<k_{\text{DH}}<0.0608$  & $0.0608<k_{\text{DH}}<0.0899$ & $0.0899<k_{\text{DH}}\leq0.2$\\
 & &$0<k_{\text{UH}}<0.0208$ & $0.0208<k_{\text{UH}}<0.0416$ &  $0.0416<k_{\text{UH}}<0.0747$ & $0.0747<k_{\text{UH}}\leq0.2$\\ \hline
 &1 & $0<k_{\text{DH}}<0.0295$ & $0.0295<k_{\text{DH}}<0.0590$ &$0.0590<k_{\text{DH}}<0.1734$ & $0.1734<k_{\text{DH}}\leq0.2$ \\ 
 & &$-$ &$-$  & $-$  & $0.1512<k_{\text{UH}}\leq0.2$\\
 \cline{2-6} 
 &3  & $0<k_{\text{DH}}<0.0337$ & $0.0337<k_{\text{DH}}<0.0674$  &  $0.0674<k_{\text{DH}}<0.1228$ & $0.1228<k_{\text{DH}}\leq0.2$   \\ 
& & $0<k_{\text{UH}}<0.0201$ & $0.0201<k_{\text{UH}}<0.0402$  &  $0.0402<k_{\text{UH}}<0.1067$ & $0.1067<k_{\text{UH}}\leq0.2$   \\ 
\cline{2-6} 
0 &5  & $0<k_{\text{DH}}<0.0345$  & $0.0345<k_{\text{DH}}<0.0690$ & $0.0690<k_{\text{DH}}<0.1051$ &$0.1051<k_{\text{DH}}\leq0.2$ \\
 & & $0<k_{\text{UH}}<0.0241$ & $0.0241<k_{\text{UH}}<0.0482$  &  $0.0482<k_{\text{UH}}<0.0907$ & $0.0907<k_{\text{UH}}\leq0.2$   \\
 \cline{2-6} 
 &7  & $0<k_{\text{DH}}<0.0349$  & $0.0349<k_{\text{DH}}<0.0698$ & $0.0698<k_{\text{DH}}<0.0951$ &$0.0951<k_{\text{DH}}\leq0.2$   \\ 
 & & $0<k_{\text{UH}}<0.0256$ & $0.0256<k_{\text{UH}}< 0.0512$  &  $0.0512<k_{\text{UH}}<0.0815$ & $0.0815<k_{\text{UH}}\leq0.2$\\ 
 \hline
&1 & $0<k_{\text{DH}}<0.0341$ & $0.0341<k_{\text{DH}}<0.0682$ &$0.0682<k_{\text{DH}}<0.1807$ & $0.1807<k_{\text{DH}}\leq0.2$\\
& &$-$ &$-$ & $-$  & $0.1606<k_{\text{UH}}\leq0.2$\\
\cline{2-6}
& 3 & $0<k_{\text{DH}}<0.0382$ & $0.0382<k_{\text{DH}}<0.0764$  & $0.0764<k_{\text{DH}}<0.1284$ & $0.1284<k_{\text{DH}}\leq0.2$ \\
0.4& & $0<k_{\text{UH}}<0.0251$ & $0.0251<k_{\text{UH}}< 0.0502$  &  $0.0502<k_{\text{UH}}<0.1138$ & $0.1138<k_{\text{UH}}\leq0.2$\\
\cline{2-6} 
&5 & $0<k_{\text{DH}}<0.0388$ & $0.0388<k_{\text{DH}}<0.0776$ &$0.0776<k_{\text{DH}}<0.1103$ & $0.1103<k_{\text{DH}}\leq0.2$  \\ 
& & $0<k_{\text{UH}}<0.0285$ & $0.0285<k_{\text{UH}}< 0.0570$  &  $0.0570<k_{\text{UH}}<0.0971$ & $0.0971<k_{\text{UH}}\leq0.2$\\
\cline{2-6} 
 &7 & $0<k_{\text{DH}}<0.0392$ & $0.0392<k_{\text{DH}}<0.0784$ &$0.0784<k_{\text{DH}}<0.1000$ & $0.1000<k_{\text{DH}}\leq0.2$  \\
 & & $0<k_{\text{UH}}<0.0295$ & $0.0295<k_{\text{UH}}< 0.0590$  &  $0.0590<k_{\text{UH}}<0.0875$ & $0.0875<k_{\text{UH}}\leq0.2$\\
 \hline 
\end{tabular}}
\end{table}

\begin{figure}[ht!]
\begin{center}
\subfigure[$\tau=-0.8$]{\includegraphics*[width=5.4cm]{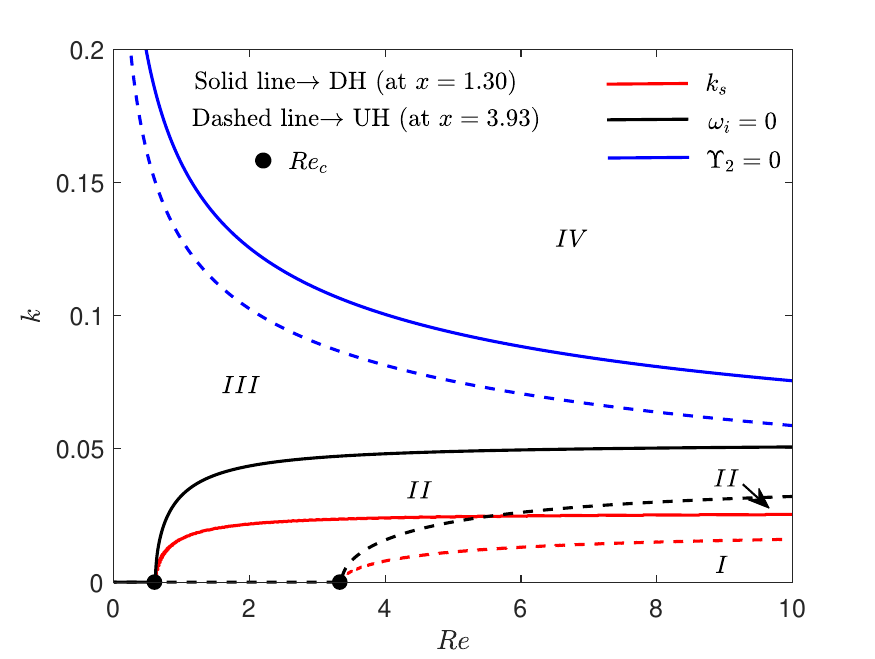}}
\subfigure[$\tau=-0.4$]{\includegraphics*[width=5.4cm]{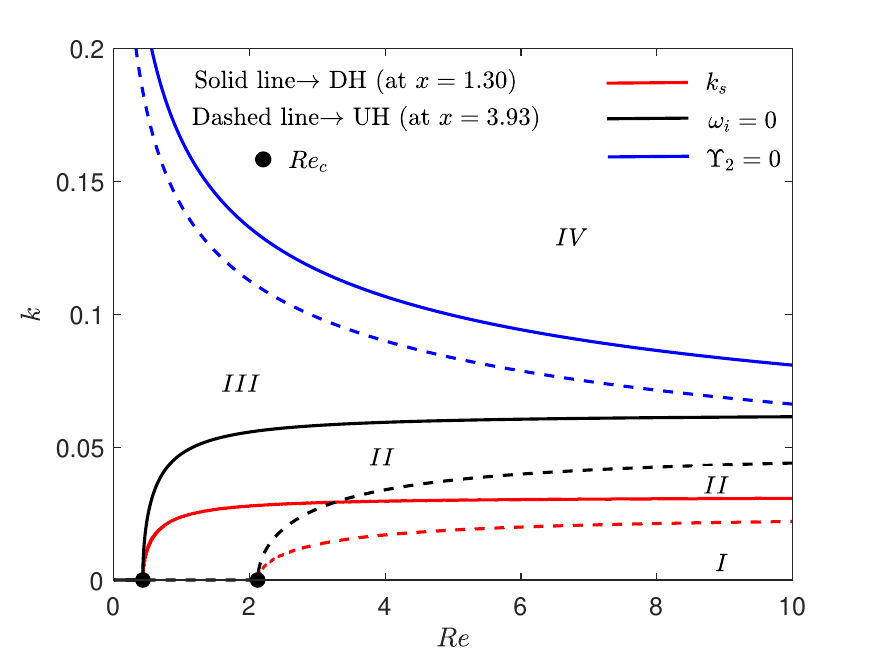}}
\subfigure[$\tau=0$]{\includegraphics*[width=5.4cm]{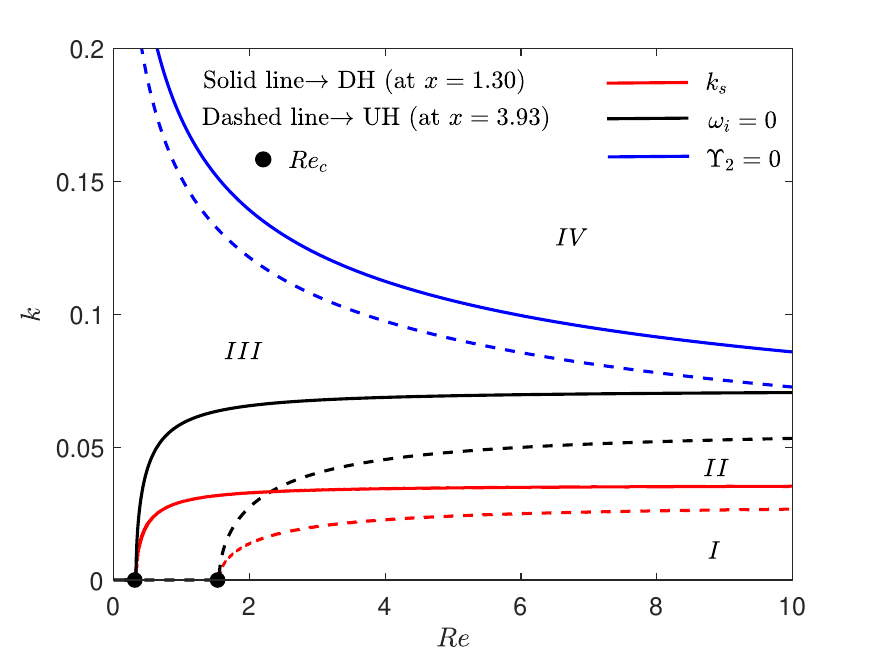}}
\subfigure[$\tau=0.4$]{\includegraphics*[width=5.4cm]{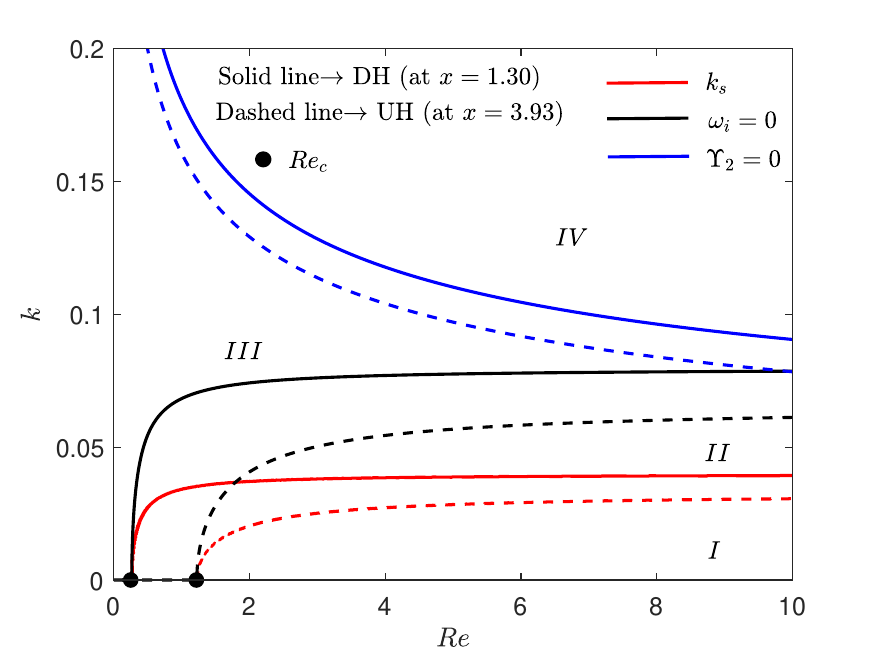}}
\subfigure[$\tau=0.8$]{\includegraphics*[width=5.4cm]{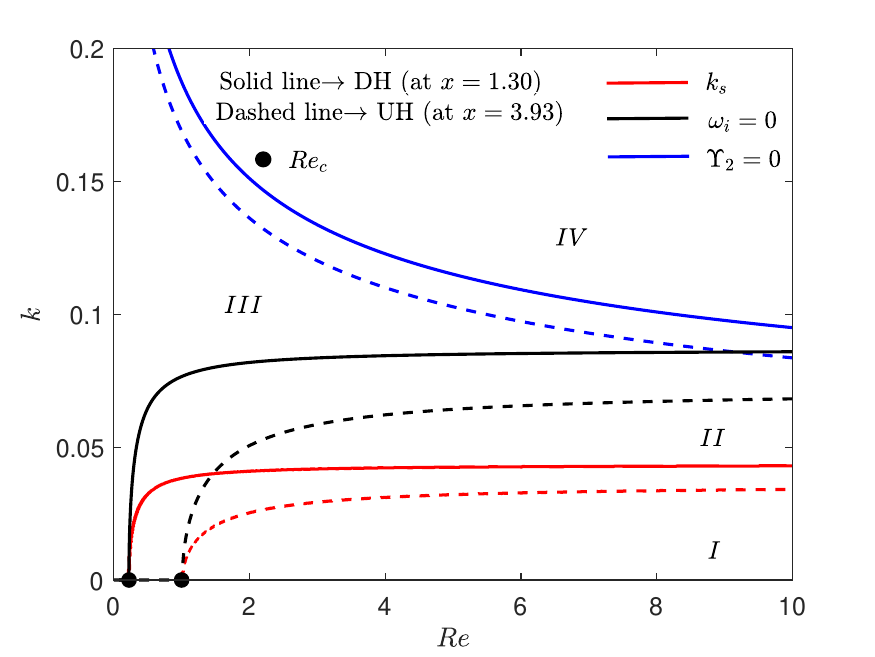}}
\end{center}
\caption{The effect of external shear (a) $\tau=-0.8$, (b) $\tau=-0.4$, (c) $\tau=0$, (d) $\tau=0.4$, and (e) $\tau=0.8$ on different flow regions for uphill (UH: $x=3.93$) and downhill (DH: $x=1.30$) portions. The different flow regions $I: \text{supercritical unstable}~ \omega_i>0,\,\Upsilon_2<0$, $II:\text{supercritical stable}~\omega_i>0,\,\Upsilon_2>0$, $III:\text{subcritical stable}~\omega_i<0,\,\Upsilon_2>0$, and $IV:\text{subcritical unstable}~\omega_i<0,\,\Upsilon_2<0$ at a point on the uphill (UH) and downhill (DH) portions. The other fixed parameters are $\xi=0.1\pi$, $K_{\rho}=0.5$, $K_{\sigma}=0.5$, $We=450$, and $\theta=60^{\circ}$, and $\epsilon=0.1$.}\label{Fig_8}
\end{figure}

\begin{figure}[ht!]
\begin{center}
\subfigure[$K_{\rho}=0.5$]{\includegraphics*[width=5.4cm]{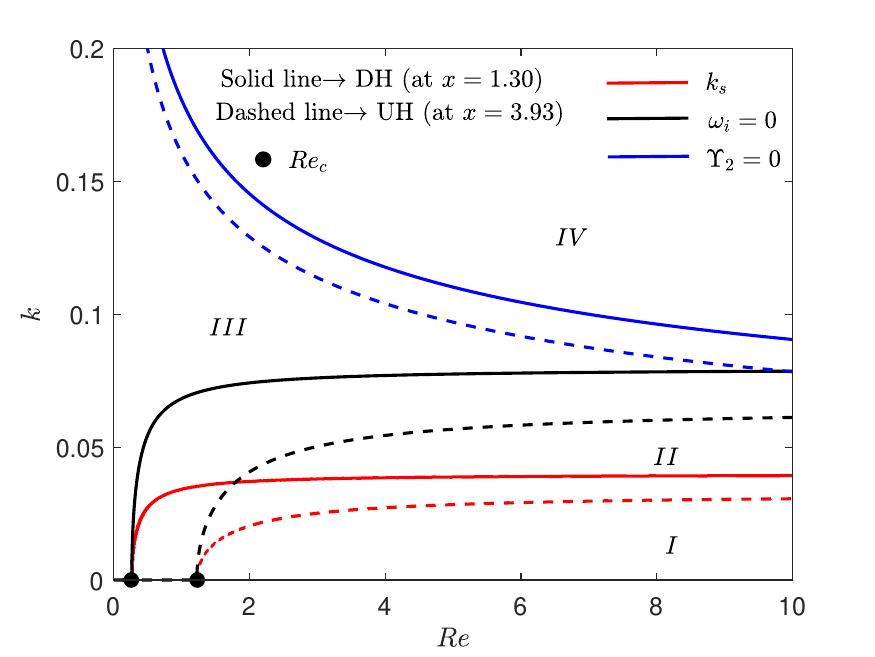}}
\subfigure[$K_{\rho}=0.6$]{\includegraphics*[width=5.4cm]{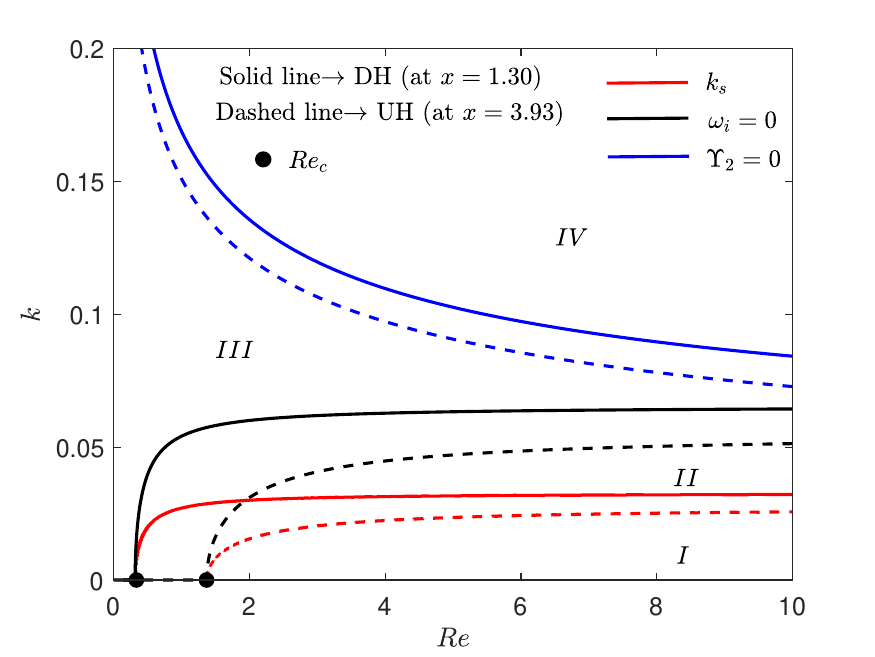}}
\subfigure[$K_{\rho}=0.7$]{\includegraphics*[width=5.4cm]{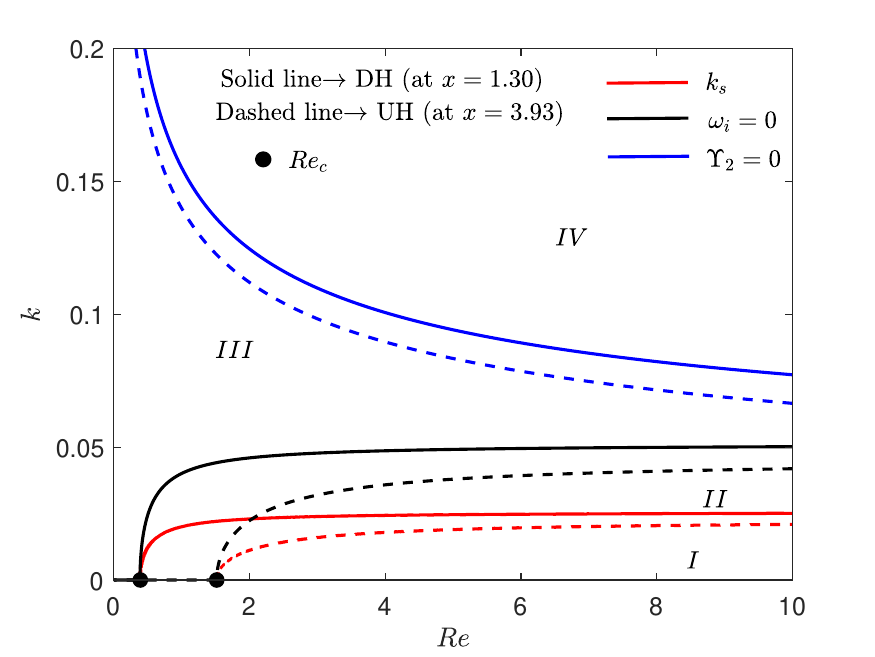}}
\end{center}
\caption{The effect of (a) $K_{\rho}=0.5$, (b) $K_{\rho}=0.6$, and (c) $K_{\rho}=0.7$ on different flow regions for uphill (UH: $x=3.93$) and downhill (DH: $x=1.30$) portions. The different flow regions $I: \text{supercritical unstable}~ \omega_i>0,\,\Upsilon_2<0$, $II:\text{supercritical stable}~\omega_i>0,\,\Upsilon_2>0$, $III:\text{subcritical stable}~\omega_i<0,\,\Upsilon_2>0$, and $IV:\text{subcritical unstable}~\omega_i<0,\,\Upsilon_2<0$ at a point on the uphill (UH) and downhill (DH) portions. The other fixed parameters are $\xi=0.1\pi$, $\tau=0.4$, $K_{\sigma}=0.5$, $We=450$, and $\theta=60^{\circ}$, and $\epsilon=0.1$. }\label{Fig_9}
\end{figure}

\begin{table}[ht!]
\caption{$k-$ range of flow states for different $Re$ with different $K_{\rho}$. $k_{\text{DH}}$ and $k_{\text{UH}}$ mark the wavenumbers for downhill and uphill, respectively. The left boundary of flow zone II is the corresponding singularity $k_s$ of  $\Upsilon_2$ and the right boundary is the critical wavenumber $k_c$, at which $\omega_i=0$. $\Upsilon_2=0$ at left boundary of the flow zone IV.}\label{Table_2}\vspace{0.2cm}
 \centering
% \small
\resizebox{16cm}{!}{
\addtolength{\tabcolsep}{5pt}
\begin{tabular}{ c| c| c c c c }
\hline
$K_{\rho}$ & \ $Re$  & I & II & III & IV  \\ 
\hline
&1 & $0<k_{\text{DH}}<0.0341$ & $0.0341<k_{\text{DH}}<0.0682$ &$0.0682<k_{\text{DH}}<0.1807$ & $0.1807<k_{\text{DH}}\leq0.2$\\
& & $-$& $-$& $-$  & $0.1606<k_{\text{UH}}\leq0.2$\\
\cline{2-6}
& 3 & $0<k_{\text{DH}}<0.0382$ & $0.0382<k_{\text{DH}}<0.0764$  & $0.0764<k_{\text{DH}}<0.1284$ & $0.1284<k_{\text{DH}}\leq0.2$ \\
0.5& & $0<k_{\text{UH}}<0.0251$ & $0.0251<k_{\text{UH}}< 0.0502$  &  $0.0502<k_{\text{UH}}<0.1138$ & $0.1138<k_{\text{UH}}\leq0.2$\\
\cline{2-6} 
&5 & $0<k_{\text{DH}}<0.0388$ & $0.0388<k_{\text{DH}}<0.0776$ &$0.0776<k_{\text{DH}}<0.1103$ & $0.1103<k_{\text{DH}}\leq0.2$  \\ 
& & $0<k_{\text{UH}}<0.0285$ & $0.0285<k_{\text{UH}}< 0.0570$  &  $0.0570<k_{\text{UH}}<0.0971$ & $0.0971<k_{\text{UH}}\leq0.2$\\
\cline{2-6} 
 &7 & $0<k_{\text{DH}}<0.0392$ & $0.0392<k_{\text{DH}}<0.0784$ &$0.0784<k_{\text{DH}}<0.1000$ & $0.1000<k_{\text{DH}}\leq0.2$  \\
 & & $0<k_{\text{UH}}<0.0295$ & $0.0295<k_{\text{UH}}< 0.0590$  &  $0.0590<k_{\text{UH}}<0.0875$ & $0.0875<k_{\text{UH}}\leq0.2$\\ \hline
 &1 & $0<k_{\text{DH}}<0.0271$ & $0.0271<k_{\text{DH}}<0.0542$ &$0.0542<k_{\text{DH}}<0.1688$ & $0.1688<k_{\text{DH}}\leq0.2$ \\ 
 & & $-$ &$-$  & $-$  & $0.1510<k_{\text{UH}}\leq0.2$ \\
 \cline{2-6} 
 & 3  & $0<k_{\text{DH}}<0.0309$ & $0.0309<k_{\text{DH}}<0.0618$  &  $0.0618<k_{\text{DH}}<0.1197$ & $0.1197<k_{\text{DH}}\leq 0.2$ \\ 
& & $0<k_{\text{UH}}<0.0205$ & $0.0205<k_{\text{UH}}<0.0410$  &  $0.0410<k_{\text{UH}}<0.1065$ & $0.1065<k_{\text{UH}}\leq0.2$   \\ 
\cline{2-6} 
0.6 &5  & $0<k_{\text{DH}}<0.0315$  & $0.0315<k_{\text{DH}}<0.0630$ & $0.0630<k_{\text{DH}}<0.1026$ &$0.1026<k_{\text{DH}}\leq0.2$ \\
 & & $0<k_{\text{UH}}<0.0237$ & $0.0237<k_{\text{UH}}<0.0474$  &  $0.0474<k_{\text{UH}}<0.0906$ & $0.0906<k_{\text{UH}}\leq0.2$   \\
 \cline{2-6} 
 &7  & $0<k_{\text{DH}}<0.0319$  & $0.0319<k_{\text{DH}}<0.0638$ & $0.0638<k_{\text{DH}}<0.0931$ &$0.0931<k_{\text{DH}}\leq0.2$   \\ 
 & & $0<k_{\text{UH}}<0.0248$ & $0.0248<k_{\text{UH}}< 0.0496$  &  $0.0496<k_{\text{UH}}<0.0815$ & $0.0815<k_{\text{UH}}\leq0.2$\\ 
 \hline
&1 & $0<k_{\text{DH}}<0.0201$ & $0.0201<k_{\text{DH}}<0.0402$ &$0.0402<k_{\text{DH}}<0.1552$ & $0.1552<k_{\text{DH}}\leq0.2$\\
& & $-$& $-$& $-$  & $0.1399<k_{\text{UH}}\leq0.2$\\
\cline{2-6}
& 3 & $0<k_{\text{DH}}<0.0240$ & $0.0240<k_{\text{DH}}<0.0480$  & $0.0480<k_{\text{DH}}<0.1099$ & $0.1099<k_{\text{DH}}\leq0.2$ \\
0.7& & $0<k_{\text{UH}}<0.0159$ & $0.0159<k_{\text{UH}}< 0.0318$  &  $0.0318<k_{\text{UH}}<0.0983$ & $0.0983<k_{\text{UH}}\leq0.2$\\
\cline{2-6} 
&5 & $0<k_{\text{DH}}<0.0246$ & $0.0246<k_{\text{DH}}<0.0492$ &$0.0492<k_{\text{DH}}<0.0942$ & $0.0942<k_{\text{DH}}\leq0.2$  \\ 
& & $0<k_{\text{UH}}<0.0189$ & $0.0189<k_{\text{UH}}< 0.0378$  &  $0.0378<k_{\text{UH}}<0.0834$ & $0.0834<k_{\text{UH}}\leq0.2$\\
\cline{2-6} 
 &7 & $0<k_{\text{DH}}<0.0250$ & $0.0250<k_{\text{DH}}<0.0500$ &$0.0500<k_{\text{DH}}<0.0854$ & $0.0854<k_{\text{DH}}\leq0.2$  \\
 & & $0<k_{\text{UH}}<0.0201$ & $0.0201<k_{\text{UH}}< 0.0402$  &  $0.0402<k_{\text{UH}}<0.0747$ & $0.0747<k_{\text{UH}}\leq0.2$\\
 \hline 
\end{tabular}}
\end{table}

\begin{figure}[ht!]
\begin{center}
\subfigure[$K_{\sigma}=0.5$]{\includegraphics*[width=5.4cm]{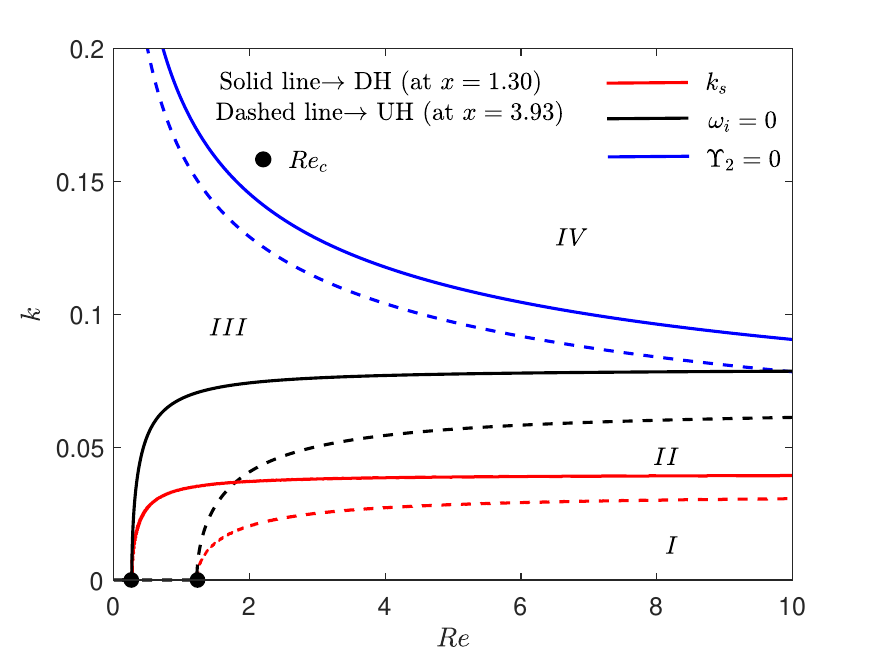}}
\subfigure[$K_{\sigma}=0.6$]{\includegraphics*[width=5.4cm]{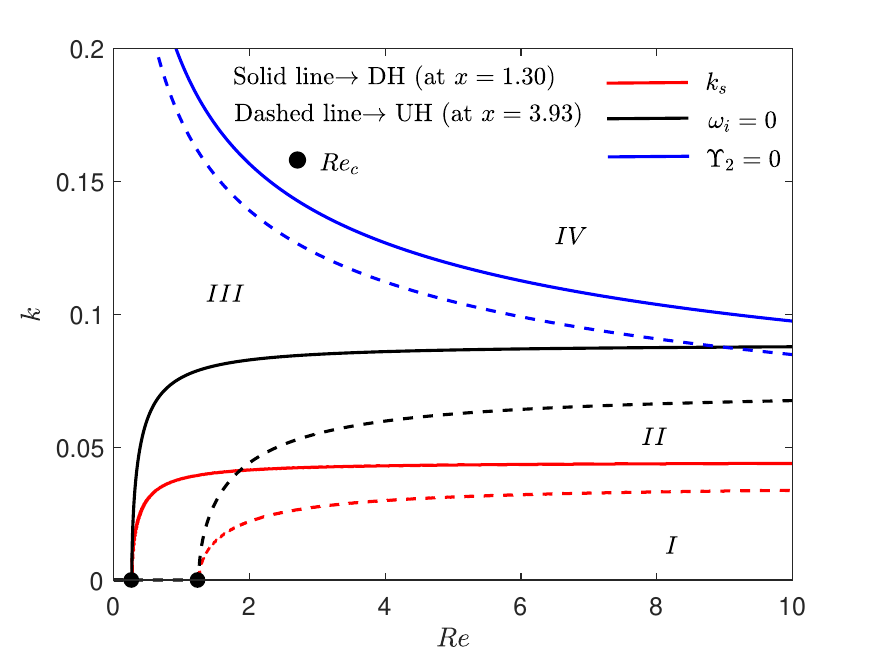}}
\subfigure[$K_{\sigma}=0.7$]{\includegraphics*[width=5.4cm]{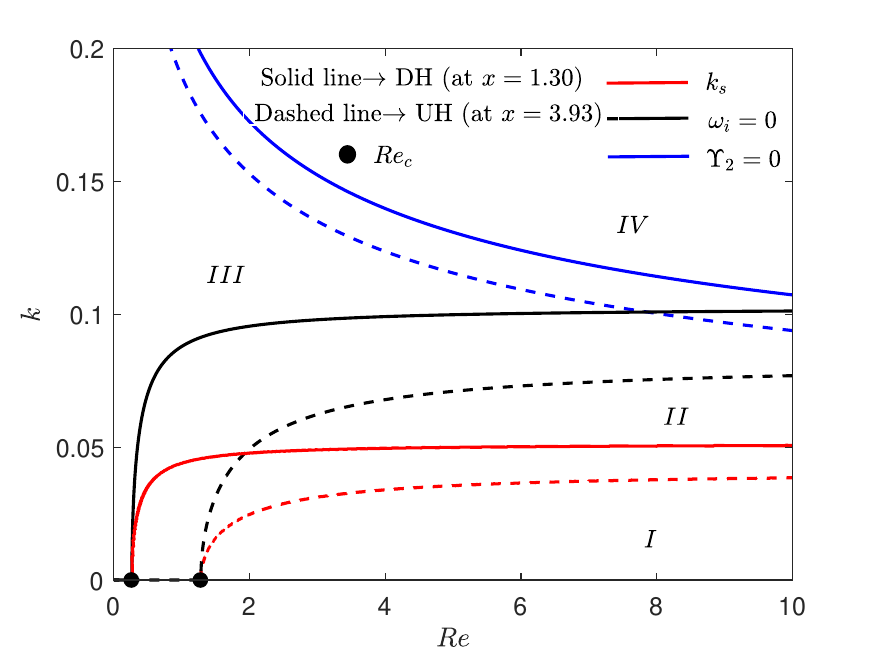}}
\end{center}
\caption{The effect of (a) $K_{\sigma}=0.5$, (b) $K_{\sigma}=0.6$, and (c) $K_{\sigma}=0.7$ on different flow regions for uphill (UH: $x=3.93$) and downhill (DH: $x=1.30$) portions. The different flow regions $I: \text{supercritical unstable}~ \omega_i>0,\,\Upsilon_2<0$, $II:\text{supercritical stable}~\omega_i>0,\,\Upsilon_2>0$, $III:\text{subcritical stable}~\omega_i<0,\,\Upsilon_2>0$, and $IV:\text{subcritical unstable}~\omega_i<0,\,\Upsilon_2<0$ at a point on the uphill (UH) and downhill (DH) portions. The other fixed parameters are $\xi=0.1\pi$, $\tau=0.4$, $K_{\rho}=0.5$, $We=450$, and $\theta=60^{\circ}$, and $\epsilon=0.1$. }\label{Fig_10}
\end{figure}

The important finding from Figs.~\ref{Fig_8}-\ref{Fig_10} is that whether the imposed shear acts in the downstream or upstream direction or the parameters $K_{\rho}$ and $K_{\sigma}$ change, the bandwidth of the supercritical stable zone ($II$) at the downhill portion is higher than the uphill portion, whereas the opposite trend is observed for the subcritical unstable zone ($IV$). This is due to the higher $Re_c$ value at the uphill portion than the downhill portion. 
\begin{table}[ht!]
\caption{$k-$ range of flow states for different $Re$ with different $K_{\sigma}$. $k_{\text{DH}}$ and $k_{\text{UH}}$ mark the wavenumbers for downhill and uphill, respectively. The left boundary of flow zone II is the corresponding singularity $k_s$ of  $\Upsilon_2$ and the right boundary is the critical wavenumber $k_c$, at which $\omega_i=0$. $\Upsilon_2=0$ at left boundary of the flow zone IV.}\label{Table_3}\vspace{0.2cm}
 \centering
% \small
\resizebox{16cm}{!}{
\addtolength{\tabcolsep}{5pt}
\begin{tabular}{ c| c| c c c c }
\hline
$K_{\sigma}$ & \ $Re$  & I & II & III & IV  \\ 
\hline
&1 & $0<k_{\text{DH}}<0.0341$ & $0.0341<k_{\text{DH}}<0.0682$ &$0.0682<k_{\text{DH}}<0.1807$ & $0.1807<k_{\text{DH}}\leq0.2$\\
& &$-$ &$-$ & $-$  & $0.1606<k_{\text{UH}}\leq0.2$\\
\cline{2-6}
& 3 & $0<k_{\text{DH}}<0.0382$ & $0.0382<k_{\text{DH}}<0.0764$  & $0.0764<k_{\text{DH}}<0.1284$ & $0.1284<k_{\text{DH}}\leq0.2$ \\
0.5& & $0<k_{\text{UH}}<0.0251$ & $0.0251<k_{\text{UH}}< 0.0502$  &  $0.0502<k_{\text{UH}}<0.1138$ & $0.1138<k_{\text{UH}}\leq0.2$\\
\cline{2-6} 
&5 & $0<k_{\text{DH}}<0.0388$ & $0.0388<k_{\text{DH}}<0.0776$ &$0.0776<k_{\text{DH}}<0.1103$ & $0.1103<k_{\text{DH}}\leq0.2$  \\ 
& & $0<k_{\text{UH}}<0.0285$ & $0.0285<k_{\text{UH}}< 0.0570$  &  $0.0570<k_{\text{UH}}<0.0971$ & $0.0971<k_{\text{UH}}\leq0.2$\\
\cline{2-6} 
 &7 & $0<k_{\text{DH}}<0.0392$ & $0.0392<k_{\text{DH}}<0.0784$ &$0.0784<k_{\text{DH}}<0.1000$ & $0.1000<k_{\text{DH}}\leq0.2$  \\
 & & $0<k_{\text{UH}}<0.0295$ & $0.0295<k_{\text{UH}}< 0.0590$  &  $0.0590<k_{\text{UH}}<0.0875$ & $0.0875<k_{\text{UH}}\leq0.2$\\ \hline
 &1 & $0<k_{\text{DH}}<0.0381$ & $0.0381<k_{\text{DH}}<0.0762$ &$0.0762<k_{\text{DH}}<0.1946$ & $0.1946<k_{\text{DH}}\leq0.2$ \\ 
 & & $-$ & $-$ & $-$  & $0.1727<k_{\text{UH}}\leq0.2$ \\
 \cline{2-6} 
 & 3  & $0<k_{\text{DH}}<0.0424$ & $0.0424<k_{\text{DH}}<0.0848$  &  $0.0848<k_{\text{DH}}<0.1383$ & $0.1383<k_{\text{DH}}\leq 0.2$ \\ 
& & $0<k_{\text{UH}}<0.0277$ & $0.0277<k_{\text{UH}}<0.0554$  &  $0.0554<k_{\text{UH}}<0.1226$ & $0.1226<k_{\text{UH}}\leq0.2$   \\ 
\cline{2-6} 
0.6 &5  & $0<k_{\text{DH}}<0.0433$  & $0.0433<k_{\text{DH}}<0.0866$ & $0.0866<k_{\text{DH}}<0.1188$ &$0.1188<k_{\text{DH}}\leq0.2$ \\
 & & $0<k_{\text{UH}}<0.0313$ & $0.0313<k_{\text{UH}}<0.0626$  &  $0.0626<k_{\text{UH}}<0.1048$ & $0.1048<k_{\text{UH}}\leq0.2$   \\
 \cline{2-6} 
 &7  & $0<k_{\text{DH}}<0.0436$  & $0.0436<k_{\text{DH}}<0.0872$ & $0.0872<k_{\text{DH}}<0.1078$ &$0.1078<k_{\text{DH}}\leq0.2$   \\ 
 & & $0<k_{\text{UH}}<0.0328$ & $0.0328<k_{\text{UH}}< 0.0656$  &  $0.0656<k_{\text{UH}}<0.0945$ & $0.0945<k_{\text{UH}}\leq0.2$\\ 
 \hline
&1 & $0<k_{\text{DH}}<0.0440$ & $0.0440<k_{\text{DH}}<0.0880$ &$-$ &$-$ \\
& & $-$&$-$ & $-$  & $0.1896<k_{\text{UH}}\leq0.2$\\
\cline{2-6}
& 3 & $0<k_{\text{DH}}<0.0489$ & $0.0489<k_{\text{DH}}<0.0978$  & $0.0978<k_{\text{DH}}<0.1525$ & $0.1525<k_{\text{DH}}\leq0.2$ \\
0.7& & $0<k_{\text{UH}}<0.0310$ & $0.0310<k_{\text{UH}}< 0.0620$  &  $0.0620<k_{\text{UH}}<0.1350$ & $0.1350<k_{\text{UH}}\leq0.2$\\
\cline{2-6} 
&5 & $0<k_{\text{DH}}<0.0500$ & $0.0500<k_{\text{DH}}<0.1000$ &$0.1000<k_{\text{DH}}<0.1309$ & $0.1309<k_{\text{DH}}\leq0.2$  \\ 
& & $0<k_{\text{UH}}<0.0354$ & $0.0354<k_{\text{UH}}< 0.0708$  &  $0.0708<k_{\text{UH}}<0.1155$ & $0.1155<k_{\text{UH}}\leq0.2$\\
\cline{2-6} 
 &7 & $0<k_{\text{DH}}<0.0502$ & $0.0502<k_{\text{DH}}<0.1004$ &$0.1004<k_{\text{DH}}<0.1187$ & $0.1187<k_{\text{DH}}\leq0.2$  \\
 & & $0<k_{\text{UH}}<0.0373$ & $0.0373<k_{\text{UH}}< 0.0746$  &  $0.0746<k_{\text{UH}}<0.1044$ & $0.1044<k_{\text{UH}}\leq0.2$\\
 \hline 
\end{tabular}}
\end{table}

For both the uphill and downhill portions of the wavy bottom, different flow zones with varying values of $\tau$ are observed (refer Fig.~\ref{Fig_8}). In the fixed bottom portion (uphill/downhill), the external shear in the downstream direction (follow Figs.~\ref{Fig_8}(c), (d), and (e)) enhances both supercritical stable and unstable zones, which are followed by the decrease in $Re_c$ with a concomitant increase in $k_c$ and $k_s$. At both bottom portions, the $k-$ range for various flow states at different $Re$ is summarized in Table.~\ref{Table_1}, analogous to Fig.~\ref{Fig_8}. In this table, at fixed $\tau$ and $Re$, the left boundary of the supercritical stable zone (II) at the bottom portion (downhill/uphill) is the singularity $k_s$ (at which $\Upsilon_2$ has the point of discontinuity), the right boundary of the same zone is the critical wavenumber $k_c$ (at which $\omega_i=0$), and the function $\Upsilon_2$ vanishes at the left boundary of zone IV. For example, when $Re=3$ and $\tau=-0.4$, for the downhill portion: left boundary of the corresponding flow zone II ( supercritical stable zone ) is $k_s=0.0291$, the singularity of the Landau coefficient $\Upsilon_2$ (i.e., $\Upsilon_2\rightarrow\infty$ as $k\rightarrow k_s$) and the right boundary is $k_c=0.0582$ at which $\omega_i=0$, and for the uphill portion: the left boundary of the corresponding zone II is $k_s=0.0134$, the singularity of the Landau coefficient $\Upsilon_2$ (i.e., $\Upsilon_2\rightarrow\infty$ as $k\rightarrow k_s$) and the right boundary of the same zone is $k_c=0.0268$ at which $\omega_i=0$.     

The continuous enlargement of the supercritical stable region (II) due to the positive external shear assures that strong surface wave instability is possible if one can impose external shear in the flow direction. The opposite scenario can be observed from Figs.~\ref{Fig_8}(c), (b), and (a), when the imposed shear acts along the back-flow direction. Another important finding from Fig.~\ref{Fig_8} is that the subcritical stable region rapidly decreases, whereas the remaining three zones increase as soon as $Re$ increases.  This affirms the instability impact of inertia force on the surface wave. Additionally, the dependency of different flow zones on the wavy bottom portions becomes very weak after a significant value of $Re$.

Fig.~\ref{Fig_9} depicts that at any bottom portion (uphill/downhill), as long as the density variation parameter $K_{\rho}$ increases, the supercritical stable zone decreases and the subcritical unstable zone increases owing to the continuous increase in the critical value $Re_c$. As a consequence, the parameter $K_{\rho}$ contributes to the stabilizing effect on the surface mode. 
% Besides, for all $K_{\rho}$ values, the bandwidth of the supercritical stable region for the downhill portion is much higher than that of the uphill portion, and in contrast, the opposite scenario is noticed in the subcritical unstable zone.   

However, with a fixed bottom region, if the surface tension variation coefficient $K_{\sigma}$ increases (see, Fig.~\ref{Fig_10}), one gets the opposite effect of $K_{\rho}$ on the sub-zones. This time, the supercritical stable zone increases, and the subcritical unstable zone reduces as $K_{\sigma}$ increases. Because of this, the surface of the liquid becomes more unstable. 
% Further, in the downhill part of the wavy bottom, the supercritical stable bandwidth is greater than in the uphill portion, whereas, in the subcritical unstable zone, the reverse happens for all values of $K_{\sigma}$. 
As noted earlier, Table.~\ref{Table_2} and Table.~\ref{Table_3} summarize the $k-$ range of all flow regions at different $Re$ with three different values of $K_{\rho}$ and $K_{\sigma}$, respectively.

\begin{figure}[ht!]
\begin{center}
\subfigure[]{\includegraphics*[width=7.2cm]{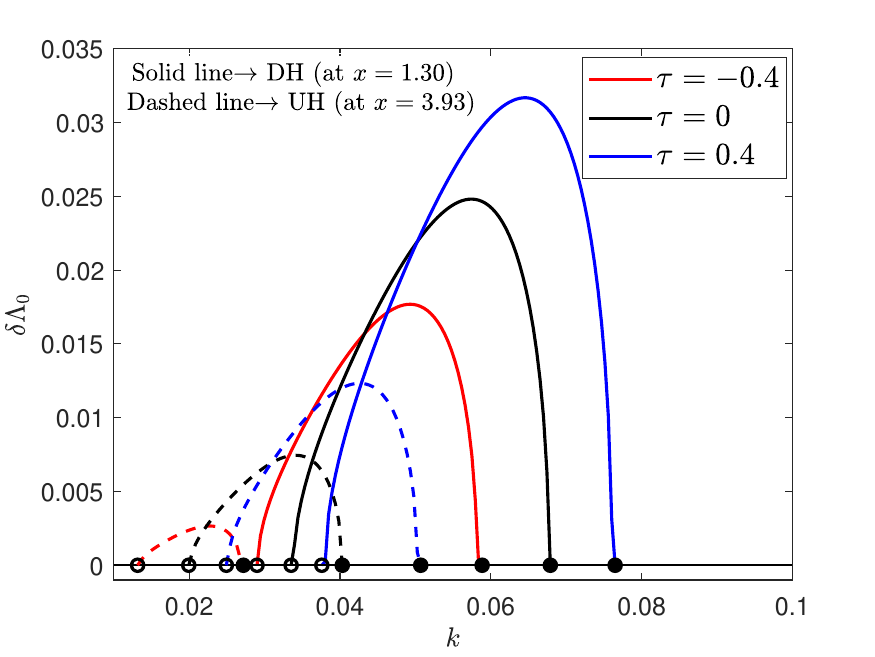}}
\subfigure[]{\includegraphics*[width=7.2cm]{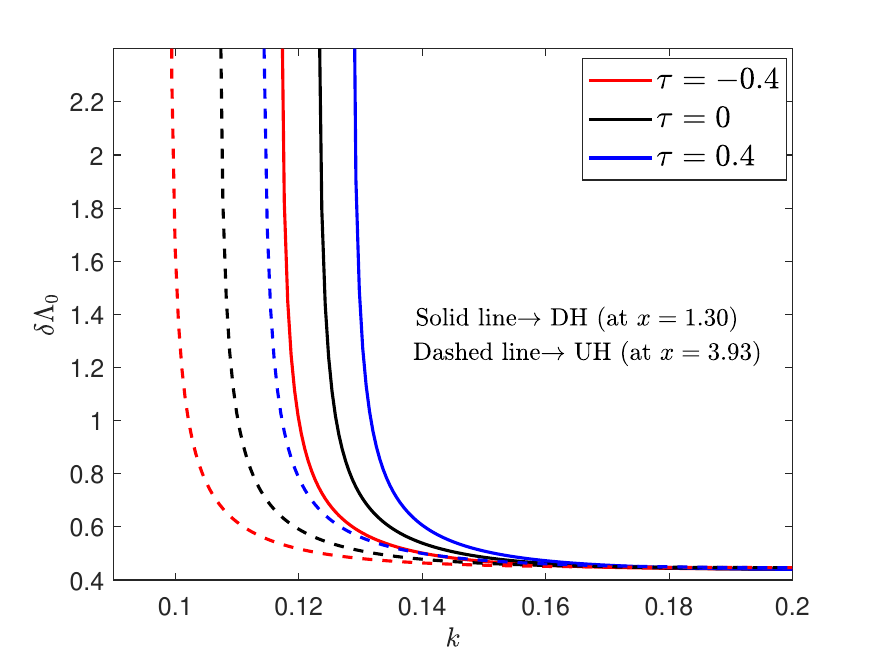}}
\end{center}
\caption{Variation in threshold amplitude ($\delta \Lambda_0$) with $k$ in the (a) supercritical stable and (b) subcritical unstable regions for different $\tau$ values at the downhill (DH: $x=1.30$) and uphill (UH: $x=3.93$) portions. The remaining parameters are $Re=3$, $\xi=0.1\pi$, $K_{\sigma}=0.5$, $K_{\rho}=0.5$, $We=450$, and $\theta=60^{\circ}$. In (a), for the downhill portion (DH): the open circles mark $\delta\Lambda_0\rightarrow0$ at $k\rightarrow~0.0291,~0.0337, ~\text{and}~0.0382$, at which $\Upsilon_2\rightarrow\infty$ and the solid circles mark $\delta\Lambda\rightarrow0$ at $k\rightarrow~0.0582,~0.0674,~\text{and}~0.0764$, at which $\omega_i=0$. For the uphill portion (UH): the open circles mark $\delta\Lambda_0\rightarrow0$ at $k\rightarrow0.0134,~0.0201,~\text{and}~0.0251$ at which $\Upsilon_2\rightarrow\infty$ and the solid circles mark $\delta\Lambda_0\rightarrow0$ at  $k\rightarrow0.0268,~0.0402,~\text{and}~0.0502$, where $\omega_i=0$. In (b), $\delta\Lambda_0\rightarrow\infty$ at $k$ value at which $\Upsilon_2\rightarrow0$. The corresponding $k$ values are given in the left boundary of the subcritical unstable zone (IV), illustrated in Table.~\ref{Table_1}.}\label{Fig_11}
\end{figure}
Fig.~\ref{Fig_11} illustrates the effect of external shear ($\tau$) on the threshold amplitude $\delta\Lambda_0$ as a function of $k$ in the supercritical stable (see, Fig.\ref{Fig_11}(a)) and subcritical unstable (see, Fig~.\ref{Fig_11}(b)) regions when the bottom portion is either uphill or downhill. It may be noted that the linearly unstable region $\omega_i>0$ does not become unbounded in the specified $k-$ range of the supercritical stable zone (see, Table.~\ref{Table_1}) but instead seeks an equilibrium state with finite amplitude as the Landau coefficient $\Upsilon_2>0$ in this zone. This is also clearly demonstrated in Fig.~\ref{Fig_11}(a).

\begin{figure}[ht!]
\begin{center}
\subfigure[]{\includegraphics*[width=7.2cm]{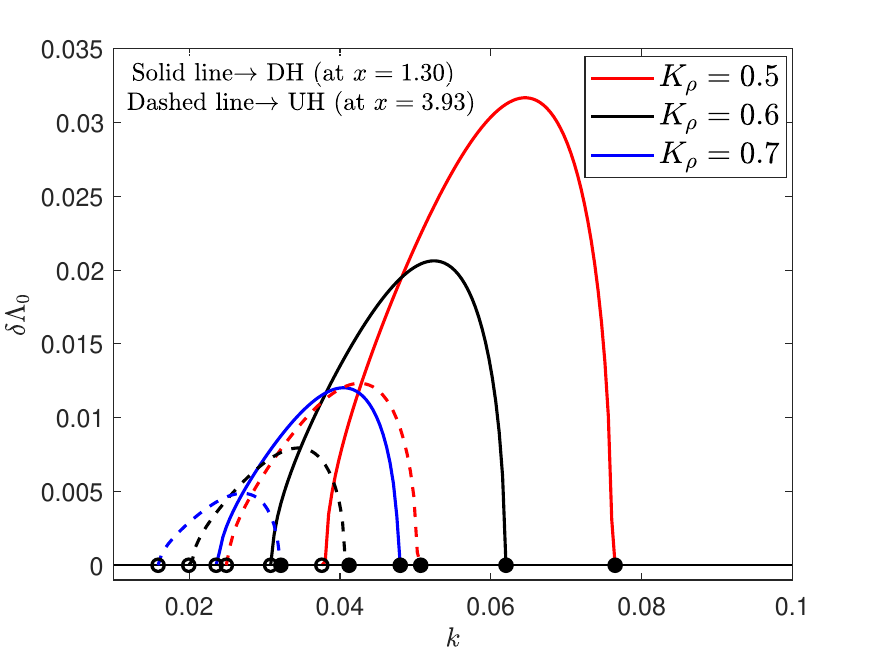}}
\subfigure[]{\includegraphics*[width=7.2cm]{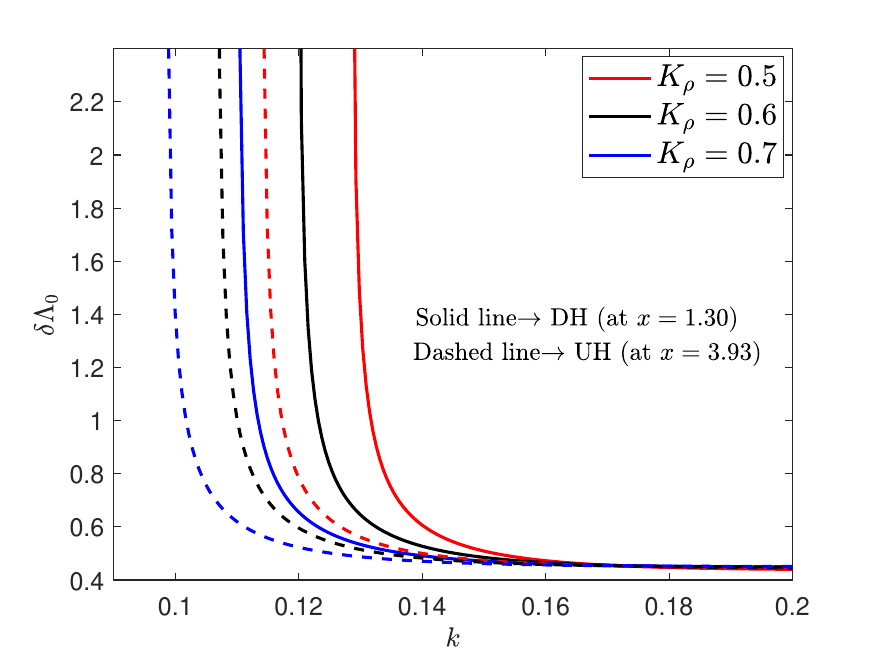}}
\end{center}
\caption{Variation in threshold amplitude ($\delta \Lambda_0$) with $k$ in the (a) supercritical stable and (b) subcritical unstable regions for different $K_{\rho}$ values at the downhill (DH: $x=1.30$) and uphill (UH: $x=3.93$) portions. The remaining parameters are $Re=3$, $\xi=0.1\pi$, $\tau=0.4$, $K_{\sigma}=0.5$, $K_{\rho}=0.5$, $We=450$, and $\theta=60^{\circ}$. In (a), for the downhill portion (DH): the open circles mark $\delta\Lambda_0\rightarrow0$ at $k\rightarrow~0.0382,~0.0309, ~\text{and}~ 0.0240$, at which $\Upsilon_2\rightarrow\infty$ and the solid circles mark $\delta\Lambda\rightarrow0$ at $k\rightarrow~0.0764,~0.0618,~\text{and}~0.0480$, at which $\omega_i=0$. For the uphill portion (UH): the open circles mark $\delta\Lambda_0\rightarrow0$ at $k\rightarrow 0.0251,~0.0205,~\text{and}~0.0159$ value at which $\Upsilon_2\rightarrow\infty$ and the solid circles mark $\delta\Lambda_0\rightarrow0$ at $k\rightarrow0.0502,~0.0410,~\text{and}~0.0318$, where $\omega_i=0$. In (b), $\delta\Lambda_0\rightarrow\infty$ at $k$ value at which $\Upsilon_2\rightarrow0$. The corresponding $k$ values are given in the left boundary of the subcritical unstable zone (IV), illustrated in Table.~\ref{Table_2}.}\label{Fig_12}
\end{figure}

As in Fig.~\ref{Fig_11}(a), the increment of nonlinear wave amplitude with the increase in positive external shear ($+\tau$) confirms the more unstable flow in the system. The higher $\tau>0$ values rapidly increase $\Upsilon_2>0$, and at the same time, $\omega_i>0$ increases (see, Eq.~\eqref{e52}), but at higher $\tau$ values, the $\Upsilon_2>0$ value is lower than at smaller $\tau$ values. This results in the rapid expansion of the threshold amplitude. In contrast, the rapid attenuation in the nonlinear wave amplitude with the increase in negative external shear ($-\tau$) affirms more stable flow in the fluid flow model. Moreover, for all $\tau$ values, the nonlinear wave amplitude at the downhill portion is much larger than the uphill portion, implying a stronger film flow instability at the downhill compared to the uphill of the wavy bottom. Besides, Fig.~\ref{Fig_11}(b) reveals the variation of nonlinear wave amplitude ($\delta\Lambda_0$) vs $k$ in the subcritical unstable zone of the film flow system over the bottom portion (downhill/uphill) with respect to the parameters $\tau$.  Basically, the nonlinear amplification rate is positive, while the linear amplification rate is negative in this zone. It means that even though the linear theory predicts stability in this zone, the disturbance is greater than the threshold amplitude, and as a result, the amplitude grows. Also, it can be seen from Fig.~\ref {Fig_11}(b) that the subcritical amplitude decreases with the increase in $k$, yielding more unstable flow. The streamwise-directed imposed shear amplifies the subcritical amplitude, which essentially makes the flow system less unstable. Physically, in the subcritical unstable zone, the negative value $\Upsilon_2<0$ decreases as long as positive imposed shear increases (see, Eq.~\eqref{e52}). As a result, the threshold amplitude in the subcritical unstable region increases with increasing $+\tau$, whereas an opposite phenomenon can be seen for the backflow-directed imposed shear ($-\tau$).

Further, Fig.~\ref{Fig_12}(a) presents that at the bottom position (uphill/downhill), the threshold amplitude in the supercritical stable zone reduces as $K_{\rho}$ increases and yields a more stable flow in the system. Whereas the
threshold amplitude in the subcritical unstable zone, as shown in Fig.~\ref{Fig_12}(b), attenuates for the higher value of $K_{\rho}$, which actually increases the film flow instability.

\begin{figure}[ht!]
\begin{center}
\subfigure[]{\includegraphics*[width=7.2cm]{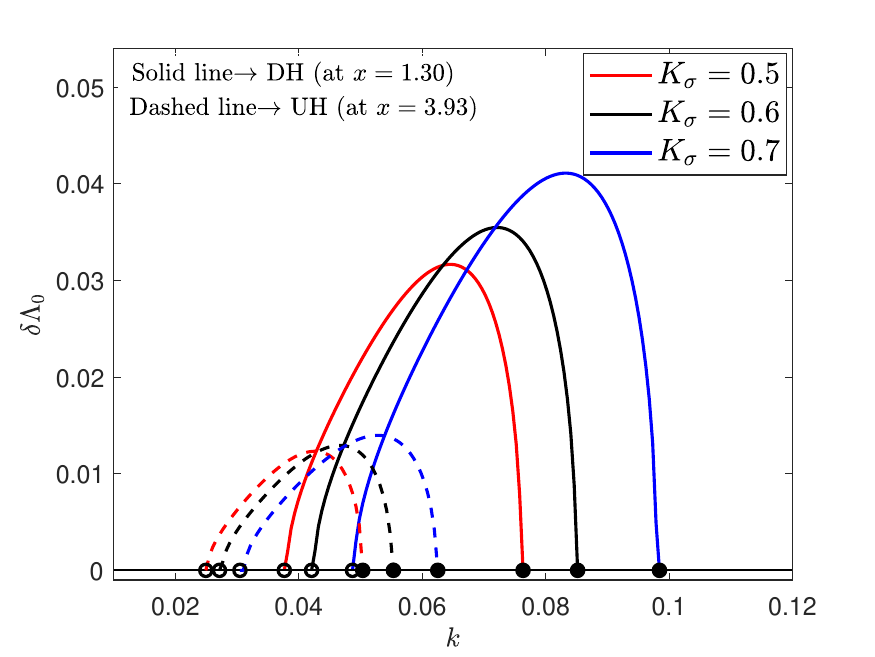}}
\subfigure[]{\includegraphics*[width=7.2cm]{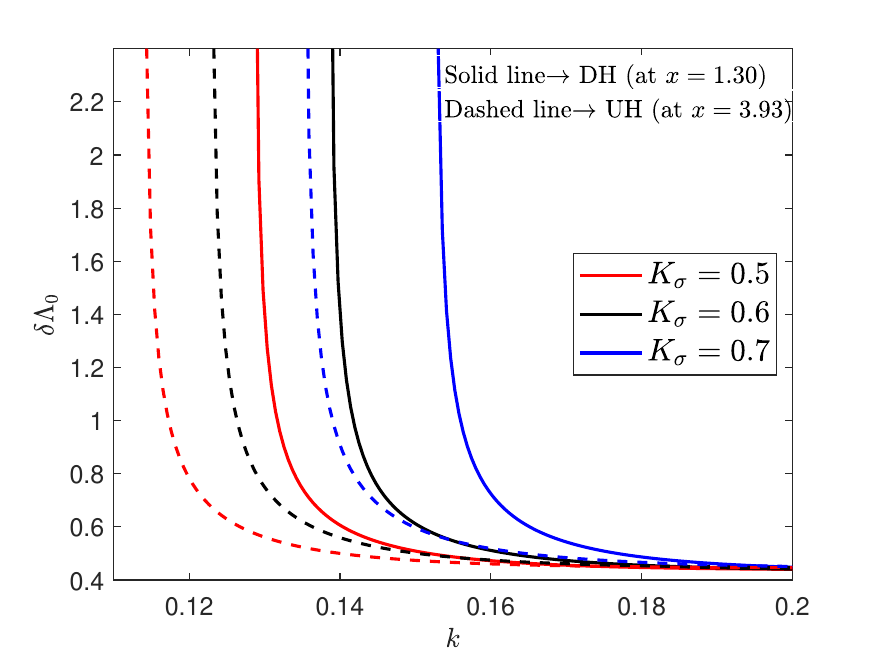}}
\end{center}
\caption{Variation in threshold amplitude ($\delta \Lambda_0$) with $k$ in the (a) supercritical stable and (b) subcritical unstable regions for different $K_{\sigma}$ values at the downhill (DH: $x=1.30$) and uphill (UH: $x=3.93$) portions. The remaining parameters are $Re=3$, $\xi=0.1\pi$, $\tau=0.4$, $K_{\rho}=0.5$, $We=450$, and $\theta=60^{\circ}$. In (a), for the downhill portion (DH): the open circles mark $\delta\Lambda_0\rightarrow0$ at $k\rightarrow~0.0382,~0.0424, ~\text{and}~0.0489$, at which $\Upsilon_2\rightarrow\infty$ and the solid circles mark $\delta\Lambda\rightarrow0$ at $k\rightarrow~0.0764,~0.0848,~\text{and}~0.0978$, at which $\omega_i=0$. For the uphill portion (UH): the open circles mark $\delta\Lambda_0\rightarrow0$ at $k\rightarrow 0.0251,~0.0277,~\text{and}~0.0310$ value at which $\Upsilon_2\rightarrow\infty$ and the solid circles mark $\delta\Lambda_0\rightarrow0$ at  $k\rightarrow 0.0502,~0.0554,~\text{and}~0.0620$, where $\omega_i=0$. In (b), $\delta\Lambda_0\rightarrow\infty$ at $k$ value at which $\Upsilon_2\rightarrow0$. The corresponding $k$ values are given in the left boundary of the subcritical unstable zone (IV), illustrated in Table.~\ref{Table_3}.}\label{Fig_13}
\end{figure}

On the contrary, the parameter $K_{\sigma}$ behaves exactly opposite to the $K_{\rho}$ on the threshold amplitude in both supercritical stable (see, Fig.~\ref{Fig_13}(a)) and subcritical unstable (see, Fig.~\ref{Fig_13}(b)) regions, and consequences strong film flow instability. The higher $K_{\sigma}$ value enlarges threshold amplitude at both uphill and downhill portions in the subcritical unstable zone, as shown in Fig.~\ref{Fig_13}(b), leading to less unstable flow.

Another important finding from Figs.~\ref{Fig_11}, \ref{Fig_12}, and \ref{Fig_13} is that the threshold amplitude is always higher at the downhill portion of the wavy bottom relative to the uphill portion despite the presence of $\tau$ or $K_{\rho}$ or $K_{\sigma}$ in the system. Also, the stability effect of the wavy bottom's downhill portion is not sensitive to the parameters $\tau$, $K_{\rho}$, and $K_{\sigma}$.

 \begin{figure}[ht!]
\begin{center}
\subfigure[Downhill (at $x=1.30$)]{\includegraphics*[width=7.2cm]{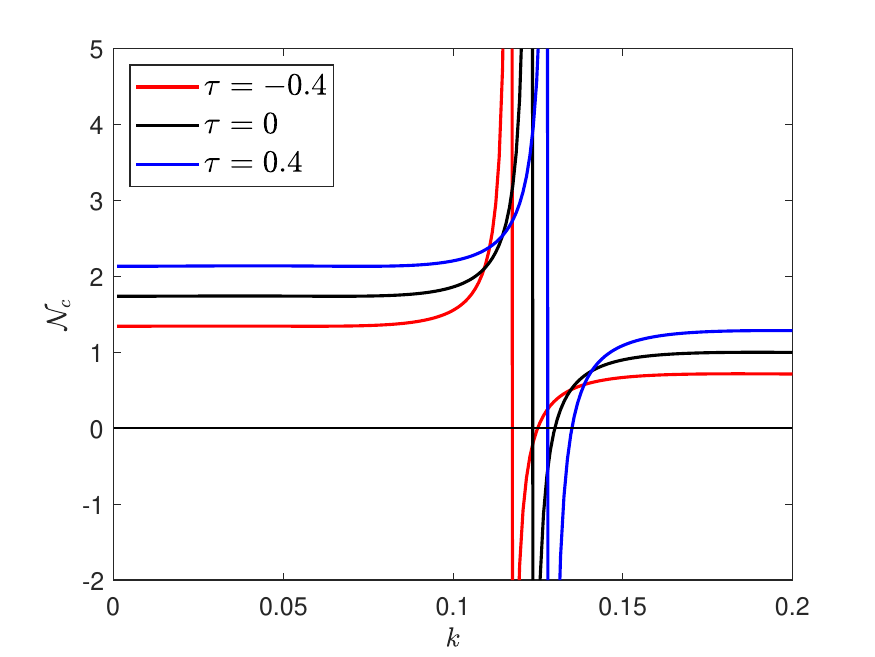}}
\subfigure[Uphill (at $x=3.93$)]{\includegraphics*[width=7.2cm]{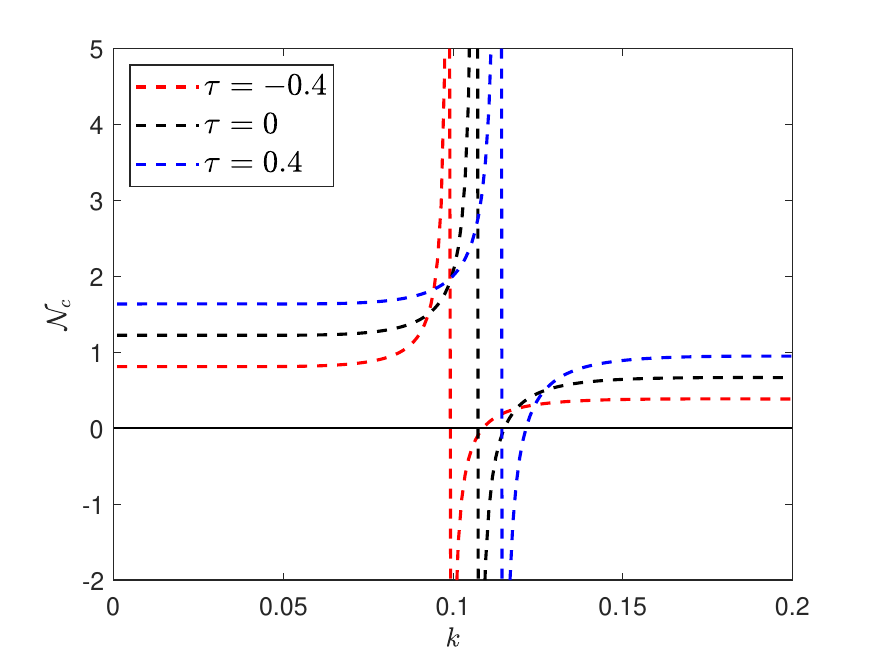}}
\end{center}
\caption{Variation of phase speed ($\mathcal{N}_c$) with $k$ for (a) downhill portion (DH: $x=1.30$) and (b) uphill portion (UH: $x=3.93$), when external shear $\tau$, alters. The remaining parameters are $Re=3$, $K_{\rho}=0.5$, $K_{\sigma}=0.5$, $\xi=0.1\pi$, $We=450$, and $\theta=60^{\circ}$. The Fig.~\ref{Fig_14}(b) is the continuation of Fig.~\ref{Fig_14}(a). At downhill portion (DH): $\mathcal{N}_c$ has the singularity at $k=0.1167,~0.1228,~\text{and}~0.1284$ at which $\Upsilon_2=0$. At uphill portion (UH): $\mathcal{N}_c$ has the singularity at $k=0.0988,~0.1067,~\text{and}~0.1138$ at which $\Upsilon_2=0$.}\label{Fig_14}
\end{figure}

\begin{figure}[ht!]
\begin{center}
\subfigure[Downhill (at $x=1.30$)]{\includegraphics*[width=7.2cm]{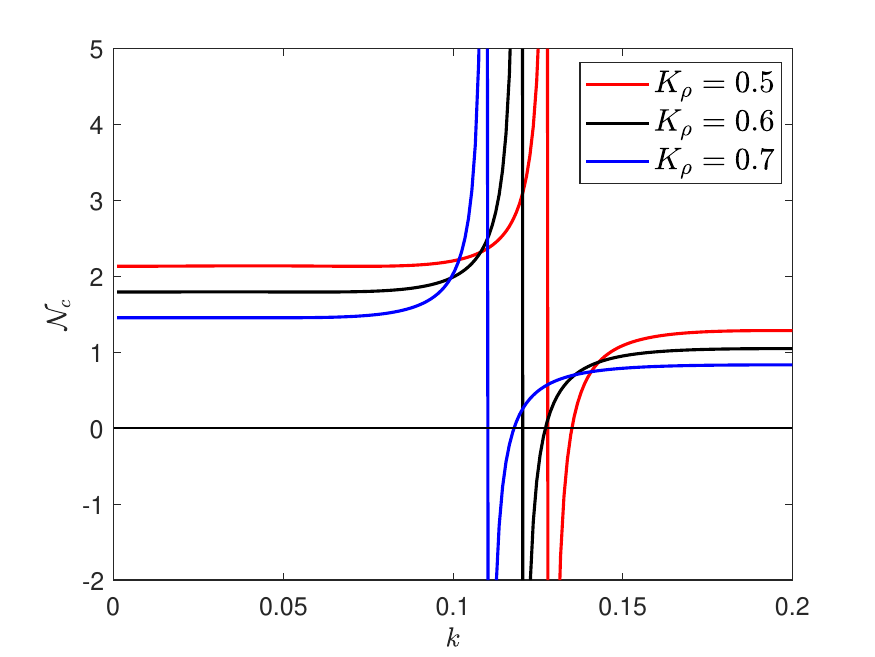}}
\subfigure[Uphill (at $x=3.93$)]{\includegraphics*[width=7.2cm]{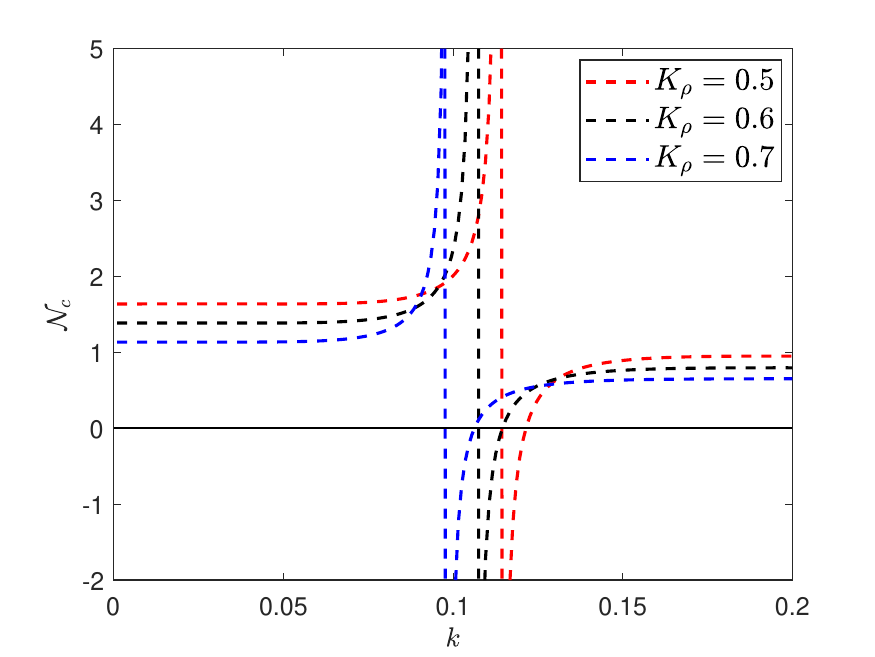}}
\end{center}
\caption{Variation of phase speed ($\mathcal{N}_c$) with $k$ for (a) downhill portion (DH: $x=1.30$) and (b) uphill portion (UH: $x=3.93$) when the parameter $K_{\rho}$ alters. The remaining parameters are $Re=3$, $\tau=0.$, $K_{\sigma}=0.5$, $\xi=0.1\pi$, $We=450$, and $\theta=60^{\circ}$. The Fig.~\ref{Fig_15}(b) is the continuation of Fig.~\ref{Fig_15}(a). At downhill portion (DH): $\mathcal{N}_c$ has the singularity at $k=0.1284,~0.1197,~\text{and}~0.1099$ at which $\Upsilon_2=0$. At uphill portion (UH): $\mathcal{N}_c$ has the singularity at $k=0.1138,~0.1065,~\text{and}~0.0983$ at which $\Upsilon_2=0$.}\label{Fig_15}
\end{figure}

\begin{figure}[ht!]
\begin{center}
\subfigure[Downhill (at $x=1.30$)]{\includegraphics*[width=7.2cm]{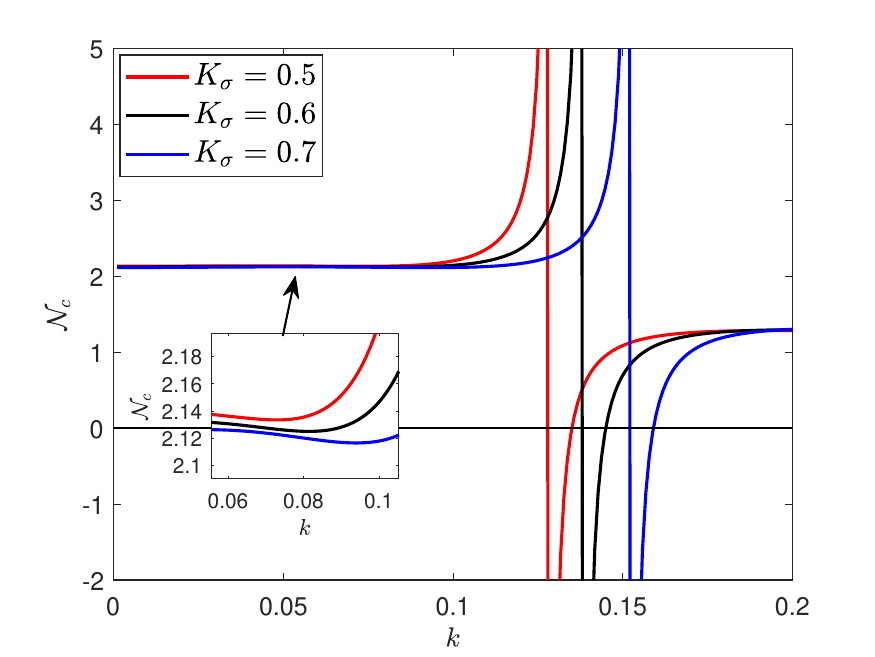}}
\subfigure[Uphill (at $x=3.93$)]{\includegraphics*[width=7.2cm]{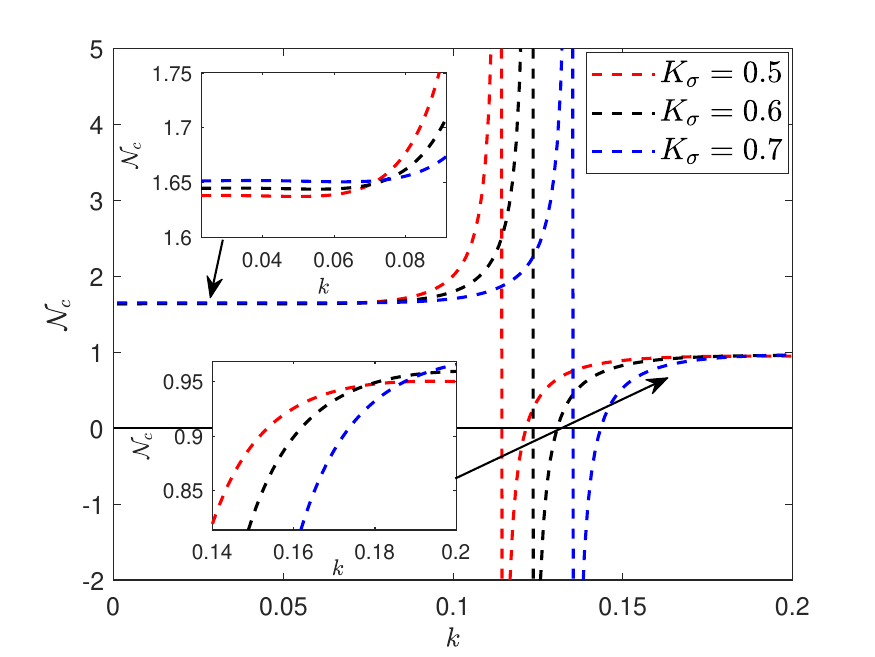}}
\end{center}
\caption{Variation of phase speed ($\mathcal{N}_c$) with $k$ for (a) downhill portion (DH: $x=1.30$) and (b) uphill portion (UH: $x=3.93$) when the parameter $K_{\sigma}$ alters. The remaining parameters are $Re=3$, $\tau=0.4$, $K_{\rho}=0.5$, $\xi=0.1\pi$, $We=450$, and $\theta=60^{\circ}$. The Fig.~\ref{Fig_16}(b) is the continuation of Fig.~\ref{Fig_16}(a). At downhill portion (DH): $\mathcal{N}_c$ has the singularity at $k=0.1284,~0.1383,~\text{and}~0.1525$ at which $\Upsilon_2=0$. At uphill portion (UH): $\mathcal{N}_c$ has the singularity at $k=0.1138,~0.1226,~\text{and}~0.1350$ at which $\Upsilon_2=0$.}\label{Fig_16}
\end{figure}

Notably, the nonlinear phase speed $\mathcal{N}_c$ depends on the wavenumber $k$ (see, Eq.~\eqref{e55}), confirming the dispersive nature of nonlinear waves. In fact, the $\mathcal{N}_c$ depends on the parameters $\tau$, $K_{\rho}$, $K_{\sigma}$, $Re$, $\xi$, $We$, and $\theta$. To discuss the correct behaviour of nonlinear wave speed, it is beneficial to demonstrate the phase speed $\mathcal{N}_c$ curve as a function for whole over the $k-$ domain, as in Figs.~\ref{Fig_14}, \ref{Fig_15}, and \ref{Fig_16}, where all four zones are visible (follow the Tables.~\ref{Table_1}-\ref{Table_3} to identify the $k-$ range of different flow zones). In  Fig.~\ref{Fig_14}, for fixed $Re=3$, the variations of the phase speed curves $\mathcal{N}_c$ with different $\tau$ values are plotted. Note that, for the downhill portion, the function $\mathcal{N}_c$ has the point of discontinuity at $k=0.1167,~0.1228,~\text{and}~0.1284$, where $\Upsilon_2=0$ and for the uphill portion, the function $\mathcal{N}_c$ has the point of discontinuity at $k=0.0988,~0.1067,~\text{and}~0.1138$, where $\Upsilon_2=0$ when $\tau$ values are $-0.4$, $0$, and $0.4$, respectively. For other $Re$ values ($Re=1,~5,~7$), one can find the discontinuity of $\mathcal{N}_c$ for both uphill and downhill of the wavy bottom from the left boundary of zone IV in Table.~\ref{Table_1}.  

The key inference that appears from this figure is that the phase speed for the subcritical unstable zone can either be positive or zero, or negative, while in the case of the other three flow regions, it is positive. Negative $\mathcal{N}_c$ indicates that the phase velocity is in the opposite direction of the energy flow (i.e., attenuation). The optimal (maximum/minimum) phase speed $\mathcal{N}_c$ attains at  $k$ value, which is close to the point of discontinuity of $\mathcal{N}_c$. At the higher phase velocity, the energy propagation ceases, which mainly emerges where the Landau coefficient $\Upsilon_2$ approaches zero. 
Further, the nonlinear phase speed is higher at the downhill portion than the uphill portion of the wavy bottom (on comparing Figs.~\ref{Fig_14}(a) and (b)). With a fixed bottom region (uphill/downhill), the external shear in the flow direction enhances the phase velocity in both supercritical stable and subcritical unstable regions except the wavenumber $k$ close to the point of discontinuity of $\mathcal{N}_c$.

When $Re=3$, Fig.~\ref{Fig_15}(a) interprets the variation of nonlinear phase speed $\mathcal{N}_c$ with the change in $K_{\rho}$ for the downhill portion, while Fig.~\ref{Fig_15}(b) is the same for the uphill portion of the wavy bottom. Here, the function $\mathcal{N}_c$ has a point of discontinuity at $k=0.1284,~0.1197,~\text{and}~0.1099$ (at which $\Upsilon_2=0$) for the downhill portion, whereas for the uphill portion, $k=0.1138,~0.1065,~\text{and}~0.0983$ (at which $\Upsilon_2=0$) are the points of discontinuity of $\mathcal{N}_c$ when $K_{\rho}$ values are $0.5$, $0.6$, and $0.7$, respectively. For different $K_{\rho}$, the corresponding singularity points of the function $\mathcal{N}_c$ at different $Re$ ($1,~5,~\text{and}~7$) are delineated in Table.~\ref{Table_2} for both bottom portions (see, the left boundary of zone IV in Table.~\ref{Table_2}). 
The singularity point of $\mathcal{N}_c$ shifts toward the left, and the phase velocity is highly positive/negative, according to the left/right neighborhood of the singular point. That means the higher value of the parameter $K_{\rho}$ slows the phase speed in the supercritical stable zone with the fixed bottom region (downhill/uphill) and results in more stable flow, whereas a similar trend is observed in the subcritical unstable zone. Also, for all $K_{\rho}$ values, the nonlinear speed in the downhill region of the liquid film is higher than that in the uphill region (On comparing Figs.\ref{Fig_15}(a) and (b)).

Further, Fig.\ref{Fig_16}(a) represents that at the downhill portion, as the parameter $K_{\sigma}$ increases, the nonlinear wave speed $\mathcal{N}_c$ decreases in the supercritical stable region, whereas a similar scenario occurs in the subcritical unstable region. However, the change in the phase speed with respect to $K_{\sigma}$ is weak enough in the supercritical stable zone, and a significant change in the same is possible near the point of discontinuity of $\mathcal{N}_c$, at which the function $\Upsilon_2=0$. 
On the other hand, at the uphill portion of the wavy bottom, the nonlinear wave speed increases as long as $K_{\sigma}$ increases in the supercritical stable region (see, the inset plot of Fig.~\ref{Fig_16}(b)) except in the neighbourhood of the discontinuity $k-$ point of $\mathcal{N}_c$, but in the subcritical unstable region we notice an opposite trend. As earlier, for all $K_{\sigma}$ values in the supercritical zone, the higher phase speed in the downhill region than the uphill region makes the flow system more unstable at the wavy bottom's downhill portion. 

\section{Conclusion} \label{CON}
The extensive stability analysis of a two-dimensional externally shear-imposed falling film over a uniformly heated wavy wall is performed. A small fluctuation in temperature causes a linear change in the density and surface tension of the liquid film. The main goal is to analyze the linear and weakly nonlinear stability behaviour of the falling film over the wavy bottom under the combined impact of imposed shear and fluid property variation.
The temperature difference between the wavy wall and the surrounding air mainly causes the heating. Here, the sinusoidal bottom of moderate steepness is considered to study the entire results. The longwave expansion theory is utilized to obtain the Benney-type equation, while the multiple-scale approach is performed to obtain the CGLE. We have presented regions of linear stability and weakly nonlinear stability. Notably, MATLAB 2020b is used to plot all the results throughout the article.

The principal observations derived from linear stability analysis confirm that the tendency of the positive/negative external shear to have a destabilizing/stabilizing nature on the liquid surface does not depend on the portion (uphill or downhill) of the wavy bottom. The surface wave instability becomes higher in the downhill portion of the wavy bottom than in the uphill portion. The higher value of $K_{\rho}$ suppresses the growth rate of the perturbed surface and helps stabilize the surface wave. The increasing $K_{\rho}$ reduces the fluid density, which weakens the flow rate of the fluid. On the other hand, the higher $K_{\sigma}$ boosts the external shear-induced surface wave instability by advancing the growth rate. The reason for this is that a higher $K_{\sigma}$  value reduces surface tension and thus weakens the bonds between fluid molecules, making the flow system more unstable. The linear instability/stability behaviour of the parameter $K_{\sigma}/K_{\rho}$ does not depend on the bottom portion. Further, in linear theory, the undulated bottom steepness plays a double role on the uphill and downhill portions.

Moreover, the weakly nonlinear theory assures the occurrence of supercritical, subcritical, unconditional, and explosive regions. No matter whether the wavy bottom portion is uphill or downhill, the downstream-directed imposed shear enhances the nonlinear wave amplitude owing to the amplification of the supercritical stable zone, while the reverse trend is possible for sturdy upstream-directed imposed shear. In addition, the supercritical stable and explosive zones decrease as $K_{\rho}$ increases, while the unconditional stable and subcritical unstable zones expand. In contrast, the supercritical stable and the explosive regions amplify, while the unconditional stable and the subcritical unstable regions shrink for a higher $K_{\sigma}$ value.

The present study proposes that the external shear can be used as a stability/instability control option for gravity-driven film flows in relevant applications by designing the substrate to be wavy, which can be modeled as substrates with constant heat. We expect all these results will provide a versatile array of active control mechanisms for fluid flows over wavy bottom structures.
% \newpage
\section*{Appendix}
\appendix
\section{Arbitrary point $P(x,\,z)$ in the Cartesian coordinate system}\label{coordinate}
Here, for an arbitrary point $S(\hat{x},\,\hat{b}(\hat{x}))$ of the wavy bottom, a local coordinate system $e_x$, $e_y$ with $e_x$ tangential and $e_z$ normal to the bottom is considered at any point $S(\hat{x},\,\hat{b}(\hat{x}))$. 
 \setcounter{figure}{16}
 \begin{figure}[ht!]
\begin{center}
\includegraphics[width=10cm]{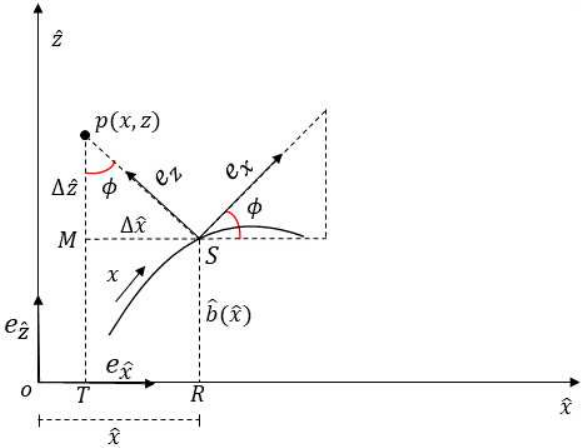}
\end{center}
	\caption{Sketch for the transformation between Cartesian and curvilinear coordinate systems. }\label{f17}
\end{figure}
Consider $P(x,\,z)$ be an arbitrary point within the fluid with $x$ as the arc length of the bottom and the distance $z$ is along $e_z$ to the bottom (i.e., $(x,z)$ is the curvilinear coordinate of the arbitrary point $P$). 
Hence, according to the Fig.~\ref{f17}, $(OT,\,TP)$ is the Cartesian coordinate of the point $P$. So,  
\begin{align}
    & OT=OR-TR=\hat{x}-\Delta \hat{x}=\hat{x}-z\sin{\phi},\\
    & TP=TM+MP=\hat{b}(\hat{x})+\Delta \hat{z}=\hat{b}(\hat{x})+z\cos{\phi}.
\end{align}
Therefore, the Cartesian coordinate of an arbitrary point $p(x,z)$ is $(\hat{x}-z\sin{\phi},\,\hat{b}(\hat{x})+z\cos{\phi})$.
As we focus on film flow over weakly undulated bottoms, this relation is always unique.
% \newpage
\section{The following zeroth-order and first-order equations of motion, along with the boundary conditions, are utilized to derive the solutions:
}\label{Equations}
\subsection{\textbf{Zeroth Order Equations}} \label{Equations11}
\allowdisplaybreaks
\begin{align}
    &u_{0x}+w_{0z}=0,\\
    &3\frac{\sin{(\theta-\phi)}}{\sin{\theta}}(1-K_{\rho}T_0)+u_{0zz}=0,\\
    & Re p_{0z}+3\frac{\cos{(\theta-\phi)}}{\sin{\theta}}(1-K_{\rho}T_0)=0,\\
    &T_{0zz}=0,\\
    &u_0=w_0=0\quad\text{and}\quad T_0=1\quad\text{at}\quad z=0,\\
    & w_0=f_t+u_0f_x \quad\text{at}\quad z=f(x,t),\\
    &T_{0z}=0 \quad\text{at}\quad z=f(x,t),\\
    &u_{0z}=\tau-Ma(T_{0x}+f_xT_{0z})\quad\text{at}\quad z=f(x,t),\\
    &p_{\infty}-p_0=\epsilon^2 We(1-K_{\sigma}T_0)(f_{xx}-\chi \kappa+\xi^2\kappa^2f)\quad\text{at}\quad z=f(x,t).
\end{align}
\subsection{\textbf{First Order Equations}}\label{Equations22}
\allowdisplaybreaks
\begin{align}
    &u_{1x}+w_{1z}+\xi\kappa~w_0+\xi\kappa~z~w_{0z}=0,\\
    &Re(u_{0t}+u_0u_{0x}+w_0u_{0z})=-Re~p_{0x}+\xi\kappa~u_{0z}+u_{1zz}-3\frac{\sin{(\theta-\phi)}}{\sin{\theta}}K_{\rho}T_1,\\
    &-Re~\xi\kappa~u_0^2=-Re~p_{1z}+w_{0zz}-3\frac{\cos{(\theta-\phi)}}{\sin{\theta}}K_{\rho}T_1,\\
    &Re~Pr\bigg(T_{0t}+u_0T_{0x}+w_0T_{0z}\bigg)=T_{1zz}+\xi\kappa~T_{0z},\\
    & u_1=w_1=T_1=0\quad\text{at}\quad z=0,\\
    &w_1=u_1f_x+\xi\kappa~f~u_0~f_x\quad\text{at}\quad z=f(x,t),\\
    &T_{1z}=0\quad\text{at}\quad z=f(x,t),\\
    &u_{1z}+\xi~\kappa\bigg(2fu_{0z}-u_0\bigg)=-\epsilon Ma(T_{1x}+f_xT_{1z})\quad\text{at}\quad z=f(x,t),\\
   &p_1=\frac{2}{Re}\bigg(w_{0z}+u_{0z}f_x\bigg)\quad\text{at}\quad z=f(x,t). 
\end{align}

% \section{Differential Operators $\mathcal{L}_0$, $\mathcal{L}_1$, and $\mathcal{L}_2$, and Nonlinear Terms $\mathcal{M}_2$ and $\mathcal{M}_3$:} \label{terms}
% \allowdisplaybreaks
% \begin{align}
% & \mathcal{L}_0=\partial_t+a_1\partial_x+b_1\partial_{x^2}+c_1\partial_{x^4}, \nonumber
% \\
% & \mathcal{L}_1=\partial_{t_1}+a_1\partial_{x_1}+2b_1 \partial _{xx_1}+4c_1 \partial_{x^3x_1}, \nonumber
% \\
% & \mathcal{L}_2=\partial_{t_2}+b_1\partial_{x^2_1}+6c_1 \partial_{x^2x^2_1},\nonumber
% \end{align}
% \allowdisplaybreaks
% \begin{align}
% & \mathcal{M}_2= a_1^{'} F_1 F_{1x}+b_1^{'} F_1 F_{1xx}+\,c_1^{'} F_1 F_{1xxxx}+b_1^{'} F^2_{1x}+c_1^{'} F_{1x} F_{1xxx} ~,\nonumber
% \\
% &\mathcal{M}_3= a_1^{'}\biggl(F_1F_{2x}+F_1F_{1x_1}+F_2F_{1x}\biggr)+b_1^{'}\biggl(F_1F_{2xx}+2F_1F_{1xx_1}+F_2F_{1xx}\biggr) \nonumber
% \\
% &+c_1^{'}\biggl(F_{2xxx}F_{1x}+3F_{1xxx_1}F_{1x}+F_{1xxx}F_{2x}+F_{1xxx}F_{1x_1}\biggr) +c_1^{'}\biggl(F_1F_{2xxxx}\nonumber\\
% &+4F_1F_{1xxxx_1}+F_2F_{1xxxx}\biggr)+b_1^{'}\biggl(2F_{1x}F_{2x}+2F_{1x}F_{1x_1}\biggr)+\frac{1}{2}a_1^{''} F^2_1 F_{1x}\nonumber
% \\
% &+\frac{b_1^{''}}{2} F^2_1 F_{1xx}+ \frac{c_1^{''}}{2} F^2_1 F_{1xxxx}+ b_1^{''} F_1 F^2_{1x}+c_1^{''} F_1 F_{1x}F_{1xxx}.\nonumber
% \end{align}
% \newpage
\section{Data set of the flow problem}
It is important to highlight the range of flow parameters in order to study the linear stability mechanism of the current flow model. In this study, different liquids are implemented in the experiment; for example, \citet{scheid2005validity} mentioned that the values of $Bi$ are $0.008$, $0.009$, $0.045$, $0.047$, and $0.02$, respectively, for water at $20^{\circ}C$, water at $15^{\circ}C$, FC-72 at $20^{\circ}C$, MD-3F at $30^{\circ}C$, and 25 ethyl-alcohol at $20^{\circ}C$ for the temperature difference $\Delta\,T=1K$ with the heat transfer coefficient $K_g=100W m^{-2}K^{-1}$ as reference values. 

\begin{table}[ht!]
\caption{Values of dimensionless parameters for current flow problem.}\label{Table4}
\vspace{0.3cm}
     \centering
\begin{tabular}{c  c  c c}
\hline
    \hline
    Dimensionless number & Symbol & Typical Values& References\\
    \hline 
    \\
    Aspect ratio  & $\epsilon$ & 0.1& \cite{sadiq2008thin} and \cite{chattopadhyay2021thermocapillary}
    \\
    Bottom steepness  & $\xi$ & $0-0.4$& \cite{mukhopadhyay2020hydrodynamics} and \cite{mukhopadhyay2020waves} 
      \\
      Reynolds number & Re & $0-10$& \cite{mukhopadhyay2020hydrodynamics}, \cite{hossain2022linear} and \cite{mukhopadhyay2021thermocapillary}
      \\
      Weber number & $We$ &  $450$& \cite{mukhopadhyay2020stability} and \cite{mukhopadhyay2020hydrodynamics} \\
      External shear force & $\tau$ & $-1\, - \,1$& \cite{hossain2022shear}, \cite{hossain2022linear}, \cite{bhat2019linear} and \cite{samanta2014shear}
      \\
      Surface tension variation w.r.t temperature&$K_{\sigma}$& $0.5\,-\,0.7$&\cite{chattopadhyay2021odd} and \cite{pascal2013long}\\
      Density variation w.r.t temperature&$K_{\rho}$& $0.5\,-\,0.7$&\cite{chattopadhyay2021odd} and \cite{pascal2013long}\\
      \\
      \hline
      \hline
    \end{tabular}
\end{table}
\begin{table}[ht!]
\caption{Typical values for water/silicon oil 50 cS.}\label{Table_5}
\vspace{0.3cm}
     \centering
\begin{tabular}{c  c  c }
\hline
    \hline
    Dimensionless number & Symbol & Typical Values\\
    \hline 
    Aspect ratio  & $\epsilon$ & 0.01
    \\
    Bottom steepness  & $\xi$ & $0-0.4$ 
      \\
      Reynolds number & Re & $0-10$
      \\
      Inverse Bond number & $Bo$ &  $0-16.2$\\
      Surface tension variation w.r.t temperature & $K_{\sigma}$ & $0-0.006$\\
      Density variation w.r.t temperature & $K_{\rho}$ & $0-0.006$\\
      Prandtl number & $Pr$ & $0-7.1$
      \\
      \hline
      \hline
    \end{tabular}
\end{table}
A fixed inclination angle $\theta=60^{\circ}$ is considered for our analysis. Moreover, the surface tension and fluid density are assumed to have declined for the temperature with the scaled formulas $1-K_{\sigma}T$ and $1-K_{\rho}T$, respectively. Since the scaled temperature difference, $T$, attains a maximum value of 1, in order for these quantities to be positive, the parameters $K_{\sigma}$ and $K_{\rho}$ must be in $[0,\,1)$. The other physical parameters that are selected for our study are summarized in Table.~\ref{Table4}.

 \begin{figure}[ht!]
\begin{center}
\includegraphics[width=7.2cm]{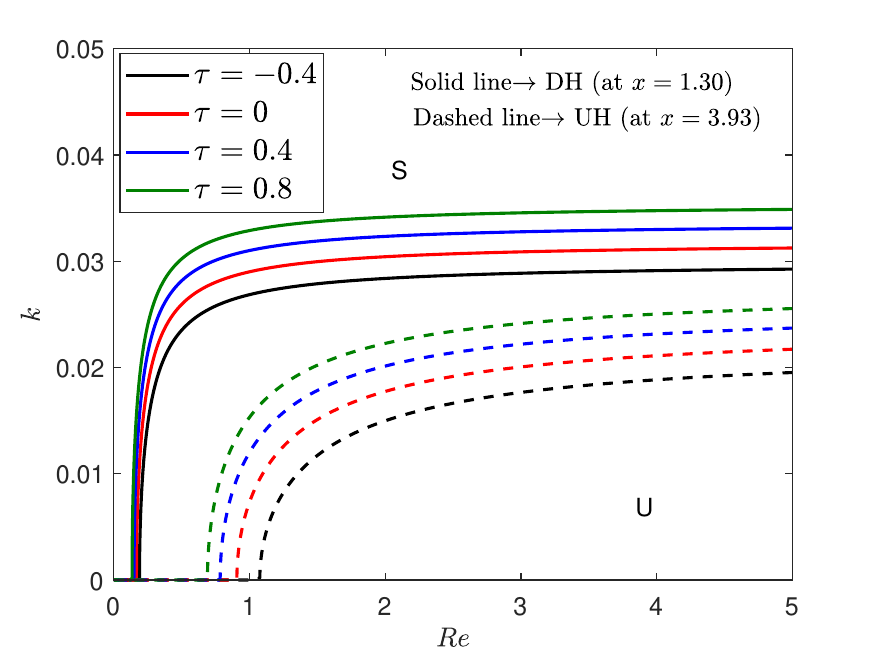}
\end{center}
	\caption{Stability boundaries in ($Re-k$) plane for different $\tau$ with uphill (UH: $x=3.93$) and downhill (DH: $x=1.30$) with  $We=4500$, $\xi=0.1\pi$, $\theta=60^{\circ}$, and $\epsilon=0.01$. }\label{f20}
\end{figure}
\begin{figure}[ht!]
\begin{center}
\subfigure[$\tau=-0.8$]{\includegraphics*[width=5.4cm]{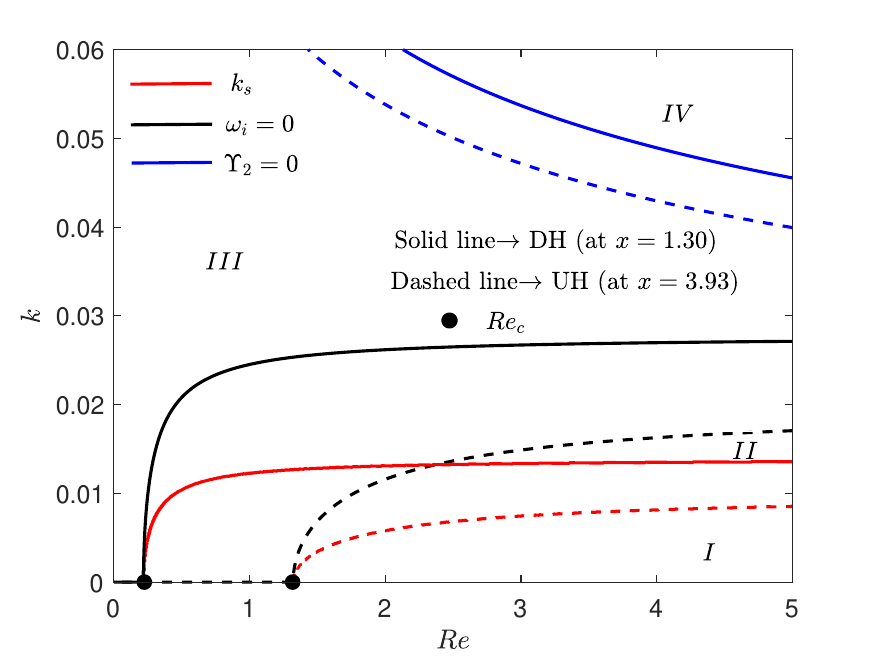}}
\subfigure[$\tau=0$]{\includegraphics*[width=5.4cm]{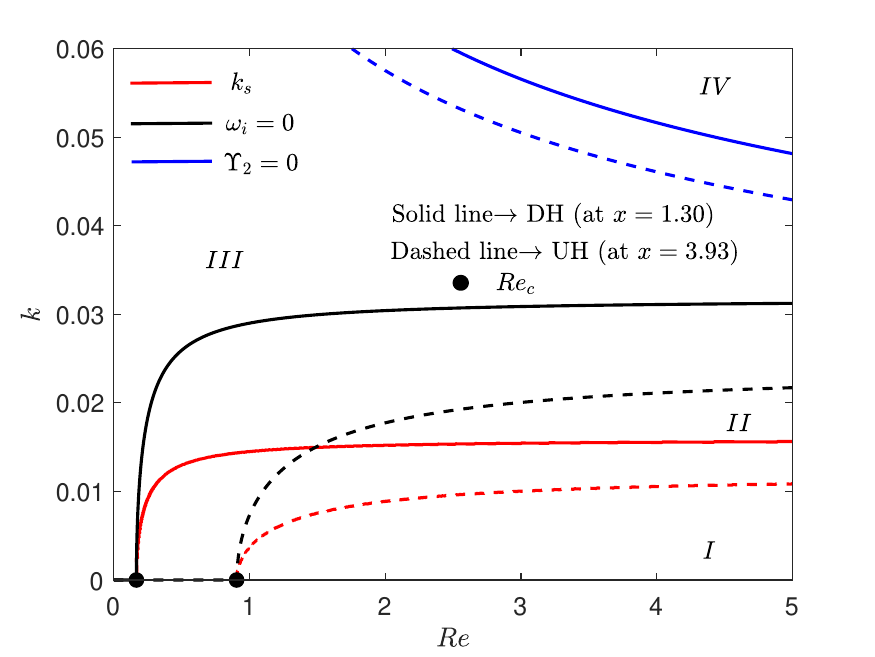}}
\subfigure[$\tau=0.8$]{\includegraphics*[width=5.4cm]{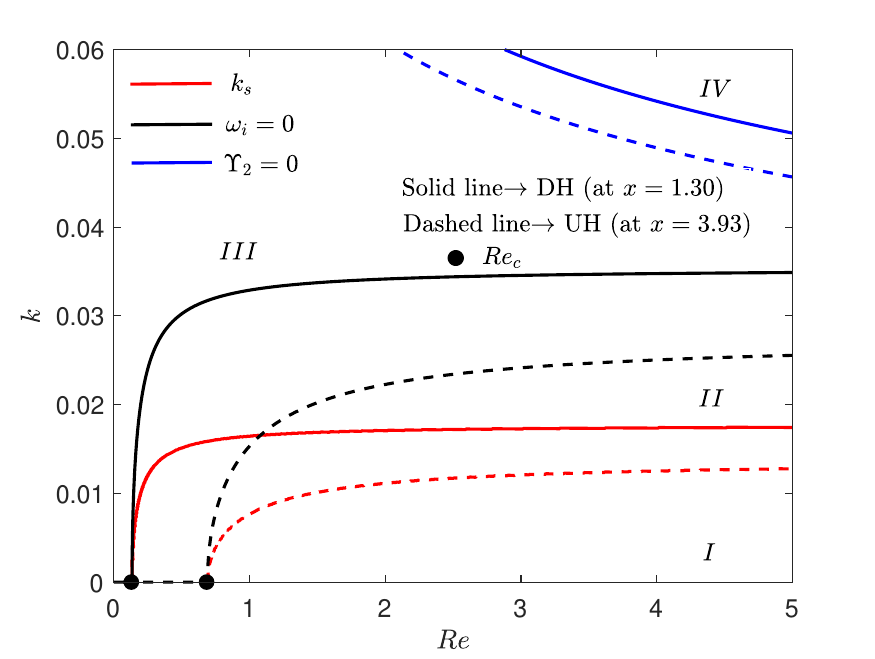}}
\end{center}
\caption{The effect of $\tau$ on different flow regions for uphill (UH: $x=3.93$) and downhill (DH: $x=1.30$) portions. The different flow regions $I: \text{supercritical unstable}~ \omega_i>0,\,\Upsilon_2<0$, $II:\text{supercritical stable}~\omega_i>0,\,\Upsilon_2>0$, $III:\text{subcritical stable}~\omega_i<0,\,\Upsilon_2>0$, and $IV:\text{subcritical unstable}~\omega_i<0,\,\Upsilon_2<0$ at a point on the uphill (UH) and downhill (DH) portions. The other fixed parameters are $\xi=0.1\pi$, $We=4500$, and $\theta=60^{\circ}$, and $\epsilon=0.01$. }\label{Fig_22}
\end{figure}
Further, to make our study realistic, we have taken typical values for the relevant physical constants for water/silicon oil $50$ cS as follows: acceleration due to gravity $g = 9.8\, ms^{-2}$, inclination angle $\theta=60^{\circ}$, density $\rho_{\infty}=10^3 kg\,m^{-3}$, kinematic viscosity $\nu=1.12\times10^{-6}\,m^2s^{-1}$, thermal conductivity $K_{T}=0.15\,Wm^{-1}K^{-1}-0.6\, Wm^{-1}K^{-1}$, specific heat at constant pressure $C_p=4182\, J\,kg^{-1}K^{-1}$, surface tension $\sigma_{\infty}=3.6\times 10^{-2}-7.34\times 10^{-2}\,Nm^{-1}$, thermal surface tension coefficient $\gamma=5\times10^{-5}\,Nm^{-1}K^{-1}$. Follow \citet{yeo2003marangoni} for the real physical data. The estimated values of the relevant parameters are given in Table.~\ref{Table_5}. 

For the values $\tau$, we have followed \citet{bhat2019linear} and have considered the values within the range $[-1,\,1]$. 
Now, we have plotted results (see, Fig.~\ref{f20}) related to linear stability for the real fluid flow data. The aim is to check whether the current analysis is valid for water/silicon oil 50 cS or not.  

Fig.~\ref{f20} confirms that the positive external shear ($+\tau$) has a destabilizing effect on the liquids with real physical data (see, Table.~\ref{Table_5}) with their respective orders. Moreover, different instability boundary lines are demonstrated in Fig.~\ref{Fig_22} for the data set in Table.~\ref{Table_5} of the realistic liquid. The external shear has a similar impact, as in Fig.~\ref{Fig_8}, on different flow zones (see, Fig.~\ref{Fig_22}) in the vicinity of the instability threshold. Thus there is no major change in the instability control effect of external shear force on wave formation physics. Hence, to understand a broad idea about the dynamics and instability mechanisms of falling film over an undulated bottom, one can use an externally imposed shear in an open-type flow as a great option for instability control of surface waves.

\section{Experimental validation of the flow model}
In the current flow problem, the critical Reynolds number $Re_c=\frac{5}{6}\cot{\theta}$ when the fluid flows over the non-heated inclined plane (i.e., $\xi$, $K_{\rho}$, and $K_{\sigma}\rightarrow0$) with negligible external shear $\tau$ at the liquid surface. 
\begin{figure}[ht!]
\begin{center}
\includegraphics[width=10cm]{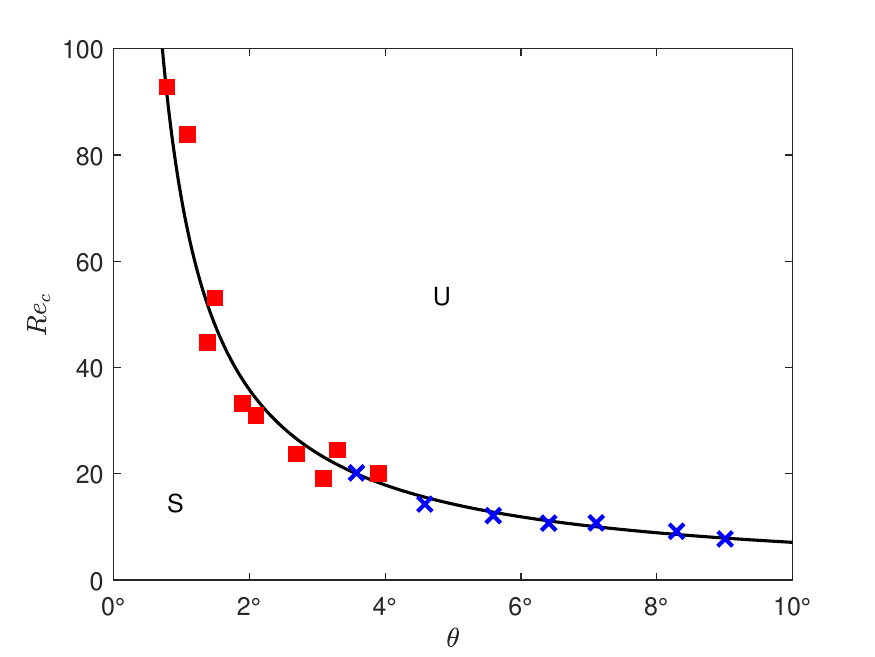}
\end{center}
	\caption{Critical Reynolds number $Re_c$ as a function of angle of inclination $\theta$ when $\xi\rightarrow0$, $\tau\rightarrow0$, $K_{\sigma}\rightarrow0$, and $K_{\rho}\rightarrow0$. Red solid squares mark the result for water and the blue cross symbols denote the result for glycerin-water solutions (see, the work of \citet{liu1993measurements}).}\label{f19}
\end{figure}
In Fig.~\ref{f19}, we have plotted $Re_c$ as a function of $\theta$ for the maximum base velocity $\displaystyle\,U_c=\frac{\rho g \sin{\theta \hat{f}^2}}{2\mu}$, which is $2/3$ times of the
free surface velocity $U_c$ that is considered in the current study. Also, the experimental data of $Re_c$ obtained by \citet{liu1993measurements} for falling liquid film over an inclined plane for both pure water and glycerin-water solution are demonstrated in Fig.~\ref{f19}. It can be concluded from Fig.~\ref{f19} that the physical approximations made in the formulation of the current stability theory are in good agreement with the experimental data derived by \citet{liu1993measurements}.

\section{Validation of weakly nonlinear stability analysis}
\begin{figure}[ht!]
\begin{center}
\includegraphics[width=10cm]{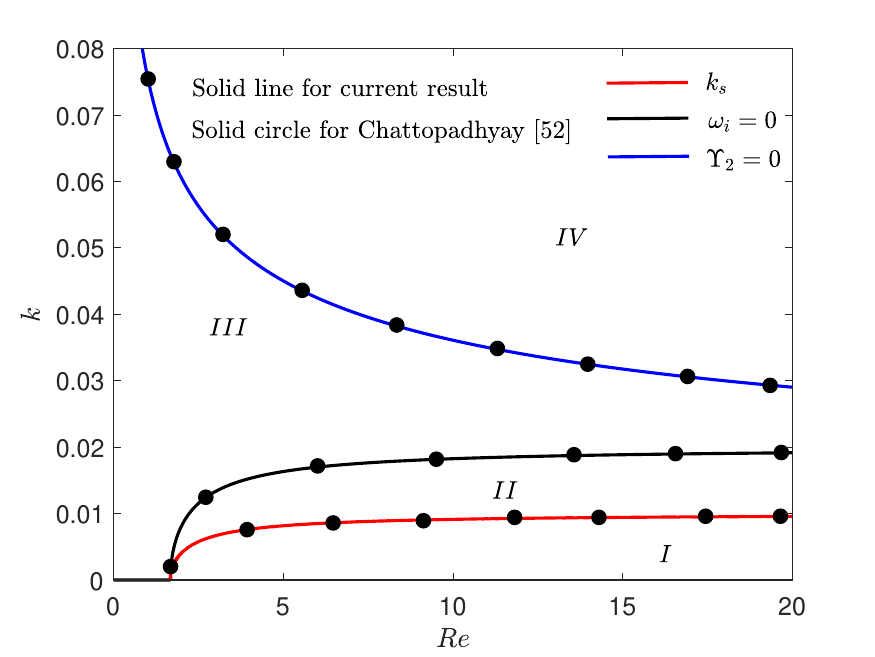}
\end{center}
	\caption{Different flow regions: $I: \text{supercritical unstable}~ \omega_i>0,\,\Upsilon_2<0$, $II:\text{supercritical stable}~\omega_i>0,\,\Upsilon_2>0$, $III:\text{subcritical stable}~\omega_i<0,\,\Upsilon_2>0$, and $IV:\text{subcritical unstable}~\omega_i<0,\,\Upsilon_2<0$. The other fixed parameters are $\xi=0$, $\tau=0$, $K_{\sigma}=0.5$, $We=4496$, and $\theta=45^{\circ}$, and $\epsilon=0.1$. The black solid circles are the results of \citet{chattopadhyay2021odd} (see, Fig.~4(a) of their paper). }\label{f18}
\end{figure}

To validate the results of weakly nonlinear stability analysis, we have compared our results, as shown in Fig.~\ref{f18}, with the results of odd-viscosity-induced falling film over a uniformly heated inclined plane with variable density (\citet{chattopadhyay2021odd}). The different flow regions near the instability threshold are plotted for the parameter values $\xi=0$, $\tau=0$, $K_{\sigma}=0.5$, $K_{\rho}=0.5$, $We=4496$, $\theta=45^{\circ}$, and $\epsilon=0.1$. It is noteworthy to point out that the boundary lines of instability match well with the result of \citet{chattopadhyay2021odd} when the odd viscosity parameter is negligible.

 \newpage
 \section*{Declaration of interests}\vspace{-0.25cm}
The authors report no conflict of interest.
% \textbf{Declaration of interests:} The authors report no conflict of interest.

\section*{Data Availability}
 The data that supports the findings of this study are available
within the article, highlighted in the related figure captions and corresponding discussions.

\bibliographystyle{unsrtnat}
\bibliography{REF}
\end{document}